\documentclass{nature}
\usepackage{hyperref}
\usepackage{epsfig}
\usepackage{color}
\usepackage[T1]{fontenc}
\usepackage{mathtools}
\usepackage{setspace}
\usepackage{graphicx,subcaption}
\usepackage{caption}
\usepackage{subcaption}
\usepackage{tocloft}
\usepackage{etoc}       
\usepackage{amsfonts}
\usepackage{amssymb}
\usepackage{amsmath}

\usepackage[super,comma,sort&compress]{natbib}

\usepackage{hyperref}
\usepackage{verbatim}
\usepackage{float}
\usepackage{placeins}
\usepackage{graphicx}

\def\mearth{{\rm\,M_\oplus}}

\renewcommand\contentsname{\huge Supplementary Information}

\newcommand{\beginextendeddata}{%
        \setcounter{table}{0}
        \renewcommand{\thetable}{Extended Data \arabic{table}}%
        \setcounter{figure}{0}
        \renewcommand{\thefigure}{\arabic{figure}}%
        \renewcommand{\figurename}{Extended Data Figure}%
     }

\newcommand{\beginsupplement}{%
        \setcounter{table}{0}
        \renewcommand{\thetable}{Supplementary \arabic{table}}%
        \setcounter{figure}{0}
        \renewcommand{\thefigure}{\arabic{figure}}%
        \renewcommand{\figurename}{Supplementary Figure}%
     }

\title{Planetesimal rings as the cause of the Solar System's planetary architecture}

\author{Andre Izidoro$^{1}$, Rajdeep Dasgupta$^{1}$, Sean N. Raymond$^{2}$,  Rogerio Deienno$^{3}$,  Bertram Bitsch$^{4}$, and Andrea Isella$^{5}$}

\begin{document}

\maketitle

\begin{affiliations}
 \item Department of Earth, Environmental and Planetary Sciences, MS 126,  Rice University, Houston, TX 77005, USA
 \item Laboratoire d'Astrophysique de Bordeaux, Univ. Bordeaux, CNRS, B18N, all{\'e}e Geoffroy Saint-Hilaire, 33615 Pessac, France
 \item Southwest Research Institute, 1050 Walnut St. Suite 300, Boulder, CO 80302, USA
 \item Max-Planck-Institut für Astronomie 
Königstuhl 17, 69117 Heidelberg, Germany
\item Department of Physics and Astronomy, Rice University, Houston, TX 77005, USA
\end{affiliations}

\begin{abstract}

Astronomical observations reveal that protoplanetary disks around young stars commonly have ring- and gap-like structures in their dust distributions. These features are associated with pressure bumps trapping dust particles at specific locations, which simulations show are ideal sites for planetesimal formation. Here we show that our Solar System may have formed from rings of planetesimals -- created by pressure bumps -- rather than a continuous disk. We model the gaseous disk phase assuming the existence of pressure bumps near the silicate sublimation line (at $T \sim$1400~K), water snowline (at $T \sim$170~K), and CO-snowline (at $T \sim$30~K). Our simulations show that dust piles up at the bumps and forms up to three rings of planetesimals: a narrow ring near 1~au, a wide ring between $\sim$3-4~au and $\sim$10-20~au, and a distant ring between $\sim$20~au and $\sim$45~au. We use a series of simulations to follow the evolution of the innermost ring and show how it can explain the orbital structure of the inner Solar System and provides a framework to explain the origins of isotopic signatures of Earth, Mars and different classes of meteorites. The central ring contains enough mass to explain the rapid growth of the giant planets' cores.  The outermost ring is consistent with dynamical models of Solar System evolution proposing that the early Solar System had a primordial planetesimal disk beyond the current orbit of Uranus.

\end{abstract}


 The parent bodies of non-carbonaceous (NC) and carbonaceous  (CC) chondrite meteorites are associated with classes of asteroids with overlapping orbital distributions~\citep{demeocarry14} and accretion ages, but different bulk chemical and isotopic compositions~\citep{kruijeretal20}. NC and CC-like meteorites have water-poor and water-bearing compositions, which may suggest that they sample parent bodies that accreted inside and outside the snowline, respectively. They also show distinct stable-isotopic compositions for several volatile and non-volatile chemical elements~\citep{kruijeretal20,grewaletal21}. This isotopic `dichotomy' implies that the mm-sized constituents of meteorites -- {\em pebbles} -- did not mix during the early accretion stages of planet formation,\citep{kruijeretal20} but their parent bodies did.  The parent bodies of NC and CC meteorites must have accreted from distinct dust and pebble reservoirs that remained disconnected for 2-4 million years~\citep{kruijeretal20,brassermojzsis20}. This is challenging given the predominant view that pebbles drift (mainly inward) quickly through the gas\citep{birnstieletal12} such that mixing of different reservoirs is unavoidable\citep{brassermojzsis20}. Yet after their assembly, asteroid-sized bodies must have dynamically mixed~\citep{walshetal11,raymondizidoro17a}.

A possible missing ingredient comes from high-resolution ALMA observations revealing that rings and gaps are ubiquitous structures in protoplanetary disks~\citep{huangetal18}.  Ring-like structures suggest that pebbles have been concentrated in pressure bumps of the disk~\citep{dullemondetal18}. At pressure bumps, the gas is locally accelerated, acting to trap drifting pebbles and increase the local solid-gas ratio, thus facilitating planetesimal formation \cite{johansenetal07}. The existence of pressure bumps at different locations of protoplanetary disks is also supported  by state-of-art disk models\citep{mulleretal21,charnozetal21}.

Here we propose a model of Solar System formation that accounts for the isotopic dichotomy and  provides a link between the sun's natal disk and the structure of young disks around other stars. We first show that, if pressure bumps exist at important condensation fronts within the gaseous disk, planetesimals form in rings rather than a continuous disk.  We use a suite of numerical simulations coupling different stages of planet formation -- from dust coagulation to the final accretion phase of terrestrial planets --  to demonstrate that the inner Solar  System's orbital architecture and meteorite data are consistent with the existence of up to three pressure bumps in the sun's natal disk. Our model accounts for the Solar System's isotopic dichotomy, the formation of the terrestrial planets, and the orbital distribution of NC and CC-like parent asteroids.

We model planetesimal formation by simulating dust coagulation, fragmentation, and turbulent mixing in a young protoplanetary disk. We model dust evolution by solving the 1D advection-diffusion equation for the dust column density\cite{birnstieletal12}.  Micrometer-sized dust grains grow to  millimeter and centimeter sized pebbles and start drifting inwards due to gas drag\citep{birnstieletal12}. As pebbles drift via drag they may also fragment. The local maximum pebble size depends on the local gas disk and grain properties, specifically  the level of turbulence in the midplane, gas density, and fragmentation velocity of icy and silicate pebbles ~\cite{birnstieletal12,gundlachblum}. We solve the advection-diffusion equation  for the largest (pebbles) and initial/smallest dust grains, assuming that the dust mass is mostly carried by largest grains~\cite{birnstieletal12}. Pebbles inside the water snowline have silicate composition and   smaller sizes relative to icy pebbles beyond the water snowline (see Methods). This difference in sizes is imposed to reflect different  threshold fragmentation velocities of ice-rich and silicate pebbles\cite{gundlachblum} (see Supplementary Information for numerical tests of the importance of this parameter). In our model, silicate pebbles that reach regions of the gas disk hotter than 1400~K are completely lost by sublimation. Ice-rich pebbles that cross the disk water snowline lose their ice, assumed to correspond to 50\% of their bulk mass, releasing smaller silicate dust grains in the disk.  Our gas disk model includes three (or two) pressure bumps. We model the gaseous protoplanetary disk by using radial power-law functions to represent the gas surface density ($\Sigma_{\rm gas}\propto r^{-x}$, where we assume $x=1$ or $x=1.5$) and temperature profiles ($T_{\rm gas}\propto r^{-\beta}$), where we set $\beta
\approx0.7-1.0$. We mimic the presence of pressure bumps by rescaling the gas surface density by gaussian and hyperbolic tangent functions at specific disk locations. The inner bump is associated with a transition in the gas disk's viscosity due to thermal ionization of the gas at temperatures higher than $T_{gas}>$1000~K~\cite{deschturneretal15,flocketal17}. It is located slightly outside the silicate sublimation line ($T_{gas}\sim 1400$~K). The central and outer pressure bumps are associated with  the gas disk water snowline at $T_{gas}\sim$170~K and CO-snowline at $T_{gas}\sim 30$~K, respectively.  The origin of pressure bumps at snowlines is usually associated with transition in the disk opacity due to different pebble sizes inside and outside the snowlines, and sublimation/recondensation of grains at these locations~\cite{pinillaetal12,mulleretal21,charnozetal21}. Pressure bumps may also be associated with long-lived disk zonal flows\cite{dittrichetal13}, although their location may not necessarily correlate with snowlines.

 \begin{figure}
\centering
\includegraphics[scale=1.2]{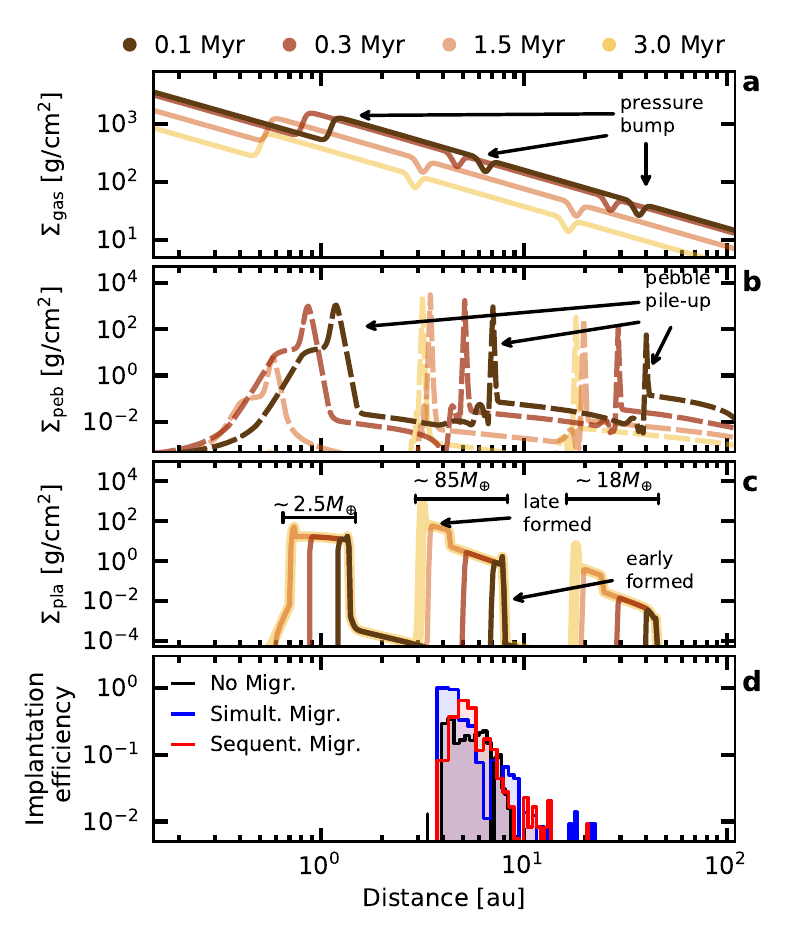}
\caption[]{ Simulated evolution of the Sun's planet-forming disk, assuming the existence of three pressure bumps: near the silicate sublimation line (at the transition between high- and low viscosity regions of the disk), at the water snowline, and at the CO snowline.  Panels a-c show snapshots of the radial distribution of gas (a), pebbles (b), and planetesimals (c). Panel (c) shows that the planetesimal production fronts (pressure bumps) move inwards as the disk cools in time. In this particular simulation we assume that Jupiter must have reached a critical mass  to block the pebble flux by $\sim$2~Myr\cite{kruijeretal20} (see Methods), avoiding the early overlapping of the inner and middle rings.  Panel (d) shows the implantation efficiency of planetesimal as a function of initial orbital distance from the Sun for three end-member scenarios modeling the growth and migration histories of the giant planets in the Solar System\citep{raymondizidoro17a}. These formation scenarios correspond to cases where Jupiter and Saturn grow near their current locations (No migration; black), grow at $\sim$10~au and $\sim$15~au and migrate inwards simultaneously (Simult. Migr.; blue), and grow at $\sim$10~au and $\sim$15~au and migrate inwards sequentially (Sequent. Migr.; red). The planetesimal formation efficiency in this simulation is $\epsilon\approx1.5\times10^{-6}$. The initial dust-to-gas ratio is $Z_0=1.5\%$, the turbulence level in the disk midplane is set by $\alpha_t\approx3.6\times10^{-5}$, and $\alpha_{\rm MRI}=3\alpha_{\nu}$.  Panel (c) also shows the total mass in planetesimals formed in each ring (see Supplementary Information for a discussion). The gas density cutoff radius (see Methods) is set $r_{\rm c}=\infty$.}
\label{fig:lowefficiency}
\end{figure}

As expected,\cite{johansenetal07,dullemondetal18} planetesimals naturally form at pressure bumps from pileups of drifting pebbles (Fig.~\ref{fig:lowefficiency}a-c). In our model, planetesimal formation takes place if the local dust-to-gas midplane density ratio becomes larger than unity~\cite{johansenetal07}, or if  the pebble flux is high enough to promote pebble concentration via assistance of zonal flows in the gas disk\citep{dittrichetal13} (Extended Data Figure \ref{fig:threebumps_examepl2}).  The pressure bump at the snowline works as an efficient barrier for icy pebbles drifting from beyond the snowline, disconnecting the inner and outer Solar System\citep{izidoroetal21} and promoting planetesimal formation. Only dust originally inside the snowline can contribute to planetesimals formation in the inner disk\cite{izidoroetal21}. The pressure bump near the silicate sublimation line  -- at the interface between weakly and strongly ionized regions of the disk -- is less efficient at trapping pebbles because they are smaller than icy pebbles and can more easily diffuse through and sublimate when they reach regions of the disk hotter than 1400~K. Yet a fraction of pebbles is trapped at the inner bump and converted into planetesimals.  As the disk evolves and cools, the  pressure bumps move inward producing planetesimals from the outside-in, forming three rings of planetesimals. The first planetesimals to form appear at $t<$100~kyr, near 1.5~au, 8~au, and 45~au. As the inner bump moves inwards, planetesimal are produced from 1.5~au down to $\sim$0.7~au (the {\em inner ring}). The bumps at the water and CO snowlines produce wider rings of planetesimals extending from $\sim$3~au to 8~au (the {\em central ring}), and $\sim$20~au to 45~au (the {\em outer ring}). Planetesimal formation in the inner disk ceases at $\lesssim$0.8 Myr (see Supplementary Information for additional examples) because most pebbles originally in the inner parts of the disk are lost by inward radial drift. Planetesimal formation  beyond the snowline proceeds while the pebble supply lasts  and the gas is still around (see Supplementary Information for a discussion about the remaining mass in pebbles beyond the pressure bumps);  we stop the simulation at 3~Myr. From $\sim$2-3~Myr to the end of the gas disk phase at $\sim$5~Myr, growing and migrating giant planet cores  would probably strongly influence planetesimal formation in the outer rings\citep{kruijeretal20}, a process that we do not model here.  The latest formed planetesimals  in the central ring forms near $\sim$3-5~au. Meteorite constraints forbid the inner and middle rings from overlapping\citep{kruijeretal20}, assuming these to be associated with NC and CC planetesimals, respectively. We rule out scenarios where the water snowline moves inwards rapidly reaching regions inside $\sim$1-3~au early in the disk history (e.g. before Jupiter's core formation; we discuss this issue in the Supplementary Information).

Our simulations systematically produce two (Extended Data Figure \ref{fig:twobumps_nominal})  or three rings of planetesimals (Figure \ref{fig:lowefficiency}). The total mass and radial distribution of planetesimal in each ring depend on parameters such as the planetesimal formation efficiency ($\epsilon$), the timing of pressure bump formation, gas viscosity in the magnetorational active zone of the disk ($T_{gas}>1000$~K\cite{deschturneretal15,flocketal17}), level of turbulence in the gas disk midplane, initial dust-to-gas ratio ($Z_0$), gas disk width, and disk surface density and temperature profiles (see Supplementary Information). In Figure \ref{fig:lowefficiency}, the inner ring contains a total of 2.5~${\rm M_{\oplus}}$ in planetesimals whereas the central ring carries $\sim$85~${\rm M_{\oplus}}$. The outer ring has a total of 18~${\rm M_{\oplus}}$ in planetesimals. If silicate pebbles are too small (e.g. when the level of turbulence in the disk midplane parameterized via $\alpha_{\rm t}$ is sufficiently high; $\alpha_{\rm t} \gtrapprox  10^{-4}-10^{-3}$) or the inner pressure bump is too weak (e.g. when the region where $T_{\rm gas}\geq1000~K$ is not sufficiently ionized and the gas $\alpha$-viscosity at this location is $\alpha_{\rm MRI} \approx \alpha_{\nu} \approx 10^{-3}$, where $\alpha_{\nu}$ corresponds to the viscosity where $T_{\rm gas}<1000~K$;  see Methods), pebbles may not pile up sufficiently in the inner pressure bump and, consequently, the efficiency of planetesimal formation is dramatically reduced in the inner ring. On the contrary, if silicate pebbles are  too large or if the inner bump is too strong, all  pebbles inside the snowline may be converted into planetesimals at the inner pressure bump. The initial gas disk temperature also strongly controls the location of each planetesimal ring. In our preferred scenario, the initial disk temperature  is typically $\sim$1000~K at 
$\sim$1-1.3~au, which sets the initial location of the inner pressure bump. An initially colder disk where the gas temperature is 1000~K at $\sim$0.1~au would lead to a planetary system unlike the Solar System, if planetesimals can efficiently form in the inner pressure bump~\cite{izidoroetal21}. The total mass in planetesimals in each ring also depends on the assumed planetesimal formation efficiency. Our preferred scenarios require planetesimals formation efficiencies varying from 
$\sim10^{-6}$ to $\sim10^{-4}$, which is consistent with values assumed in previous studies~\cite{drazkowskaalibert17}.

We now model the evolution of the Solar System starting from a system with three rings of planetesimals like the one from Fig~\ref{fig:lowefficiency}.  We consider constraints from the terrestrial and giant planets, the asteroid belt, and meteorite measurements.  We also use the inner Solar System's orbital architecture to further constrain the properties of the initial planetesimal rings.  Our investigation proceeds from the inside out, from the growth of the terrestrial planets to the asteroid belt, giant planets and outer Solar System.

We model the growth of planetary embryos within the inner ring of planetesimals, assuming planetesimals to be originally born with a characteristic diameter of $\sim$100~km\cite{simonetal16}.  We compute planetary growth via pebble and planetesimal accretion using semi-analytical prescriptions calibrated from the results of N-body numerical simulations\citep{chambers06,izidoroetal21}. We feed our calculations with the evolving pebble flux and planetesimal surface density from our dust coagulation simulations (e.g. Figure \ref{fig:lowefficiency}) and follow the growth of planetesimals in simulations producing inner rings with different total mass in planetesimals.  Our simulations show that in the inner ring  where the total mass in planetesimals is $\sim$2.5$M_{\oplus}$ (e.g. Figure \ref{fig:lowefficiency}), planetesimals grow to roughly Mars-mass planetary objects in $\sim$1-3~Myr (see Supplementary Information). These planetary embryos grow slowly and are unlikely to migrate substantially via type-I migration\citep{izidoroetal21}. In more massive disks the inner rings produce more planetesimals; for instance, a disk with a higher pebble flux (or  higher planetesimal formation efficiency) produces an inner ring of 20$M_{\oplus}$ and forms Earth-mass planets in less than $\sim$0.5~Myr (see Supplementary Information). Such massive planets would migrate quickly\citep{tanakaetal02} and likely reach the innermost parts of the disk\cite{lambrechtsetal19,izidoroetal21}. For this reason, simulations producing inner rings containing more than a few Earth-masses in planetesimals are not consistent with the current Solar System. Our simulations also show that the growth of terrestrial planetary embryos in  rings around $\sim$1~au  takes place via planetesimal accretion rather than pebble accretion, regardless of the ring's total mass in planetesimals (see Supplementary Information). The contribution from pebble accretion to the growth of protoplanetary embryos in the inner ring is negligible. This is because of the  pressure bump at the snowline, which efficiently disconnects the outer and inner disk~\cite{izidoroetal21}. Pebbles in the inner disk (inside the snowline) are lost via radial drift before they can be efficiently accreted by growing planetesimals at $\sim$1~au\citep{izidoroetal21}.

\begin{figure}
\centering
\includegraphics[scale=0.8]{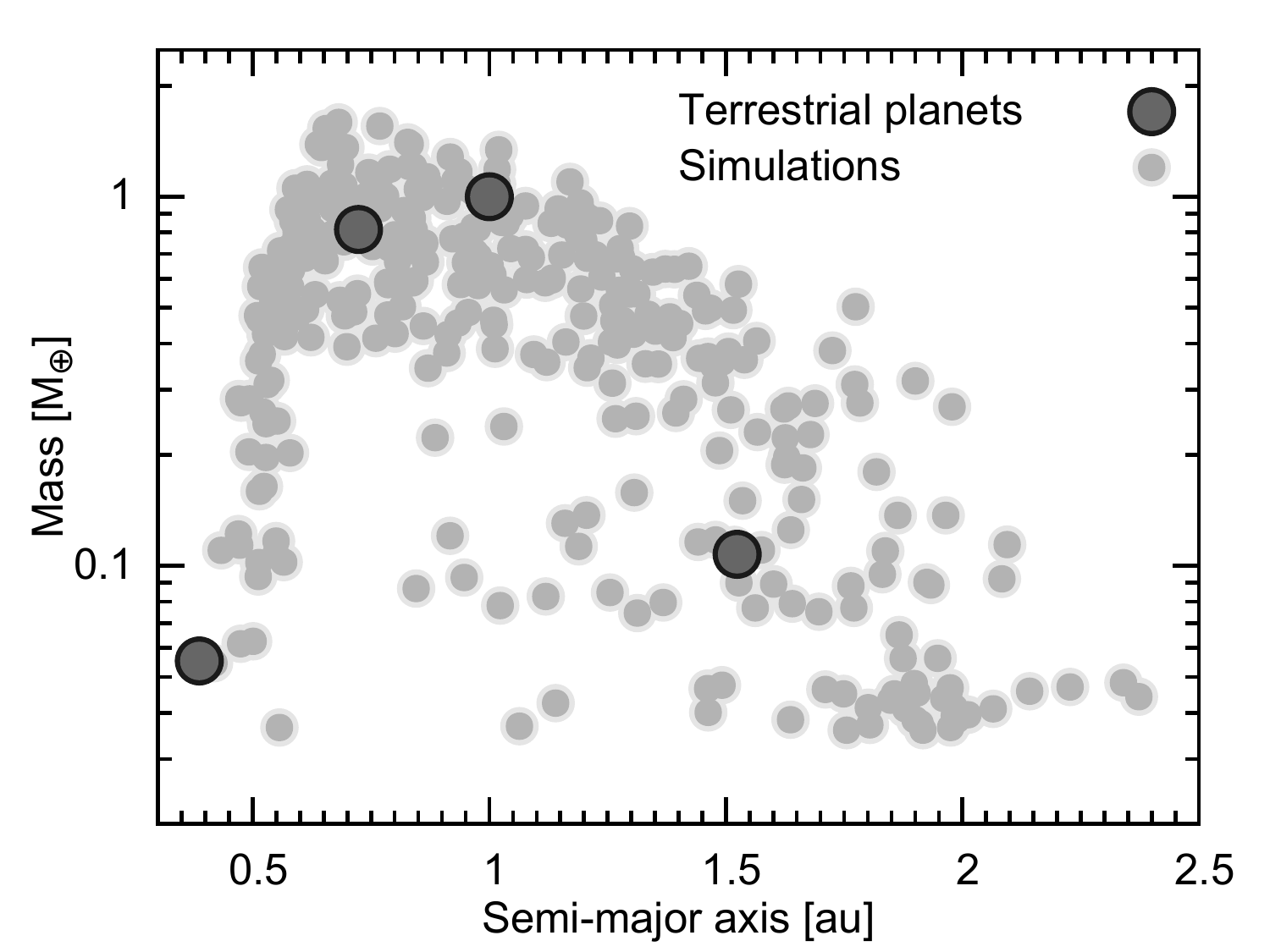}
\caption{Final distribution of terrestrial planets produced from 80 simulations modeling the late stage of accretion of terrestrial planet formation from a ring-like distributions of planetary objects  around 1~au. The vertical axis shows mass and the horizontal axis planet's orbital semi-major axis. Light-grey dots show simulated planets while dark-grey show the real terrestrial planets. In these simulations, Jupiter and Saturn are considered fully formed and in resonant and almost circular and coplanar orbits\cite{morbidellicrida07}.}
    \label{fig:terrestrial}
\end{figure}

We performed N-body simulations of late-stage terrestrial accretion in the inner ring. The ring was assumed to extend from 0.7~au to 1.5~au and contain $\sim$2.5~$M_{\oplus}$ (as in Figure \ref{fig:lowefficiency}) in planetesimals and planetary embryos following different surface density profiles, which accounts for potential different structures of the inner ring produced in our simulations.  Figure \ref{fig:terrestrial} shows the terrestrial planets that formed after 200~Myr of integration produced in 80 simulations with slightly different initial conditions (see Supplementary Information). The distribution of terrestrial planets in our simulations is in good agreement with the masses and orbits of the terrestrial planets in the Solar System. Our results are also consistent with  previous formation models suggesting that the low-mass of Mars relative to the Earth's mass reflects Mars' formation in a mass-depleted region\cite{hansen09}. Previous simulations\citep{hansen09,izidoroetal15b,levisonetal15b,raymondizidoro17b} had adopted ad-hoc initial conditions, generally assuming that the asteroid belt was born mostly empty (the {\em low-mass asteroid belt} model).  Here we have demonstrated the validity of this model from the earliest phases of the disk to planetesimal and terrestrial planet formation.  Our model also addresses another Solar System mystery: the lack of planets inside Mercury's orbit. In our nominal simulation, the inner ring of planetesimals  has a sharp edge at about 0.7~au due to sublimation of silicate pebbles at temperatures higher than 1400~K preventing planetesimal formation inside $\sim$0.7~au\cite{morbidellietal16,izidoroetal21}.

Our model sheds lights on the isotopic dissimilarities between Earth and Mars. Martian meteorites show distinct isotopic signatures compared to Earth for several elements as O, Ti, Ni, Mo, Nd, Cr, and V \citep{warrenetal11,dauphasetal14,wittmannetal15}. Difference in  mass-independent isotopic compositions between Earth and Mars suggest that these two planets accreted from broadly distinct materials from the sun's natal disk\citep{lodders00}. We investigate the source regions of protoplanetary objects constituting our final simulated planets in Figure \ref{fig:terrestrial}.  Figure \ref{fig:feeding} shows that, in our best Solar System analogues (see Supplementary Information), Earth and Venus analogues have broadly similar feeding zones, with most of their constituents (by mass) originating inside 1~au. In contrast,  on average, only $\sim$30\% of the mass of our Mars-analogues comes from inside 1~au, with some Mars-analogues accreting entirely from material originally beyond 1.2~au. Although there is some variation in the feeding zones of Mars-analogues, with some analogues also accreting  mass from the inner parts of the inner ring~\citep{hansen09}, these planets generally incorporate more mass from the outer parts of the inner ring (see Extended Data Figure \ref{fig:feeding_2rings} and Supplementary Information).  We therefore propose that the origin of the distinct compositions of Earth and Mars is linked with how planetesimals form in the inner ring. Planetesimal formation in the inner ring starts at $\sim$1.5~au.  As the disk cools, and the inner bump moves towards the sun, the planetesimal production front also moves inwards. The moving inner bump is continuously fed by drifting pebbles from the outer parts of the inner disk while the supply lasts. For instance, at 0.04~Myr, planetesimals have started to form at $\sim$1.5~au and the total mass in pebbles between 2 and 5~au is  $\sim$12$M_{\oplus}$. These pebbles will eventually reach the inner bump. Planetesimals forming at different locations in the inner ring are made from pebbles arriving at different times at the bump, coming originally from different locations. Thus, our results suggest that the inner Solar System NC-like dust reservoir also had initially a subtle isotopic gradient that is reflected today in the bulk (isotopic) compositions of Earth and Mars\citep{dauphas17,brasseretal17}.

  \begin{figure}
\centering
\includegraphics[scale=1]{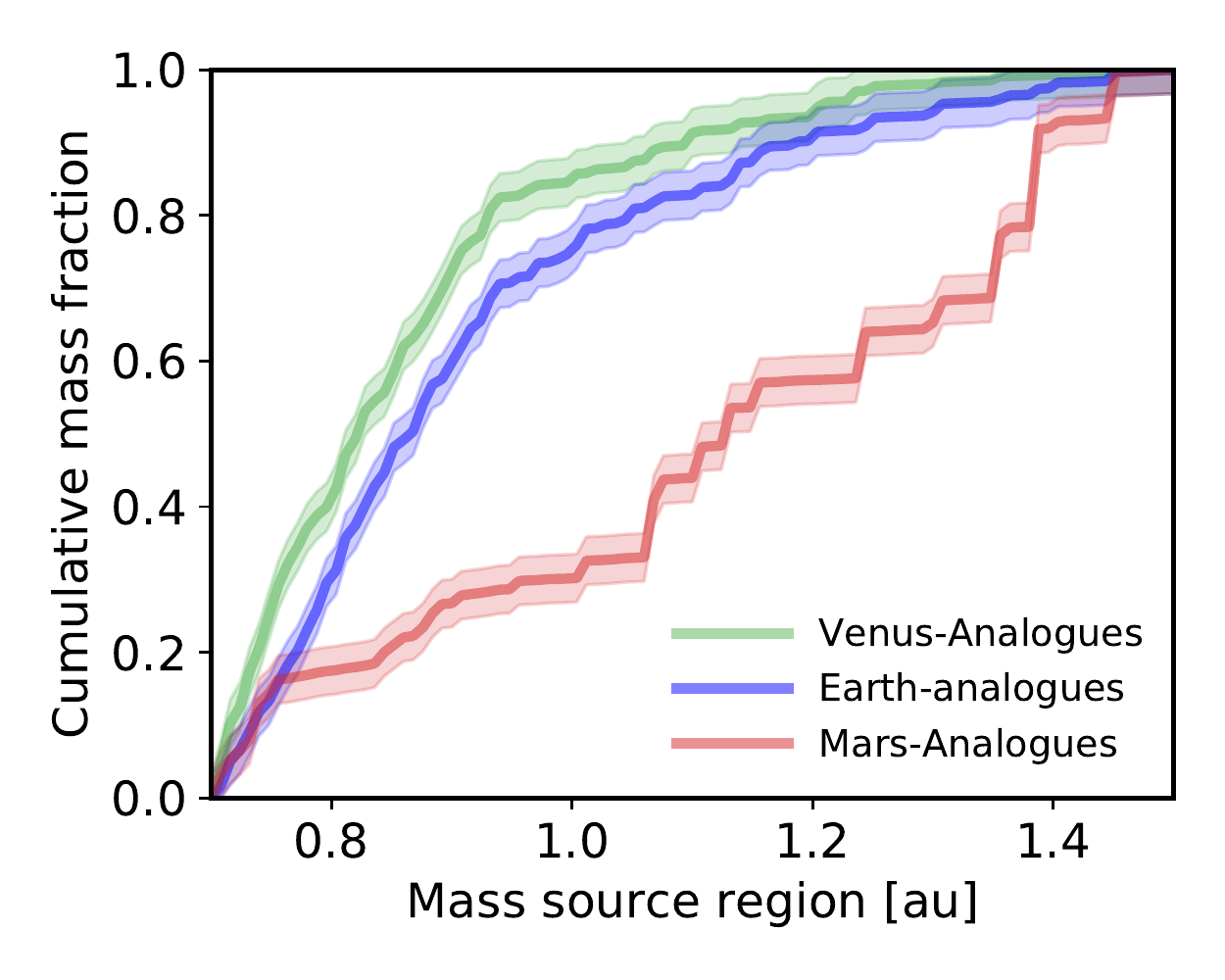}
\caption{Cumulative mass fraction distributions representing the feeding zones of terrestrial planets in 17 simulations that formed good analogues to the Solar System terrestrial planets (see Supplementary Information for definition). Thin green, blue, and red curves represent Venus, Earth, and Mars analogues. Shaded regions encompassing each thin line represent 95\% confidence bands derived from the Kolmogorov-Smirnov statistic (see Extended Data Figure \ref{fig:feeding_2rings} for the effects of  different giant planet configuration and planetesimal distributions in the inner ring, and Supplementary Information for examples of individual planetary systems). Each selected planetary system contains one  Venus, Earth, and Mars-analogue.}
    \label{fig:feeding}
\end{figure}

In our planetesimal formation simulations the main asteroid belt (from 1.8 to 3.2~au) is either born with very little mass in planetesimals of NC-like composition or completely empty (e.g. in Fig. \ref{fig:lowefficiency} the total mass in asteroidal planetesimals is  $\sim10^{-4}M_{\oplus}$ -- see Supplementary Information for parameter tests). While the present-day belt contains very little total mass, it has a complex orbital and compositional structure.\cite{demeocarry14} The belt is dominated by two main groups, the C-type and S-type asteroids\cite{demeocarry14}.

A fraction of planetesimals from the terrestrial planet-forming region are scattered outward and implanted into the asteroid belt on stable orbits\cite{bottkeetal06,raymondizidoro17b}. We associate these implanted planetesimals (as well as the small number of planetesimals that formed within the belt) with NC-like asteroids (the parent bodies of S-type/siliceous asteroids). We use our N-body simulations of the late stage of accretion of terrestrial planets to track the trapping efficiency of planetesimals from the inner ring into the belt over the first 200~Myr of the Solar System history. Planetesimals in the inner ring are scattered by massive planetary embryos onto orbits that cross the asteroid belt region. In order to be captured into the belt, a scattered planetesimal must have its motion decoupled from the interacting embryo by an external perturbation that increases the pericenter of the object before it is ejected\citep{bottkeetal06,raymondizidoro17b}. External perturbations may come from other planetary embryos in the ring or be due to orbital resonances with the giant planets in the asteroid belt. 

\begin{figure}
\centering
\includegraphics[scale=0.85]{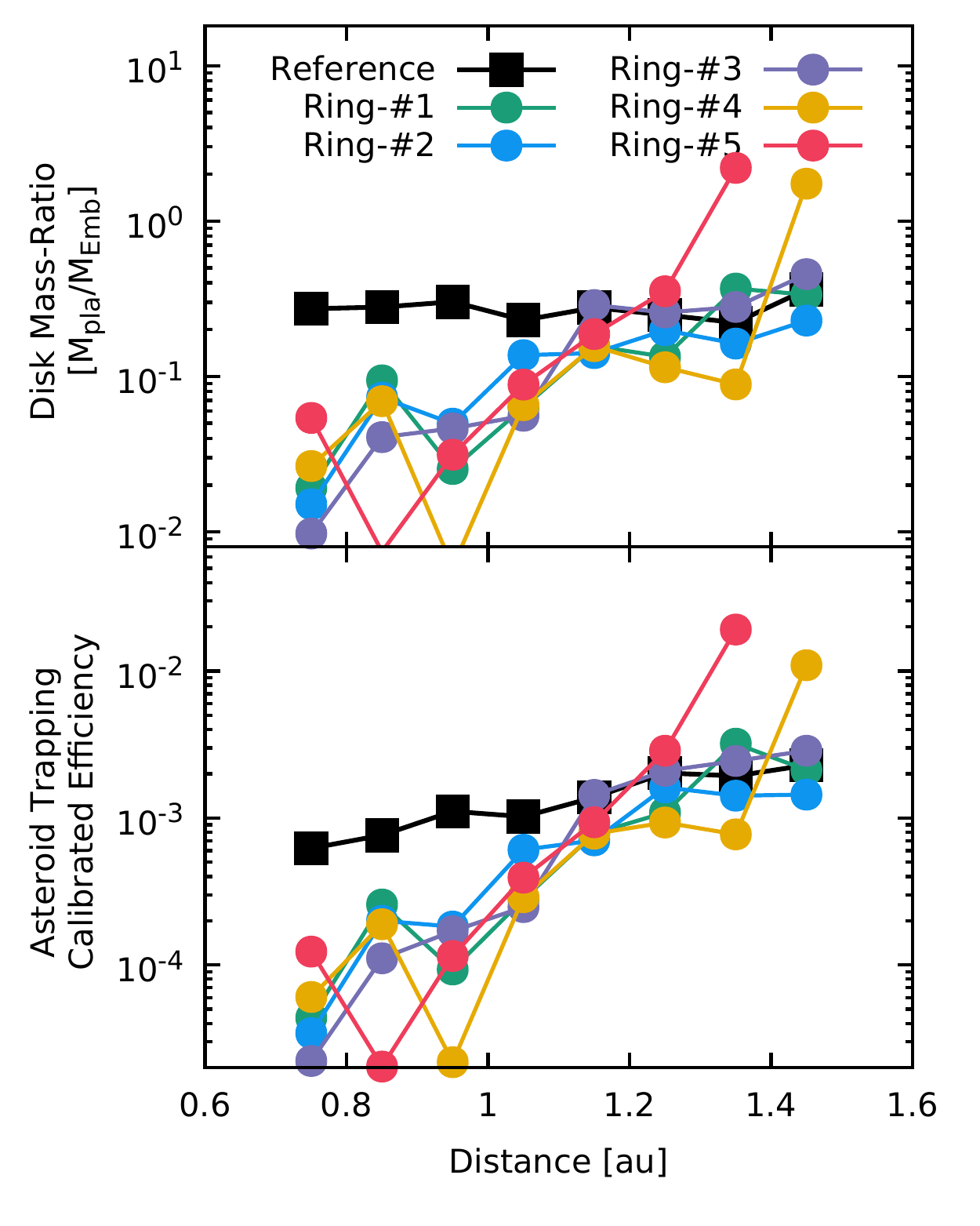}
\caption{Planetesimal implantation efficiency into the asteroid belt for different inner ring configurations {\bf Top:} Planetesimal-embryo disk mass ratio as a function of orbital distance. Blue, yellow, green and purple shows the planetesimal-embryo disk mass ratio at 5~Myr (timing of the gas disk dispersal), in simulations modeling the growth of planetesimal in rings around 1~au.  In all cases, the ring extends from $\sim$0.7 to $\sim$1.5~au, as in Figure \ref{fig:lowefficiency}, and starts with $\sim$2.5$M_{\oplus}$ in 100~km sized (diameter) planetesimals. Planetary objects are defined as objects more massive than the Moon. Color-coded lines also show rings with different initial planetesimal surface density profiles, proportional to $r^0$ (green), $r^{-1}$ (blue), $r^{-2}$ (purple), $r^{-5.5}$ (pink), and $(-200(r/{\rm au}-1)^2+24){~\rm g/cm^2}$ (yellow).  The black curve shows the disk mass ratio in our reference simulation. {\bf Bottom:} Calibrated planetesimal trapping efficiency into the asteroid belt as function of orbital distance. Colors represent different rings as in the top panel.  The black line-dot curve shows the reference planetesimal trapping efficiency into the asteroid belt derived from our simulations modeling the late stage of accretion of terrestrial planets, after 200~Myr of integration.}
    \label{fig:implantation}
\end{figure}

Most implanted NC-like asteroids originated in the Mars region.  In our N-body simulations of terrestrial accretion we imposed an initial planetesimal/embryo mass ratio of 30\% that was constant throughout the inner disk. This choice is consistent with previous simulations \cite{chambers01,walshetal11} and allows us to sample the trapping efficiency as a function of starting orbital radius anywhere within the inner ring. However, detailed calculations show that the planetesimal/embryo mass ratio should not be constant across the inner ring\citep{kokuboida02}.  Rather, at the end of the gaseous disk phase the total mass in planetesimals near the inner edge of the terrestrial ring should be much smaller than that at the outer edge (compare black and colorful lines in top panel of Figure \ref{fig:implantation}; see also Supplementary Information).  This difference exists because planetary embryos grow faster in the inner parts of the ring due to shorter dynamical timescales, consequently consuming planetesimals faster than embryos in the outer parts\cite{kokuboida02}. We account for this by re-scaling our derived trapping efficiency (bottom panel of Figure \ref{fig:implantation}; black-curve)  by the radially-dependent planetesimal-embryo mass ratio. While the non-normalized  trapping efficiency is only a factor of 2 lower at 0.8~au compared to 1.4~au, the re-scaled trapping efficiency is a factor of $\sim$10-100 higher  in the  outer parts of the ring than in the inner parts (colored lines at the bottom panel of Figure \ref{fig:implantation}).

The non-uniform implantation of planetesimals into the NC-like asteroid belt is consistent with Mars' distinct chemical composition relative to Earth. S-type asteroids are linked with ordinary chondrite meteorites.\cite{busbinzel02} Martian meteorites and ordinary chondrites are similar in isotopic signatures of Cr, Ti, O, and V\cite{brasseretal17}, and Mars is thought to be mostly made of material akin to ordinary chondrites\citep{dauphas17,brasseretal17}.  This compositional similarity is naturally explained  by the high implantation efficiency of planetesimals from the Martian region relative to Earth region (e.g. $\sim$1.2-1~au), and the different feeding zones of the two planets (Figure \ref{fig:feeding}).   Enstatite chondrites, associated with the less abundant E-type asteroids\citep{demeocarry14}, may sample planetesimals originally inside $\sim$1~au. This is consistent with isotopic models for Earth's formation  suggesting enstatite chondrites as the major constituent\cite{dauphasetal14,dauphas17}.

The C-type asteroids were likely implanted from the giant planet-forming region during the gaseous disk phase.  If we assume that pebbles and planetesimals beyond the snow line have CC-like compositions, we can compare their orbital evolution with that of present-day carbonaceous asteroids (C-type), which are broadly distributed across the entire main asteroid belt between 1.8~au and 3.2~au.\citep{demeocarry14} CC-like planetesimals formed exterior to 3-4 au (Fig.~\ref{fig:lowefficiency}) and must have interacted with the gas giant planets.\citep{raymondizidoro17b,walshetal11} As Jupiter and Saturn's cores grew and migrated, they scattered nearby planetesimals onto eccentric orbits; a fraction were implanted into the asteroid belt due to the effects of gas drag in the disk.\citep{raymondizidoro17b} The implantation efficiency of planetesimals from the central ring into the belt  is a steep function of orbital distance, with the innermost planetesimals in the central ring having the highest probability of being implanted (Fig.~\ref{fig:lowefficiency}d). This implies that late formed planetesimals, typically born $\sim$2-3~Myr after the start of our simulations, have the highest implantation probabilities. In our model, these correspond to the parent bodies of CC-like  chondrites whose accretion ages are estimated at least 2~Myr after the formation of the calcium-aluminum inclusions~\citep{kruijeretal20}. Planetesimal beyond the snowline formed during the first 1 million years are also implanted into the asteroid belt but with much smaller probabilities. These objects are broadly  consistent with the formation ages of parent bodies of iron meteorites\citep{kruijeretal20}.  Note that the implantation of planetesimals from the middle ring into the belt and the trapping of planetesimals from the inner ring into the belt region are dictated by different mechanisms. Whereas the implantation of planetesimals from the central ring is assisted by gas drag, the trapping of planetesimals from the inner ring is virtually independent of gas drag and due to gravitational interactions. 

Our model can match the inner Solar System's chemical and isotopic constraints.  In our simulations, planetesimal in the inner ring form during the first $\sim$0.5-1~Myr of the Sun's disk lifetime (see Supplementary Information). The decay of short-lived radioactive isotopes as $^{26}$Al is expected to trigger melting and differentiation in early formed planetesimals\citep{urey55}. We have proposed that ordinary chondrites sample planetesimals accreted near Mars' current location. Ordinary chondrites are usually linked to undifferentiated parent bodies\cite{vernazzaetal15}, which should have accreted after most $^{26}$Al has been extinct (e.g. $>$2~Myr) or thought to be the undifferentiated shells of (partially) differentiated planetesimals\cite{weisselkins-tanton13,neumannetal18}. Chondrule ages in ordinary chondrites suggest that their parent bodies accreted 2-3Myr after CAIs\citep{kruijeretal20}. Chondrules in ordinary chondrites may sample material from the outermost chondritic layers of  early formed planetesimals\citep{sandersscott12}, with onion-shell structures\citep{weisselkins-tanton13,neumannetal18}. These layers may be hot and molten\cite{moskovitzgaidos11} but undifferentiated\citep{neumannetal18} and get excavated by impacts generating chondrules\citep{asphaugmovshovitz11}. Jupiter's  formation is also expected to trigger the formation of chondrules in the disk\cite{deschconnoly02}, if it interacts with a population of planetesimals existing in the belt (e.g. Figure \ref{fig:lowefficiency}; or beyond its orbit in the case of CC-planetesimals). Chondrules generated via different processes may either reacrete onto first generation planetesimals or trigger the formation of a second generation of planetesimals if local solid-to-gas ratio becomes sufficiently large to promote planetesimal formation at the very end of the gas disk lifetime\citep{yangetal17} (see Supplementary Information for estimates of dust production in our simulations modeling the growth of planetesimals to planetary embryos).  Our results suggest that ordinary chondrites in reality most likely sample differentiated, partially differentiated and perhaps even undifferentiated parent bodies\citep{weisselkins-tanton13,neumannetal18}.

A model with multiple rings of planetesimals also matches the outer Solar System. The integrated mass in the giant planets' cores is $\sim 60-100\mearth$.\cite{kunitomoetal18}  The ring of planetesimals associated with the water snow line, extending from 4 to $\sim$10~au  (Fig.~\ref{fig:lowefficiency}), typically contains $\sim40-100 \mearth$ (see Supplementary Information) and $\sim0-100 \mearth$ in leftover pebbles.  Massive cores form quickly by pebble accretion onto the most massive planetesimals\cite{levisonetal15b}.  Simulations that match the ice giants' mass distribution invoke a phase of giant collisions between $\sim 5\mearth$ cores whose inward migration was blocked by the near fully-grown Jupiter and Saturn.\cite{izidoroetal15c} We can envision the growth of the ice giant cores either from the outer parts of the middle ring of planetesimals or the inner parts of the outer ring of planetesimals.

It is well-accepted that the giant planets underwent a dynamical instability (see Supplementary Information for a discussion of our model in the context of giant planet migration and solar system evolution), which spread out and excited the giant planets' orbits,\cite{tsiganisetal05} and can explain the orbital distributions of many small body populations.\cite{nesvorny18}  The instability may have been triggered by dynamical interactions between the giant planets, whose orbital configuration was more compact at early times, and an outer disk of planetesimals.\cite{tsiganisetal05}  Matching the giant planets' orbits requires a primordial outer disk containing $\sim 10-30 \mearth$.\cite{deiennoetal17}  The outer planetesimal disk in Fig.~\ref{fig:lowefficiency} contains $\sim18 \mearth$ and fits nicely with dynamical models.  

The cold classical Kuiper belt is a population of objects on low-eccentricity, low-inclination orbits extending from $\sim 42$~au out to $\sim 45$~au.  The cold classicals were likely never strongly scattered by the planets\cite{nesvornyetal20} and may thus represent the outermost planetesimals that formed around the Sun.  While all Kuiper belt objects (KBOs) show red colors, the cold classicals have far redder colors than other KBOs with similar orbital radii but with larger eccentricities and inclinations\cite{gladmanetal08}.  It has been proposed that KBO colors correlate with their formation distance from the Sun.\cite{nesvornyetal20}.  Our multiple-ring model naturally explains the color dichotomy, as the cold classicals would represent remnants from the distant parts of the outermost planetesimal ring whereas more dynamically excited KBOs originated in the outer parts of the central ring or the inner parts of the outermost ring.

It is legitimate to wonder how generically applicable our model is to the formation of other planetary systems. Observations and statistical analysis suggest that planets with sizes between those of Earth and Neptune (1 and 4$R_{\oplus}$) are common around other stars\citep{fressin2013}. These planets are typically refereed as super-Earths. Super-Earths with orbital period shorter than 100 days have been measured to orbit at least 30\% of the sun-like stars \citep{mayoretal11,fressin2013}. Mercury's orbital period is  $\sim$88 days, yet no planet inside Mercury's orbit exists. Why? We have demonstrated that  our simulations can naturally lead to the formation of two classes of planetary systems. In systems where a strong pressure bump forms early at the snowline, pebbles from the outer disk are trapped at the bump and prevented from drifting inwards to the inner disk. This efficient disconnection of the inner and outer system can explain the solar system isotopic dichotomy as well as the lack of massive planets (e.g. super-Earths) in the terrestrial region and potentially inside Mercury's orbit. In this particular scenario, planetary embryos growing in the terrestrial region grow at most to (a few) Mars-mass planetary embryos and avoid large scale radial migration\cite{izidoroetal21}, forming a systems of terrestrial planets like those in the solar system.  On the other hand, in systems where the pressure bump forms late (Extended Data Figure \ref{fig:growbumptocomparenominal}) or is not as strong (more leaky), the inner system is invaded by tens to hundreds of Earth masses in pebbles from the outer disk. In such disks, Earth-mass (or more massive) planetary embryos are likely to form rapidly near 1~au and undergo large-scale radial migration, eventually reaching the inner edge of the disk and forming systems of hot super-Earths~\cite{lambrechtsetal19,izidoroetal19}. The fundamental question that emerges is: Why do some systems form efficient pressure bumps (e.g. the Solar System) whereas others do not (e.g. super-Earths systems)? We propose that this may be linked to intrinsic characteristics of the protoplanetary disk, such as the level of turbulence (viscosity) in the disk controlling pebble sizes ($\alpha_{\rm t}$), or the existence of favorable conditions for the early formation of giant planets (e.g. Jupiter) at specific locations of the disk, as at the disk water snowline\citep{drazkowskaalibert17,mulleretal21}. Pressure bumps in gaseous disks with smaller pebbles  may be far more  leakier than those in disks with larger ones\cite{pinillaetal12}, potentially allowing a lot of mass in pebbles from the outer disk to be delivered to the terrestrial region. In our Solar System, Jupiter's early core  formation may have also played a decisive role in preventing that from happening\cite{lambrechtsetal19}. If a giant planet core promptly forms in the pressure bump at the snowline\citep{morbidelli20}, it may induce the formation of another pressure bump beyond its orbit~\citep{lambrechtsetal14}. The Solar System architecture, which seems to be an unusual outcome of planet formation~\citep{raymondetal18c}, probably reflects all these conditions.

\begin{figure}
\centering
\vspace{-2cm}
\includegraphics[scale=.75]{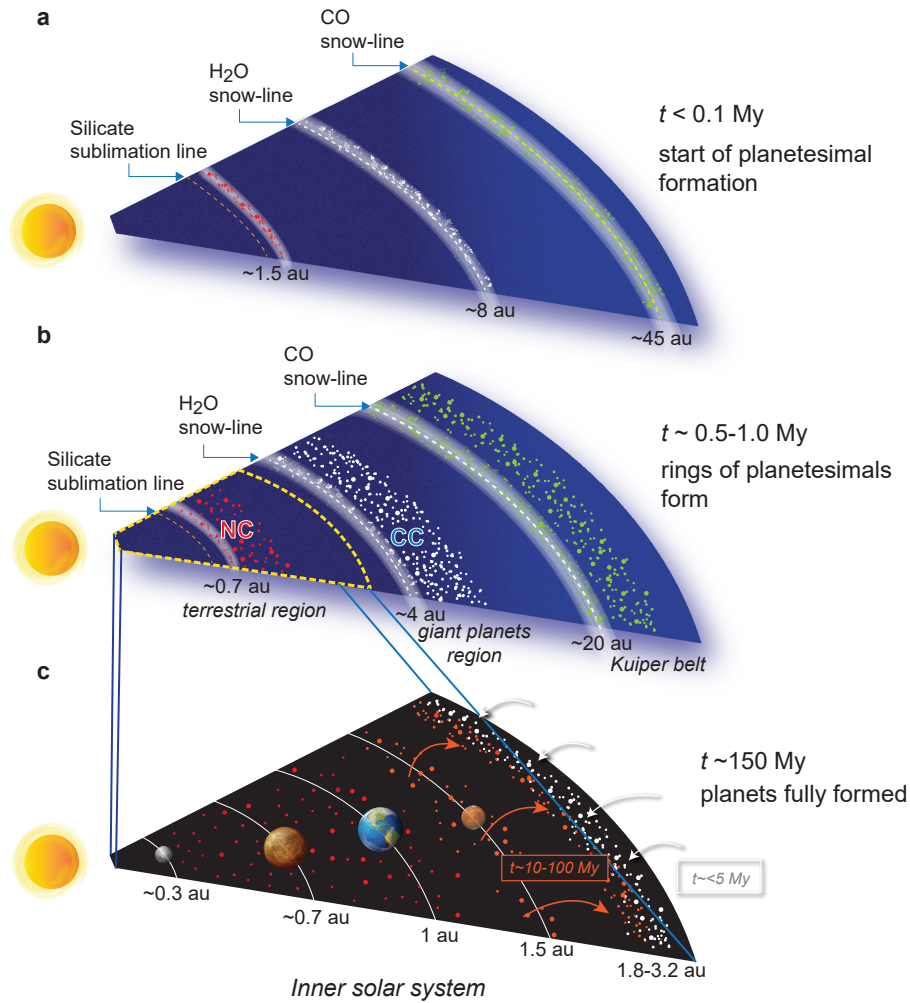}
\vspace{-5cm}
\caption{A schematic showing migration of three pressure bumps (grey shaded annular bands) associated with the CO snowline (yellow-green dashed line), ${\rm H_2O}$ snowline (white dashed line), and silicate sublimation line (orange dashed line) in the Sun’s natal disk. \textbf{Top:} pebbles pile up at each of the pressure bumps, which lead to planetesimal  formation in the three distinct rings. Planetesimals at each of the rings are assumed to have distinct compositions as shown by three different colors (green: CO-snow-line; white: ${\rm H_2O}$-snow-line; red-orange: silicate sublimation line). \textbf{Middle:} the narrow, innermost ring form NC-like planetesimals and the wider, intermediate ring form CC-like planetesimals. The outer ring is the expected location of the primordial Kuiper-belt. \textbf{Bottom:}  zoomed-in view of the innermost Solar System after the dissipation of the gaseous disk with the final configuration of the rocky planets and the asteroid belt; it highlights that the asteroid belt is populated by CC planetesimals scattered inward by giant planets and NC planetesimals preferentially from the Mars-region (orange colored circles) and no planetesimals are formed inside the orbit of Mercury.
}
    \label{fig:cartoon}
\end{figure}

Our model demonstrates that the present-day Solar System could form from three rings of planetesimals (Figure \ref{fig:cartoon}). This diverges from standard models that assume a continuous disk of planetesimals\cite{kokuboida02}, and is reminiscent of the ring-like structures observed in disks around young stars.\cite{huangetal18}  Our model can simultaneously explain the orbits and masses of terrestrial planets, the distribution of different types of asteroids in the main asteroid belt, and link different classes of meteorites to the main building blocks of Earth and Mars.  The terrestrial planets formed from a narrow ring in the terrestrial region, which naturally accounts for the lack of planets inside Mercury's orbit and the low mass of Mars. The gas giant planets formed from a wide central ring located beyond the asteroid belt, probably via planetesimals and pebble accretion. The asteroid belt was never much more massive than it is today. The outermost planetesimal ring located beyond the current orbit of Uranus, was sculpted during the giant planets growth and their subsequent dynamical evolution, producing the current Kuiper-belt. 

\section*{Methods}

Our model couples different stages of planet formation, from dust evolution in a young gaseous disk to the final stage of accretion of terrestrial planets. We have performed simulations modeling:

\begin{itemize}
    \item[-]  Dust evolution (growth and fragmentation) in a gaseous disk;
    \item[-] Planetesimal formation;
    \item[-] Growth from planetesimals to planetary embryos in rings around 1~au;
    \item[-] The late stage of accretion of terrestrial planets - namely growth from planetary embryos to planets.
    \item[-] Asteroid implantation from inside-out during the terrestrial planet formation; 
\end{itemize}
We next describe the different ingredients of our model individually.

\subsection{Gas disk model -}

We model the radial distribution of gas and disk temperature using power-law  profiles.  Gas disk dissipation and cooling are mimicked via exponential decay timescales. Although simplistic, our approach allows us to easily disentangle the effects of different parameters of the model. In nominal simulations, our 1D underlying protoplanetary disk extends from 0.1~au to $\sim$120~au  and  is represented by a simple power-law disk with radial profile given as\cite{pinillaetal17}
\begin{equation}
\Sigma_{\rm ref}(r) = 1700 \left(\frac{r}{1~{\rm au}}\right)^{-1}  \times \exp\left[-\frac{r}{r_{\rm c}}\right]  {\rm \frac{g}{cm^2}}.
\label{eq:mmsn}
\end{equation}

In Eq. \ref{eq:mmsn}, $r$ is the heliocentric distance, $r_{\rm c}$ is the cutoff radius, and au represents the astronomical unit -- average distance from Earth to the Sun. Following previous studies ~\cite{pinillaetal12,dullemondetal18,morbidelli20,izidoroetal21}, we mimic the presence of the pressure bumps in the disk by rescaling the original gas disk profile given by Eq. \ref{eq:mmsn} with a gaussian  and hyperbolic tangent functions. Each of these functions account for one of the pressure bumps in the disk. Our gas disk radial profile reads
\begin{equation}
\Sigma_{\rm gas}(r) = \Sigma_{\rm ref}(r)  \exp\left[-A_{\rm H2O}\exp\left(-\frac{(r-r_{\rm H2O})^2}{w(r_{\rm H2O})^2}\right)\right] \exp\left[-A_{\rm CO}\exp\left(-\frac{(r-r_{\rm CO})^2}{w(r_{\rm CO})^2}\right)\right] \frac{\alpha_{\nu}}{S},
\label{eq:1pb}
\end{equation}
where
\begin{equation}
S(r) = 0.5\left(\alpha_{\rm mri} - \alpha_{\nu}\right)\left(1 - \tanh{\left({\frac{T_{\rm mri} - T_{\rm gas}}{50~K}}\right)}\right) + \alpha_{\nu}.
\label{eq:innerbump}
\end{equation}
In Eq. \ref{eq:1pb}, $r_{\rm H2O}$, $r_{\rm CO}$, $A_{\rm H2O}$, $A_{\rm CO}$, $w_{\rm H2O}(r_{\rm H2O})$, and $w_{\rm CO}(r_{\rm CO})$ are free parameters used to represent the pressure bumps location, amplitude, and width. $H_{gas}(r_{\rm H2O})$ and  $H_{gas}(r_{\rm CO})$ are the gas disk scale height at the water ($T_{\rm gas}=170$~K) and CO ($T_{\rm gas}=30$~K) snowlines , respectively. Following previous studies\cite{pinillaetal12,izidoroetal21}, we set $A_{\rm H2O}=0.5$ and $w_{\rm H2O}(r_{\rm H2O})=H_{\rm gas}(r_{\rm H2O})$. We choose for the pressure bump at the snowline, a configuration that provides an efficient disconnection of the inner and outer Solar System pebble reservoirs\cite{pinillaetal12,izidoroetal21}, as constrained by the Solar System isotopic dichotomy\cite{kruijeretal20}. For simplicity, we assume an equivalent configuration for the pressure bump at the CO-snowline, i.e., $A_{\rm CO}=0.5$ and  $w_{\rm CO}(r_{\rm CO})=H_{\rm gas}(r_{\rm CO})$. $\alpha_{\nu}$ is the gas disk viscosity in regions of the disk where  $T_{\rm gas}< T_{\rm mri} = 1000$~K. In our model, $T_{\rm mri}$ defines the threshold temperature for thermal ionization of the gas disk\cite{deschturneretal15}. We set $\alpha_{\nu}=10^{-3}$ in all our simulations\citep{drazkowskadullemond18}. We assume that in regions of the disk where $T_{\rm gas}> T_{\rm mri} =1000$~K the disk viscosity increases  due to the presence of magneto-rotational instabilities fostered by thermal ionization of the gas~\citep{deschturneretal15,flocketal17,uedaetal19}. Previous studies have assumed that the gas disk viscosity in  strongly-ionized regions of the disk may be a factor of $\sim$10-100 larger than in weakly ionized regions\cite{uedaetal19}. In our nominal simulations, we assume a conservative value corresponding to $\alpha_{\rm mri}=3\alpha_{\nu}$.  At the transition between the high and low viscosity regions of the disk, the gas surface density drops to keep the gas accretion rate constant as a function of radius.  This creates a pressure bump in the disk slightly outside the silicate sublimation line of the disk~\citep{uedaetal19}.  Note that $\alpha_{\rm mri}$ and  $\alpha_{\nu}$  set the shape and ``strength'' of the inner pressure bump. We explore the effects of this parameter in our model in simulations presented in the Supplementary Information.

The initial gas disk temperature is modeled as a simple power law disk as
\begin{equation}
T_{\rm gas}(r) = 1200 \left(\frac{r}{1~{\rm au}}\right)^{-\beta}~K,
\end{equation}
where the power-law index $\beta$ is assumed to be  either equal to  0.7 or 1~\citep{idaetal16b}, i.e.,  we  test our model against two  disk temperature profiles. For  $\beta=0.7$, the temperature profile places the CO-snowline initially at $\sim$200~au. As our nominal disk extends up to $\sim$120~au, simulations with $\beta=0.7$ contain only two pressure bumps. In cases where $\beta=1.0$, the pressure bump at the CO-snowline is initially at $\sim$41~au.

Our initial disk temperature profiles are broadly consistent with models of  disk formation from the collapse  of the progenitor molecular core\cite{zhangjin15,drazkowskadullemond18}. The time-zero of our simulations probably corresponds to the first $\sim$10-100~kyr from the beginning of the colapse, when the disk has already expanded and starts to cool off in time\citep{zhangetal15,drazkowskadullemond18}. In all simulations, we assume that the gas disk dissipates following an exponential decay with e-fold timescale of 2~Myr. The gas disk temperature also decays exponentially with time -- but with different timescales  as the disk evolves -- to better mimic the evolution of the disk temperature seen in simulations modeling disk evolution\cite{zhangjin15,baillieetal19,bitschetal14}. In Figure \ref{fig:lowefficiency}, where $\beta=1$,  the disk temperature starts to drop with a short e-fold timescale of 0.75~Myr at  0.05~Myr. From 0.5~Myr to 2~Myr, we increase this timescale to 5~Myr to roughly mimic the evolution of the snowline in more sophisticated disk models\cite{zhangjin15,baillieetal19}. From 2~Myr to 3~Myr, we assume that the disk temperature is not evolving with time. This approach is used to mimic Jupiter's core formation in the disk\cite{kruijeretal20} (see also Supplementary Information for parameter tests). Once Jupiter's core forms, perhaps near/at the snowline pressure bump\cite{morbidelli20}, it prevents pebbles from beyond its orbits to drift inwards, and may even heat-up the disk shifting the location of the snowline further out\cite{ziamprasetal20}. We do not include these effects in this work, but they are not expected to impact qualitatively our main conclusions. The final position of the snowline in our disks is broadly consistent with the results of more sophisticated disk models\cite{zhangjin15,baillieetal19}. Note that our disk is initially much hotter than that  considered in our previous study modeling the effects of a pressure bump on the formation of the Solar System terrestrial planets\citep{izidoroetal21}. In Izidoro et al\cite{izidoroetal21}, the disk snowline was initially  at 5~au and the transition in the disk viscosity due to thermal ionization would be at $\sim$0.1~au. The disk models invoked here and that of Izidoro et al\citep{izidoroetal21} represent two possible end-member scenarios. The main advantage of our new model is that it provides a simple solution to one of major problems of the scenario of Izidoro et al\citep{izidoroetal21}, our new model naturally accounts for lack of planets inside Mercury's orbit in the Solar System, a problem that remained unsolved in Izidoro et al\cite{izidoroetal21}.

\subsection{Dust evolution calculations -}

We model coagulation, fragmentation, drift, and turbulent mixing of dust grains in a 1D gaseous disk. Our dust evolution code is based on previous studies\citep{birnstieletal12, birnstieletal15,drazkowskaetal16}. We assume that the underlying gas disk profile is given as
\begin{equation}
\Sigma_{\rm dust} = Z_0 \Sigma_{\rm gas},
\end{equation}
where $Z_0$ represents the initial dust-to-gas ratio. In our simulations we explored different values for $Z_0$, ranging from 0.25\% to 1.5\%\citep{asplundetal09}.

To model dust evolution, we solve the 1D-advection-diffusion   equation for the column dust density \citep{birnstieletal12}
\begin{equation}
\begin{split}
\frac{\partial{\Sigma_{\rm dust}}}{\partial{t}} & +  \frac{1}{r}\frac{\partial}{\partial{r}}\left\{ r \left[\overline{v}_{\rm r,dust}\Sigma_{\rm dust} -  D_{\rm dust}  \frac{\partial}{\partial r }\left(\frac{\Sigma_{\rm dust}}{\Sigma_{\rm gas}}\right)\Sigma_{\rm gas} \right]  \right\}   =  \frac{\partial \Sigma_{\rm pla}}{\partial t}.
\end{split}
 \label{eq:continuity}
\end{equation}
We calculate the dust diffusivity as 
\begin{equation}
D_{\rm dust} =  \frac{\alpha_t c_{\rm s} H_{gas}}{1+St^2},    
\end{equation}
where $\overline{v}_{\rm r,dust}$ represents the mass weighted radial velocity of dust\cite{birnstieletal12}. Our model follows the evolution of two dust-size species, namely the largest and the smallest dust grain sizes~\citep{birnstieletal12}. The initial size of dust grains is set, $a_0=10^{-6}$~m. We refer to the largest dust grains as pebbles. Our dust coagulation model is based on previous studies~\citep{birnstieletal12,drazkowskaalibert17}, and detailed described in a previous paper~\cite{izidoroetal21}.  We assume that dust grains and pebbles beyond the snowline have a one-to-one ice-silicate composition. Water ice-particles that drift inwards and cross the water snowline are assumed to sublimate, loosing 50\% of their mass corresponding to the ice-component. This effect is particularly important when the  pressure bump at the water snowline is not present since the beginning of the simulation, and a significant mass in pebbles from beyond the water ice-line drift inside it before the bump forms. Silicate pebbles in the inner disk are assumed to sublimate when crossing the silicate sublimation line ($T_{\rm gas}=1400~K$). In our simulations, we use a radial log-grid with 400 cells, which ensure numerical convergence\cite{deschetal18}. Following previous studies~\citep{pinillaetal21,izidoroetal21},
our model assumes that the threshold fragmentation velocities of ice pebbles beyond the snowline is 10~m/s, whereas silicate pebbles fragment at velocity of 1m/s. We use a smoothing function to model the transition in the threshold velocity at the snowline~\cite{izidoroetal21}.  The turbulence level at the disk midplane is represented by $\alpha_t$ and the gas disk viscosity by $\alpha_{\nu}$, as adopted in similar studies\citep{drazkowskadullemond18,pinillaetal21}.  In our  simulations, we assume that $\alpha_t=\alpha_{\nu}/\Delta$, where $\Delta$ is a simple dimensionless free-parameter, which we assumed to range from 1 to 40\citep{drazkowskadullemond18}. We discuss the impact of these parameters in our model in the Supplementary Information. Note that for these levels of turbulence in the disk midplane, planetesimals are unlikely to form during the disk build-up\cite{drazkowskadullemond18}. Therefore, our simulations are envisioned to correspond to the timing when the gas in-fall from the molecular cloud has ceased, and the disk is fully formed. In this work,  we do not account for the re-condensation of vapor diffusing back to regions beyond the sublimation line when drifting pebbles cross a giving sublimation front~\citep{drazkowskadullemond18,schneideretal21} because we do not solve the advection-diffusion equation of the gas. We also solve the advection-diffusion equation  for the dust component considering  two dust-size populations~\cite{birnstieletal12} rather than a ``continous'' of grain sizes. We do not expect these assumptions to impact the broad qualitative results of our paper.

\subsection{Planetesimals formation -}

We account for planetesimal formation  via the sink term on the right-hand side of Eq. \ref{eq:continuity}, which takes the following form
\begin{equation}
\frac{\partial \Sigma_{\rm pla}}{\partial t} = 
\begin{dcases} 
-\varepsilon\frac{ \Sigma_{\rm peb}}{T_{\rm K}},& \text{if }  \frac{\rho_{\rm peb}}{\rho_{\rm gas}} \geq 1 \text{ and } St \geq St_{min}\\
- \frac{\varepsilon}{d}|v_{\rm r,peb}|\Sigma_{\rm peb},& \text{ if }  \dot{M}_{\rm peb,crit} \leq 2\pi r \Sigma_{\rm peb} |v_{\rm r,peb}| \text{ and } \frac{\rho_{\rm peb}}{\rho_{\rm gas}} < 1 \text{ and } St \geq St_{min}\\
 0, & \text{otherwise.}
\end{dcases}
\label{eq:planetesimal}
\end{equation}

In our model, planetesimals formation takes place if one of the two conditions are met: i) if the density of pebbles at the disk midplane is larger than unity\citep{johansenetal07,drazkowskaetal16,drazkowskaalibert17}; or if the pebble flux is high enough to foster particle concentration in the presence of zonal flows in the disk~\cite{dittrichetal13}. In addition, we assume that pebbles are only converted into planetesimals if their Stokes number ($St$) are larger than $St_{min}$, which is assumed to be  $St_{min}=10^{-3}$ in our nominal simulations\cite{drazkowskadullemond18,uedaetal19,lenzetal20} (see Supplementary Information for the effects of these parameter on our results). The Stokes number is given by $St = \frac{\pi a \rho}{2\Sigma_{\rm gas}}$, where $a$  and $\rho$ are the pebble size and bulk density. Silicate and ice pebbles are assumed to have bulk densities of 3${\rm g/cm^3}$ and  1${\rm g/cm^3}$, respectively.  In Eq. \ref{eq:planetesimal}, $v_{\rm r,peb}$ and $\Sigma_{\rm peb}$ represent the radial velocity and surface density of pebbles in the disk, respectively. Pebbles are the largest grain sizes at one given location of the disk. $\epsilon$ is the planetesimal formation efficiency which is taken as a free parameter in our model. In our nominal simulations\cite{drazkowskaalibert17} we set $\epsilon\approx10^{-4}-10^{-6}$. $T_{\rm K}$ is the orbital period.  $\rho_{\rm peb}$ and $\rho_{\rm gas}$ represent the pebble and gas midplane densities. $d$ represents the width of pebble-trap caused by zonal flows in the disk. We set $d=5H_{\rm gas}$\cite{lenzetal19}.

The pebble radial velocity is given by\citep{okuzumietal12,uedaetal19}
\begin{equation}
v_{\rm r,peb} = -\frac{St}{St^2 +  (1 + Z')^2}2\eta v_{\rm k} + \frac{1+Z'}{St^2 + (1 + Z')^2 }v_{\rm r,gas}
\end{equation}
where $Z'$ represents the local gas-to-dust ratio, which evolves with time.
The pressure support parameter is calculated as
\begin{equation}
\eta = -\frac{1}{2}\left( \frac{c_{\rm s}}{v_{\rm k}}\right)^2\frac{\partial \ln{P}}{\partial \ln{r}}, 
\label{eq:eta}   
\end{equation}
where $v_{\rm k}$ is the keplerian velocity. The nominal gas radial velocity is calculated as\citep{lyndenbellpringle74}

\begin{equation}
v_{\rm r,vis} \approx -\frac{\nu}{r}
\label{eq:vrgas}
\end{equation}

where the gas disk viscosity\citep{shakurasunyaev73} is $\nu=\alpha_{\nu} c_s H_{\rm gas}$\cite{drazkowskadullemond18}. $c_s$ represents the sound speed, $P$ is the midplane gas disk pressure calculate in the isothermal limit, and  $H_{\rm gas}$ is the disk scale height, derived from the temperature profile assuming that the gas disk is in vertical hydrostatic equilibrium. $h_{gas}=H_{\rm gas}/r$ is the gas disk aspect ratio. We include the effects of back-reaction of the dust on the gas disk as~\cite{drazkowskaalibert17,uedaetal19}, 
\begin{equation}
v_{\rm r,gas} = \frac{StZ'}{St^2 +  (1 + Z')^2}2\eta v_{\rm k} + \left[1 - \frac{(1+Z')Z'}{St^2 + (1 + Z')^2 }v_{\rm r,gas}\right]v_{\rm r,vis}
\label{eq:vrgas_back}
\end{equation}

The critical pebble flux given in Eq. \ref{eq:planetesimal} is defined as\cite{lenzetal19}
\begin{equation}
 \dot{M}_{\rm peb,crit} = \frac{m_{\rm c}}{\varepsilon\tau_{\rm c}}.  
  \label{eq:criteria2}
\end{equation}
The collapsing dust mass $m_{\rm c}$ is set
\begin{equation}
m_{\rm c } =\frac{4}{3} \pi l_{\rm c}^3 \rho_{\rm Hill}.
\label{eq:critflux}
\end{equation}
$\tau_{\rm c}$ represents the characteristic lifetime of traps due to zonal flows that promote pebble concentration until they can collapse. We set $\tau_{\rm c}=100T_{\rm K}$~\citep{baistone14,lenzetal20}.
$\rho_{\rm  Hill}$ represents the   hill density~\citep{gerbigetal19} defined as
\begin{equation}
    \rho_{\rm  Hill} = \frac{9}{4\pi}\frac{M_{\odot}}{r^3},
\end{equation}
where $M_{\odot}$ is the central star mass. 
The critical length scale can be derived by equating the diffusion and collapse timescale of the clump and yields\citep{lenzetal19}
\begin{equation}
l_{\rm c} = \frac{2}{3}\sqrt{\frac{\alpha_{\rm t}}{St}}H_{\rm gas}.    
\end{equation}
We have also performed simulations neglecting the contribution of zonal flows for planetesimal formation (see Supplementary Information).

\subsection{Growth from planetesimals to planetary embryos -}

We model planetary growth from planetesimals  in the inner ring near 1~au. We invoke semi-analytical calculations to model pebble  and planetesimal accretion onto planetesimals\citep{ormelklahr10,chambers06,lambrechtsetal14,johansenlambrechts17,izidoroetal21}. We use this approach to first infer how planetesimals grow in the inner ring. Subsequently, we  perform N-body numerical simulations, as will be explained later. Our analytical prescriptions for pebble and planetesimal accretion are described in Izidoro et al~\citep{izidoroetal21}. In all our simulations, we assume  that planetesimals have characteristic diameters of 100~km, which is consistent with typical sizes of planetesimals formed via streaming instability simulations~\cite{simonetal16}. We compute the mass growth rate of planetesimals using as input the pebble flux and planetesimal surface density provided by our simulations modeling the dust evolution and coagulation, and planetesimal formation. Planetary embryos forming in our inner ring of planetesimals predominately grow  via planetesimal accretion rather than pebble accretion (see Supplementary Information). This result is consistent with that of Izidoro et al\cite{izidoroetal21}, although considering a different disk model. 

 We have also performed numerical N-body simulations to model planetary growth from planetesimals in the inner ring using the LIPAD code\cite{walshlevison19,deiennoetal19}. In light of the results of our experiments modeling planetesimal growth using analytical prescription, we have neglected the contribution of pebble accretion in our numerical N-body simulations. The obvious choice of planetesimal ring configuration to model the subsequent growth from planetesimals would be that of Figure \ref{fig:lowefficiency}. However, it is important to keep in mind that neither the slope nor the total mass in planetesimals  in the ring are unique. In Figure \ref{fig:lowefficiency}, the total mass in planetesimals in the inner ring is $\sim$2.5$M_{\oplus}$, and the ring has an almost flat radial surface density profile  with very sharp edges at 0.7~au and 1.5~au, where the planetesimal surface density drops significantly.  Some of our simulations produce almost flat slopes ($\Sigma_{\rm pla}\approx r^0$; e.g. Figure \ref{fig:lowefficiency}) but others produce  radially decreasing steep surface density profiles ($\Sigma_{\rm pla}\approx r^{-2}-r^{-5.5}$) or even upside-down U-shape profiles (see Supplementary Information), although these rings may contain roughly the same total mass in planetesimals. Inspired by the diversity of the inner ring profiles produced in our simulations, we conducted five additional high resolution N-body simulations considering different scenarios, namely inner rings with planetesimal surface density proportional to  $r^{0}$, $r^{-1}$, $r^{-2}$,  $r^{-5.5}$, and $(-200(r/{\rm au}-1)^2+24)g/cm^2$. Simulations where the total mass in planetesimals in the inner ring is larger than $3-5M_{\oplus}$ produce high-mass  planets that do not match the real terrestrial planets (e.g. see Supplementary Information). Motivated by the results of previous studies\citep{hansen09,raymondizidoro17a}, we model the formation of terrestrial planets in the Solar System in rings where the total mass in planetesimals is $\sim$2.5$M_{\oplus}$. We model the growth of planetesimals to planetary embryos. In these simulations, the gas disk dissipates at 5~Myr, when we stop the simulation. We use the result of these simulations - namely the mass distribution of planetary embryos and planetesimal at the end of the gas disk phase - to calibrate the planetesimal implantation efficiency from the inner ring into the asteroid belt produced in our simulations of the late stage of accretion of terrestrial planets, as explained next.

\subsection{Growth from planetary embryos to planets -}
We also performed 80 N-body numerical simulations of the late stage of accretion of terrestrial planets. We assumed a distribution of planetary embryos and planetesimals in  a ring between $\sim$0.7~au and $\sim$1.5~au. When performing this particular set of simulations we considered two  surface density profiles for the inner ring, broadly consistent with the range of ring slopes produced in our dust coagulation and planetesimal formation simulations. We assumed rings with surface density  of planetesimals proportional to $r^{-1}$ and $r^{-5.5}$. We assumed that the gas disk have already dissipated at the beginning of our simulations. Jupiter and Saturn are assumed to be fully formed and near their current orbits, but in resonant and almost circular and coplanar orbits\cite{morbidellicrida07}. Our simulations start with a distribution of planetesimals and planetary embryos. Planetesimals carry about 30\% of the local disk total mass and the remaining fraction is carried by equal mass planetary embryos. This approach is commonly considered  in classical simulations of terrestrial planet formation~\cite{chambers01,raymondetal04,obrienetal06} and in our work comes with the advantage of allowing us to track the implantation efficiency of planetesimals from all locations of the inner ring. The initial ring mass is 2.5$M_{\oplus}$ (see Supplementary Information).  Planetesimals are considered to be non-self interacting objects but to interact with the star and planetary embryos. Planetary embryos gravitationally feel each other, planetesimals and the central star.  We numerically integrate our systems for 200~Myr.

\subsection{Asteroid implantation from inside-out during the terrestrial planet formation -} At the end of our simulations of the late stage of terrestrial planet formation, we compute the implantation efficiency of planetesimals from the inner ring into the asteroid belt region by combining the results from all our simulations.  We consider implanted planetesimals from the inner ring that have at the end of the simulation perihelion distance $q>1.8$~au, orbital eccentricities below 0.3, and orbital inclinations below 25 degrees. This yields our reference implantation efficiency, as given in Figure \ref{fig:implantation}. We rescale the implantation efficiency derived from these simulations using the planetesimal-embryo  disk mass ratio at the end of our simulations modeling the growth of planetary embryos from planetesimals. This is important because planetesimals are rapidly accreted in the inner regions of the ring by growing embryos and, in reality, the planetesimal-embryo disk mass ratio is not constant across the inner ring (see also Supplementary Information).

\vskip .2in
\noindent{\bf Data availability}\\
Simulation data that support the findings of this study or were used to make the plots are available from the corresponding author upon reasonable request. Source data associated to the main figures of the manuscript are available at \url{https://andreizidoro.com/simulation-data}.

\noindent{\bf Code availability}\\ Dust evolution simulations were performed using a  modified version of the code Two-pop-py \citep{birnstieletal12}, publicly available at \url{https://github.com/birnstiel/two-pop-py}, with modifications described in Izidoro et al\cite{izidoroetal21}. N-body simulations modeling the growth of planetesimals to planetary embryos were performed using LIPAD\cite{levisonetal12}. This is a proprietary software product funded by the Southwest Research Institute that is not publicly available. It is based on the N-body integrator SyMBA, which is publicly available at \url{https://www.boulder.swri.edu/swifter/}. Simulations of the late stage of accretion of terrestrial planets were performed using the {\tt Mercury} N-body integrator\cite{chambers99}, publicly available at \url{https://github.com/4xxi/mercury}.

\begin{addendum}
 \item [Acknowledgments] We thank the reviewers, Brad Hansen and Eiichiro Kokubo, for the very constructive comments and suggestions that helped to improve the paper. A.~Iz., R.~Da., and A.~Is.  acknowledge NASA grant 80NSSC18K0828 for financial support during preparation and submission of the work. B.B., thanks the European Research Council (ERC Starting Grant 757448-PAMDORA) for their financial support. R.~De. acknowledges support from NASA Emerging Worlds program, grant 80NSSC21K0387. S.~N.~R. thanks the CNRS's PNP program for support. A.~Iz. thanks Maxime Maurice for numerous inspirational discussions, and  the Brazilian Federal Agency for Support and Evaluation of Graduate Education (CAPES), in the scope of the Program CAPES-PrInt, process number 88887.310463/2018-00, International Cooperation Project number 3266.
\item [Author Contributions] A.~Iz. conceived the project in discussions with R.~Da. and B.~B.. A.~Iz. performed numerical simulations modeling dust evolution and planetesimal formation. S.~N.~R., R.~De., and A.~Iz.  conducted N-body numerical simulations. A.~Iz. analyzed the results of numerical simulations and led the writing of the manuscript. R.~Da. helped with the cosmochemical implications of the model and constructed Figure 5. All authors discussed the results and commented on the manuscript.
\item[Competing Interests] The authors declare that they have no competing financial interests.

\item[Correspondence] Correspondence and requests for materials should be addressed to izidoro.costa@gmail.com (Andre Izidoro)
\end{addendum}

\clearpage
\noindent{\bfseries Extended Data Figures}\setlength{\parskip}{12pt}%

\beginextendeddata

\begin{figure}[H]
\centering
\includegraphics[scale=.4]{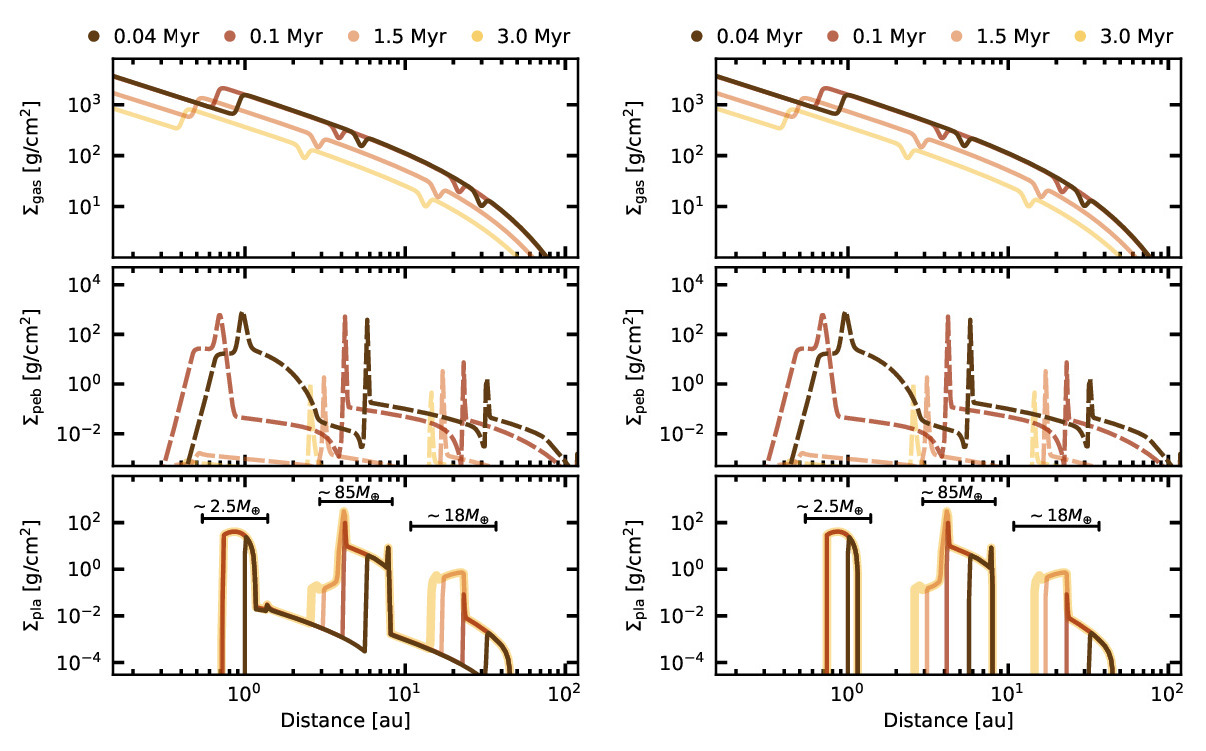}
\caption{Gas, pebbles and planetesimals surface densities in two simulations with three pressure bumps. {\bf Left}: simulation including the effects of planetesimal formation via  zonal flows\cite{lenzetal19}, see Eq. \ref{eq:planetesimal}. {\bf Right:} simulation neglecting  the effects of planetesimal formation via  zonal flows\cite{drazkowskaalibert17,drazkowskadullemond18}. Top, middle, and bottom panels of each figure show the evolution of gas, pebbles, and planetesimals surface densities, respectively. The initial dust-to-gas ratio is $Z_0=1.3\%$, $\epsilon=1\times10^{-4}$, $\alpha_t=\alpha_{\rm \nu}/27$. The final rings contain 2.5~${M_{\oplus}}$ (inner), 85~${M_{\oplus}}$ (central), and 18~${M_{\oplus}}$ (outer) in planetesimals. In both simulations $r_{\rm c}=25$~au.}
\label{fig:threebumps_examepl2}
\end{figure}

\begin{figure}[H]
\centering
\includegraphics[scale=0.4]{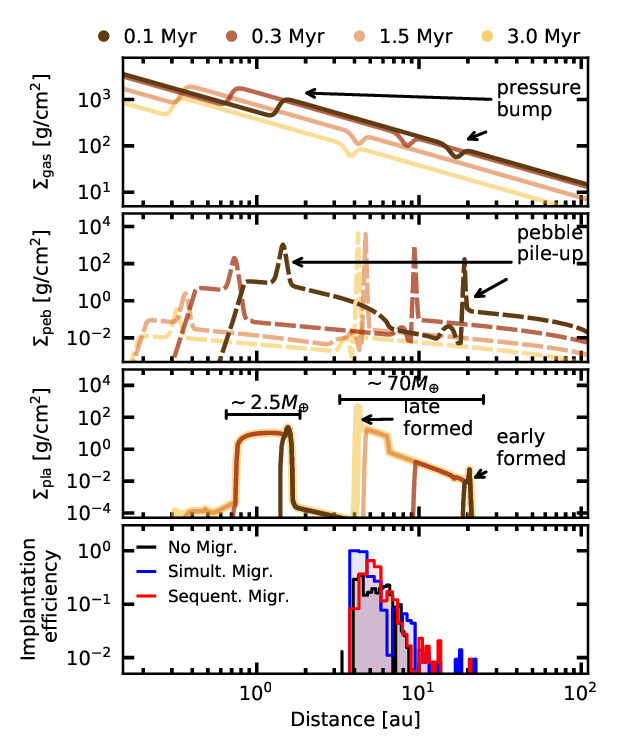}
    \caption{Final distribution of planetesimals in a simulation with two pressure bumps ($\beta=0.7$). Top and middle panels show the evolution of the gas and pebble surface densities, respectively. The planetesimal formation efficiency in this simulation is $\epsilon=7.5\times10^{-7}$. The initial dust-to-gas ratio is $Z_0=0.01$, $\alpha_t=\alpha_{\nu}/40$, $\alpha_{\rm MRI}=3\alpha_{\nu}$, and $r_{\rm c}=\infty$.}
    \label{fig:twobumps_nominal}
\end{figure}

\begin{figure}[H]
\centering
\includegraphics[scale=.25]{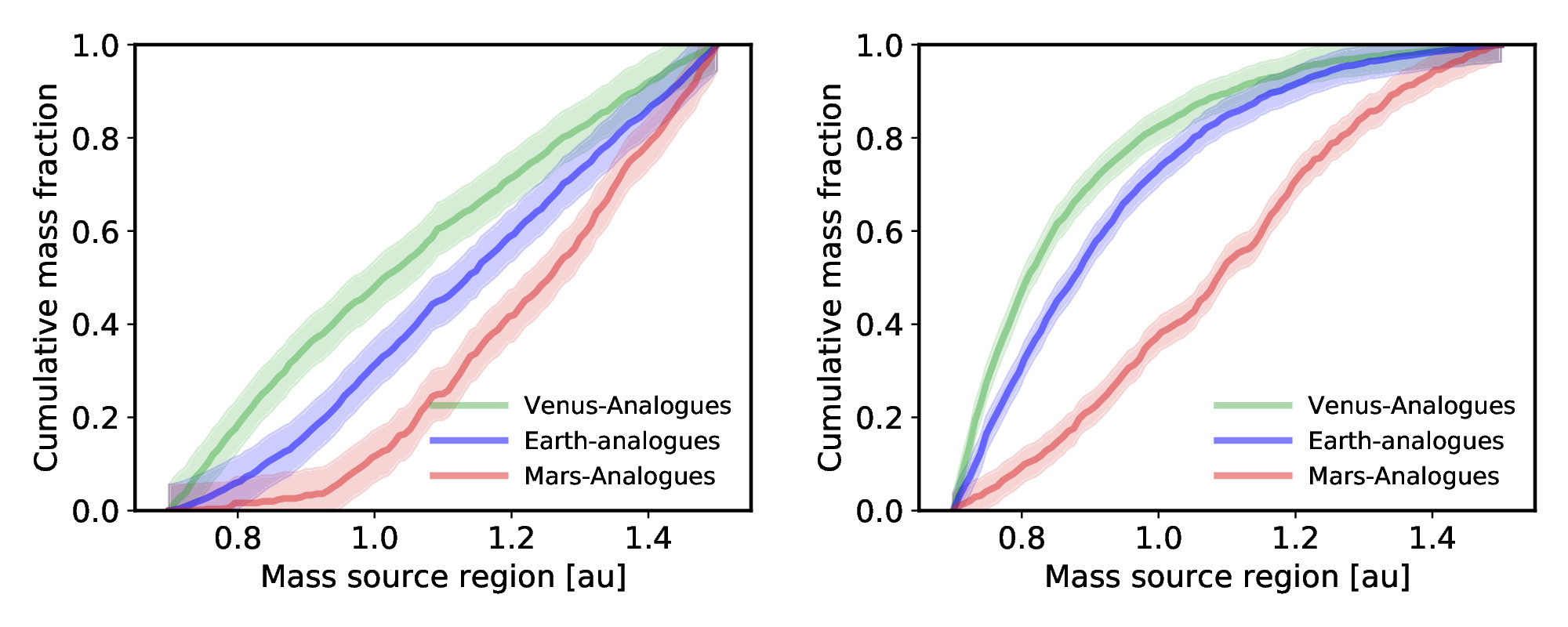}
\caption{Cumulative mass  fraction distributions  representing  the  feeding  zones  of  terrestrial planets in simulations with Jupiter and Saturn in their current orbits. {\bf Left:} Inner planetesimal rings with surface density profiles given by  $\Sigma_{\rm pla}\propto r^{-1}$. Curves are computed from 6 solar system analogues. {\bf Right:} Inner planetesimal rings with surface density profiles given by   $\Sigma_{\rm pla}\propto r^{-5.5}$. Curves are computed from 12 solar system analogues.  Thin green, blue and red curves represent Venus, Earth, and Mars analogues. Shaded regions encompassing each thin line represent 95\% confidence bands derived from the Kolmogorov-Smirnov statistic. Each selected planetary system contains one single Venus,  Earth,  and Mars-analogue.}
\label{fig:feeding_2rings}
\end{figure}

\begin{figure}[H]
\centering
\includegraphics[scale=0.4]{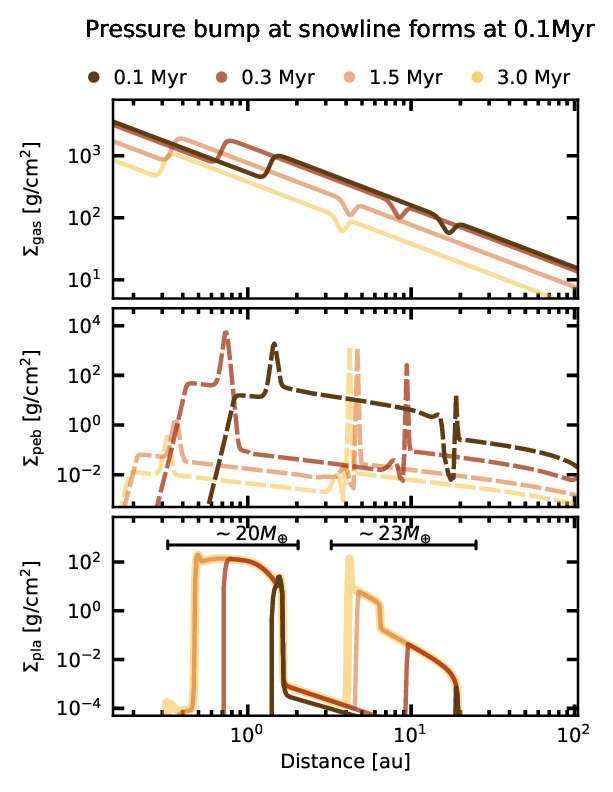}
    \caption{Simulation using the same parameters of simulation shown in Extended Data Figure \ref{fig:twobumps_nominal}, but considering that the bump at the snowline forms later, at $\sim$0.1~Myr after the beginning of the simulation. Planetesimal formation efficiency is set $\epsilon=7.5\times10^{-7}$.}
    \label{fig:growbumptocomparenominal}
\end{figure}

\renewcommand\contentsname{\huge Supplementary Information}



\beginsupplement

In this supplementary material we provide additional information on our methods and present the results of a simulations testing the effects of different parameters of our model.

\tableofcontents

\newpage

\section{Effects of different parameters of the model on the formation of two or three rings of planetesimals} 

In this section we present the results of  additional simulations testing the effects of several poorly constrained parameters of our model. We test the effects of the disk initial dust-to-gas ratio ($Z_0$), planetesimal formation efficiency ($\epsilon$), turbulent viscosity level in the disk midplane ($\alpha_t$),  minimum stokes number required to trigger planetesimal formation ($St_{min}$), timing of the bump formation\cite{Seguracoxetal20} at the snowline, and disk temperature profile and evolution. Our goal here is to illustrate the versatility of the model by showing that different combinations of parameters and assumptions may lead to very similar results.

We start by discussing the results of two simulations with three pressure bumps, similar to that shown  in Figure \ref{fig:lowefficiency} of the main text, but following a gas disk that cools faster with time and with a slightly different initial gas surface density profile. We use  $\beta=1$ and $r_{\rm c}=25$~au  (see Eq. \ref{eq:mmsn}). In both simulations of Extended Data Figure \ref{fig:threebumps_examepl2}, the disk temperature starts to drop with a short e-fold timescale of 0.75~Myr at  0.02~Myr. From 0.07~Myr to 2~Myr, we increase this timescale to 5~Myr. From 2~Myr to 3~Myr the disk dissipates with a timescale of 10~Myr. Note that the resulting disk temperature evolution is broadly consistent with that of more sophisticated disk models\cite{zhangetal15,baillieetal19}.  We assume a slightly lower initial dust-to-gas ratio $Z_0=1.3\%$\cite{asplundetal09} compared to that of Figure \ref{fig:lowefficiency} (main text), $\alpha_{t}=\alpha_{\nu}/27 \approx 3.7\times10^{-5}$, and $\epsilon=10^{-4}$.

In our model, we combine two different criteria for planetesimal formation. Planetesimal formation is either triggered when the local dust-to-gas ratio is higher than 1\citep{youdinshu02,drazkowskaalibert17,drazkowskadullemond18} or when a pebble flux larger than a critical value lead to  pebble concentration via zonal flows\cite{gerbigetal19,lenzetal19}.  In Extended Data Figure \ref{fig:threebumps_examepl2}, we show the dependence of our results on the planetesimal formation model.  Extended Data Figures \ref{fig:threebumps_examepl2}a and \ref{fig:threebumps_examepl2}b, show the results of two identical simulations, but in Extended Data Figure \ref{fig:threebumps_examepl2}b we neglect planetesimal formation via zonal flows. The formation of planetesimals between $\sim$1.5 and $\sim$3.0 au in Extended Data Figure \ref{fig:threebumps_examepl2}a (see also Figure \ref{fig:lowefficiency}) arises from our assumption that planetesimal formation also happens via assistance of short-lived zonal flows \cite{lenzetal19}. 

Whereas planetesimals born inside and outside the water snowline never overlap in Extended Data  Figure \ref{fig:threebumps_examepl2}b, this is not  true in Extended Data  Figure \ref{fig:threebumps_examepl2}a. In our simulations including the effects of zonal flows for planetesimal formation,  the water snow line sweeps early formed planetesimals born just inside the snowline as the disk cools and the water snowline moves inwards. Swept planetesimals are envisioned to have NC-like composition. We argue that this is unlikely to violate the solar system isotopic dichotomy if the surface density of NC-planetesimals produced via zonal flows is exceptionally low, i.e., virtually negligible compared to the local density of overlapping CC-planetesimals to provide significant mixing. In our nominal simulations,  the surface density of NC-planetesimals formed beyond 2-3~au  via zonal flows is $\sim10^{-2}-10^{-4}{\rm g/cm^2}$ or lower, whereas that of CC-planetesimals is typically at least two orders of magnitude higher. This strongly suggests, however, that planetesimal formation via zonal flows\cite{lenzetal19} must be sufficiently inefficient in the solar system, in particular inside the water snowline to not violate meteorite constraints when the bump at the water snowline moves inward. This issue may be also mitigated if Jupiter's core form sufficiently early\cite{Kruijeretal17,kruijeretal20} or if the bump at the snowline runs out of pebbles before sweeping NC-planetesimals. These different scenarios are accounted in Figures \ref{fig:lowefficiency} (main text) and Extended Data Figure \ref{fig:threebumps_examepl2}b. In Figure \ref{fig:lowefficiency}, we assume that Jupiter must have reached a critical mass  to block the pebble flux by $\lesssim$2~Myr\cite{Kruijeretal17,kruijeretal20}, avoiding the early overlapping of the inner and middle rings. In Extended Data Figure \ref{fig:threebumps_examepl2}b, the pressure bump at the snowline runs out of pebbles at $\sim$3~Myr, and, consequently, planetesimals stop forming in the middle ring avoiding any overlap of NC and CC-planetesimals. The  total mass in leftover pebbles beyond the water snowline in our simulations varies from $\sim$0 to $\sim100~M_{\oplus}$, depending on the setup of the model (e.g. $Z_0$, $r_{\rm c}$, disk size, etc). Some fraction of leftover pebbles would have been consumed by growing planetesimals and giant planet cores in the central and external rings\cite{lambrechtsjohansen12} potentially anticipating   the instant -- relative to the time that it happens in our simulations -- when the region  beyond the snowline runs out of pebbles. This process is not modeled in this work.

\begin{figure}[h]
\centering
\includegraphics[scale=1.1]{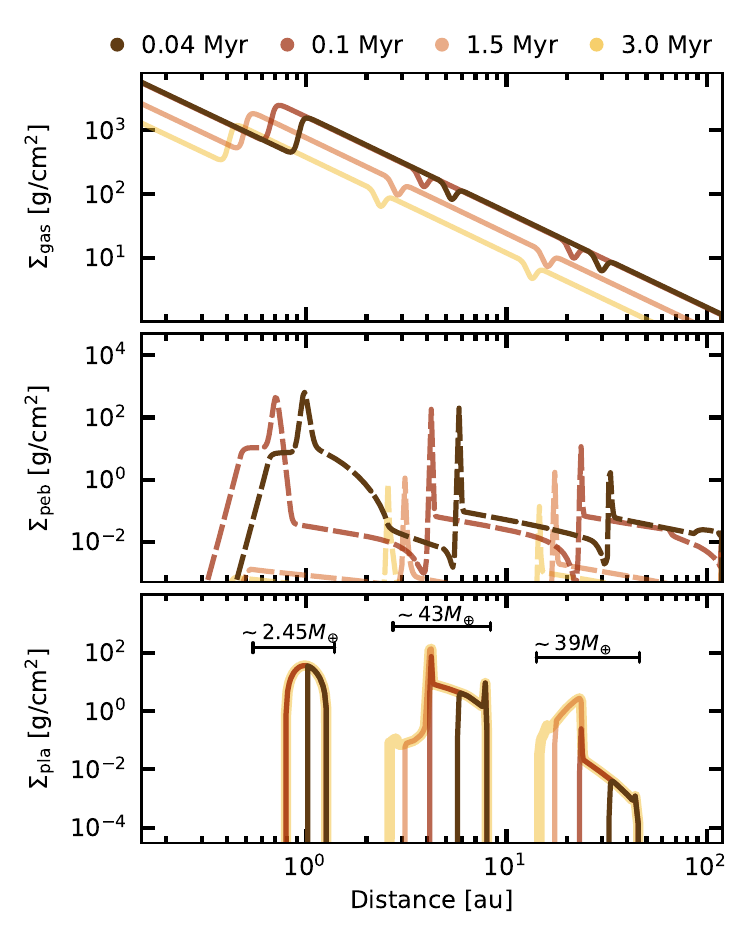}
 \hspace{0.7cm}
\caption{Final distribution of planetesimals in a simulation with three pressure bumps but with $\Sigma_{\rm ref}\propto r^{-1.5}$ (compare with our nominal disk profile given by Eq. \ref{eq:mmsn} in the main paper) and $r_{\rm c}=\infty$. The initial dust-to-gas ratio is $Z_0=1.5\%$, $\epsilon=2\times10^{-4}$, $\alpha_t=\alpha_{\rm \nu}/27.5$. The final rings contain 2.44~${M_{\oplus}}$ (inner), 43~${M_{\oplus}}$ (central), and 39~${M_{\oplus}}$ (outer) in planetesimals. We neglect planetesimal formation via zonal flows in this case.}
\label{fig:threebumps_examepl3}
\end{figure}

Supplementary Figure \ref{fig:threebumps_examepl3} shows the result of one of our simulations where $\Sigma_{\rm ref}\propto r^{-1.5}$ rather than $\Sigma_{\rm ref}\propto r^{-1}$, as assumed in our nominal simulations. The final mass in planetesimals in the inner ring is very similar to that of our nominal simulation (Extended Data Figure \ref{fig:threebumps_examepl2}). The planetesimal mass in the central/outer ring of Extended Data Figure \ref{fig:threebumps_examepl2} is  a factor of two lower/higher than in the middle/outer ring of Supplementary Figure \ref{fig:threebumps_examepl3}.  Our model is generally robust, and different combinations of plausible disk parameters may lead to broadly similar results.

\subsection{Two rings of planetesimals -} To demonstrate the dependence of our results on the disk temperature profile, we have performed simulations invoking an initially hotter protoplanetary disk, compared to that assumed in Figure \ref{fig:lowefficiency}. In simulation of Figure \ref{fig:lowefficiency}, the disk CO-snowline is initially at $\sim$40~au. In this section we discuss the results of simulations with disk temperature profile index equal to $\beta=0.7$.  In this set of simulations, the CO-snowline is initially at $\sim$200~au, i.e. beyond the outer edge of our disk, which is set  at 120~au. Although the disk cools with time, the amount of dust/pebbles remaining the in outer parts of the disk when the CO-snowline eventually gets inside 120~au is typically small. In this new set of simulations, we have assumed for simplicity that the gas disk includes only two pressure bumps ($A_{\rm CO}=0$ in Eq. \ref{eq:1pb}). In all simulations of this section we set $r_{\rm c}=\infty$.

Extended Data Figure \ref{fig:twobumps_nominal} shows  the result of one simulation with two pressure bumps\footnote{We have indeed verified that including the effects of a pressure bump at the CO-snowline in a simulation as that of Extended Data Figure \ref{fig:twobumps_nominal}, would produce a  third (outer) ring of planetesimals with only $\sim$0.01$M_{\oplus}$.}. Similar to Figure \ref{fig:lowefficiency}, planetesimals in the inner disk are clearly concentrated in a ring around $1$~au.


In our simulations, the total mass in planetesimals around 1~au may vary from  $<$0.1~${\rm M_{\oplus}}$ to $>$20~${\rm M_{\oplus}}$.  The same is true for the ring generated by the bump at the snowline, where the total mass in planetesimals varies from $<$40~${\rm M_{\oplus}}$ to $>$400~${\rm M_{\oplus}}$. In simulations with three pressure bumps, the final mass in planetesimals in the outer ring depends strongly on the disk extension and the pressure bump strength (amplitude and width).  Next we present the results of a suite of simulations exploring the effects of different parameters in simulations with two pressure bumps. We use the parameters  considered in Extended Data Figure \ref{fig:twobumps_nominal} as reference/nominal to discuss our parameter tests.

\begin{figure*}
\vspace{-1cm}
\centering
\begin{subfigure}[t]{.47\textwidth}
    \centering
\includegraphics[scale=.85]{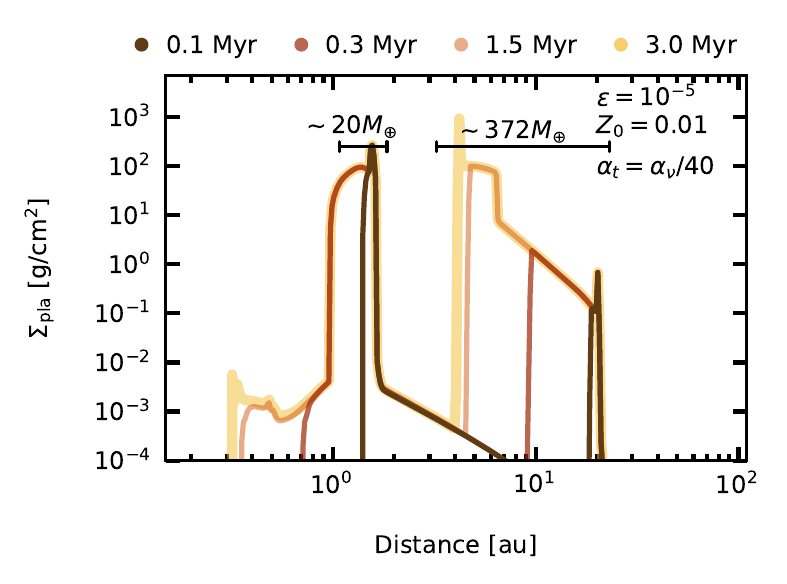}
\caption{$\epsilon$ is a factor of  $\sim$13 larger than that of Extended Data Figure \ref{fig:twobumps_nominal}.}
 \end{subfigure}
 \hspace{0.7cm}
 \begin{subfigure}[t]{.47\textwidth}
    \centering
\includegraphics[scale=.85]{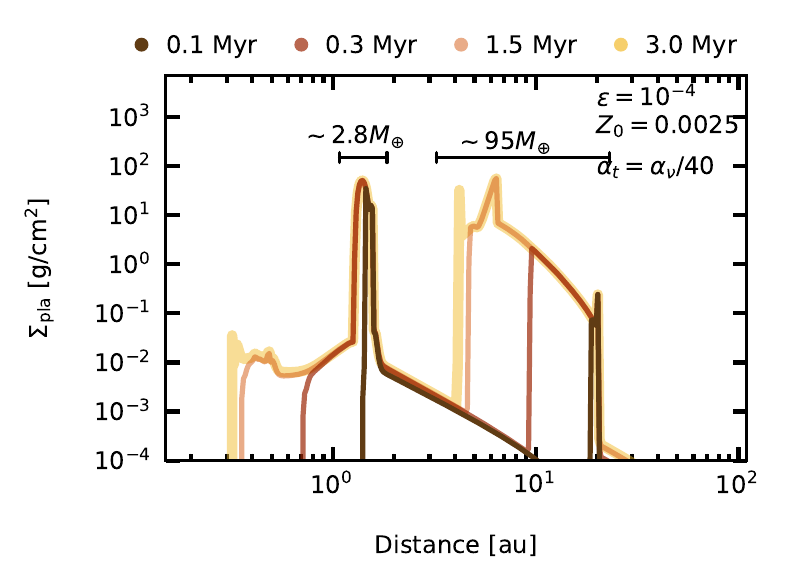}
 \caption{$\epsilon$ is factor of $\sim$100, and $Z_0$ a  factor of 4 smaller than those used in Extended Data Figure \ref{fig:twobumps_nominal}.}
 \end{subfigure}
 
  \begin{subfigure}[t]{.47\textwidth}
    \centering
\includegraphics[scale=.85]{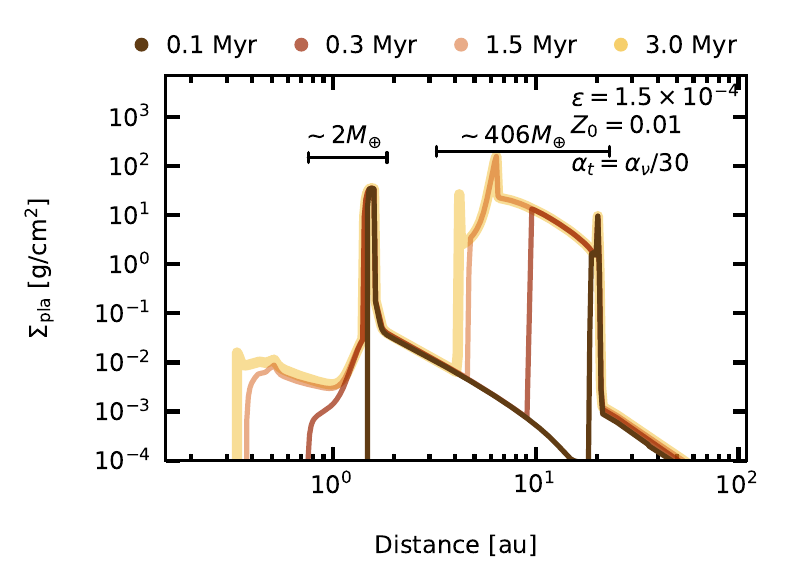}
\caption{$\epsilon$ is a factor of $\sim$200 larger, and $\alpha_{\rm t}$ is $\sim$33\% larger than those used in Extended Data Figure \ref{fig:twobumps_nominal}.}
 \end{subfigure}
  \hspace{0.7cm}
  \begin{subfigure}[t]{.47\textwidth}
    \centering
\includegraphics[scale=.8]{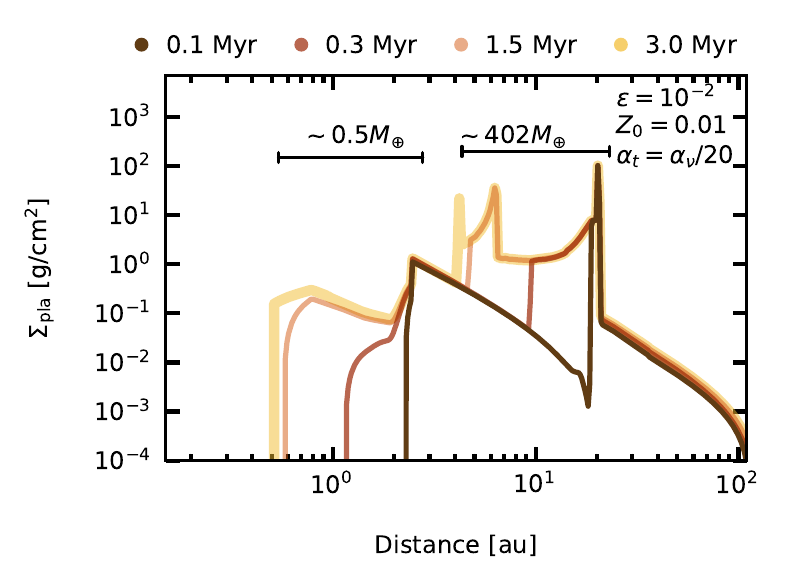}
\caption{$\epsilon$  is a factor of $\sim$1000 larger, and $\alpha_{\rm t}$ is a factor 2 larger than those used in Extended Data Figure \ref{fig:twobumps_nominal}.}
 \end{subfigure}

\begin{subfigure}[t]{.47\textwidth}
    \centering
\includegraphics[scale=.85]{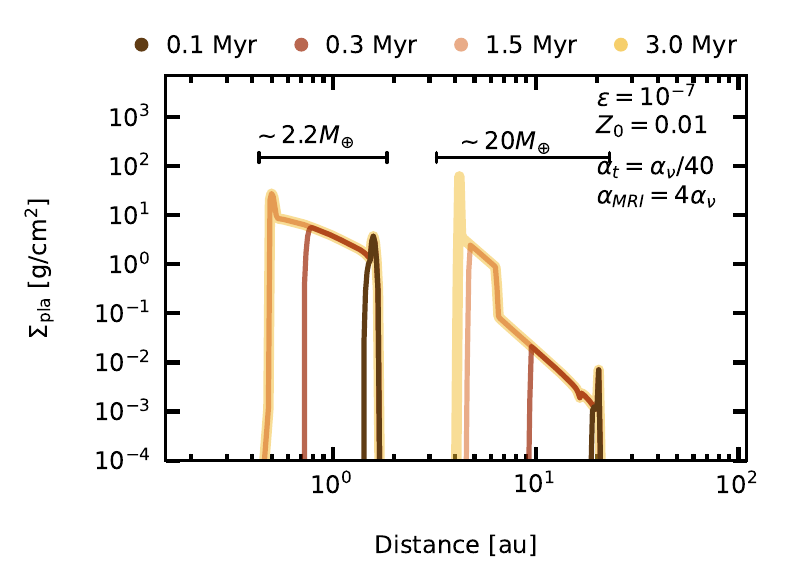}
\caption{$\epsilon$ is a factor of $\sim$7 smaller than that used in Extended Data Figure \ref{fig:twobumps_nominal}, and $\alpha_{\rm MRI}=4\alpha_{\nu}$ whereas in Extended Data Figure \ref{fig:twobumps_nominal} we set $\alpha_{\rm MRI}=3\alpha_{\nu}$. }
 \end{subfigure}
  \hspace{0.7cm}
  \begin{subfigure}[t]{.47\textwidth}
    \centering
\includegraphics[scale=.85]{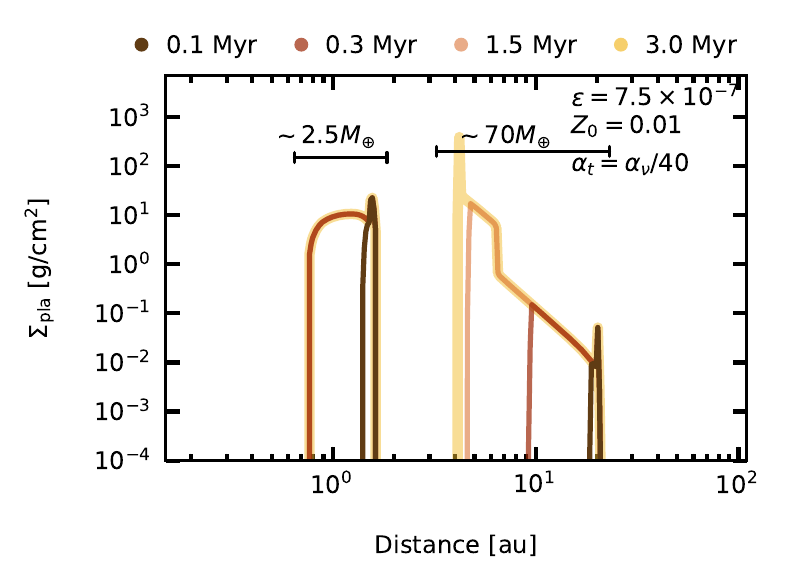}
\caption{Disk parameters are the same as in Extended Data Figure \ref{fig:twobumps_nominal}, but neglecting planetesimal formation via zonal flows\cite{lenzetal19}. In this case, planetesimals do not form in the belt}.
 \end{subfigure}
 \caption{Example of planetesimal rings produced in simulations with different planetesimal formation efficiency ($\epsilon$), initial dust-to-gas ratio ($Z_0$), disk viscosity in the highly ionized region of the disk ($\alpha_{\rm MRI}$), and level of turbulence in the disk midplane $\alpha_{t}$. In all cases, the bump at the snowline exists since the beginning. When a specific parameter of the model is not explicitly defined in the panel, it is the same of that of Extended Data Figure \ref{fig:twobumps_nominal}.}
 \label{fig:starwithtbump}
 \end{figure*}

Supplementary Figure \ref{fig:starwithtbump}a shows that increasing the planetesimal formation efficiency ($\epsilon$) by a factor of $\sim$10 relative to our nominal value considered in Extended Data Figure \ref{fig:twobumps_nominal} increases the total mass in planetesimals in the inner ring from $\sim$2.5$M_{\oplus}$ to  $\sim$20$M_{\oplus}$, and the total mass in planetesimals in outer ring from $\sim$70$M_{\oplus}$ to  $\sim$372$M_{\oplus}$.  Supplementary Figure \ref{fig:starwithtbump}b shows that two rings of planetesimals are also formed  in simulations where the initial dust-to-gas ratio ($Z_0$)  is reduced by a factor of 4 relative to the nominal value,  while the planetesimal formation efficiency is increased by a factor of $\sim$100 ($\epsilon=10^{-4}$ is in fact the value typically invoked in previous studies\cite{drazkowskaalibert17,uedaetal19}).  Supplementary Figure \ref{fig:starwithtbump}c shows that by increasing the turbulent level of the gas disk midplane, represented by the parameter $\alpha_t$, by 33\%, relative to the nominal value, requires a planetesimal formation efficiency equal to $\epsilon=1.5\times10^{-4}$ in order to produce a ring of planetesimals with $\sim$2.0$M_{\oplus}$. The turbulence level in the disk midplane controls the size of pebbles in the disk~\cite{birnstieletal10}. The outer planetesimal ring, in this particular case, is very massive and carries $\sim$400$M_{\oplus}$ in planetesimals. If one  increases the turbulence level at the disk midplane by a factor of 2 relative to the nominal simulation, planetesimal formation becomes very inefficient in the inner disk, even if we increase the efficiency of planetesimal formation by a factor of $>$1000 (Supplementary Figure \ref{fig:starwithtbump}d). In Supplementary Figure \ref{fig:starwithtbump}-e, we decrease planetesimal formation efficiency by a factor of 7 and  increase the disk viscosity in the strongly-ionized region of the disk ($T_{gas}>1000$~K). This makes the inner pressure ``stronger'', meaning that pebbles can pile-up more efficiently at the bump which compensate the imposed reduction in the planetesimal formation efficiency. The final inner planetesimal ring is as massive as that of Extended Data Figure \ref{fig:twobumps_nominal},  containing $\sim$2.2$M_{\oplus}$. Supplementary Figure \ref{fig:starwithtbump}-f shows the distribution of planetesimals in a simulation with the same parameters of Extended Data Figure \ref{fig:twobumps_nominal} but neglecting the formation of planetesimal via zonal flows\cite{lenzetal19}.  Efficient planetesimal formation via zonal flows\citep{lenzetal19} may not be required to explain the formation of the inner solar system.

\subsection{Effect of the timing of the pressure bump formation - }

In  our nominal simulations, we assume that the inner bump is always present since the beginning and we test how the formation of the pressure bump at the snowline at different times would impact our results. Extended Data Figure \ref{fig:growbumptocomparenominal} shows the result of a simulation identical to that of Extended Data Figure \ref{fig:twobumps_nominal}, but with the bump at the snowline growing at $\sim$0.1~Myr. We mimic the bump formation by using linear interpolation between the original power law disk ($A=0$; see methods in the main paper) and the disk profile with the bump ($A=0.5$). The pressure bump is assumed to grow in a timescale of 20~kyr. Until the pressure bump form,  pebbles originally  beyond the snowline can drift inwards and enter into the inner disk (Extended Data Figure \ref{fig:growbumptocomparenominal}).  Two rings of planetesimals form as in Extended Data Figure \ref{fig:twobumps_nominal}, however the inner disk is very massive and contains  $\sim20M_{\oplus}$ in planetesimals. The outer ring is less massive than that in our nominal simulation and contains $23M_{\oplus}$ in planetesimals. These are consequences of the pebbles in the outer disk being lost by inward drift, and eventually promoting planetesimal formation in the inner disk. We explore the effects of the timing of the bump formation only for  simulations with two pressure bumps. Disks with three pressure bumps would produce broadly equivalent results.

\begin{figure*}
\vspace{-1cm}
\centering
\begin{subfigure}[t]{\textwidth}
    \centering
\includegraphics[scale=1.]{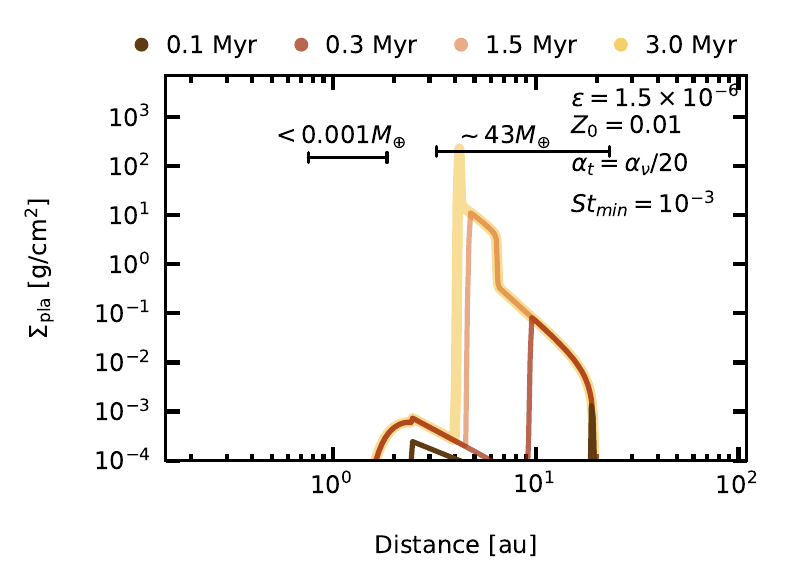}
\caption{$\epsilon$ and $\alpha_{\rm t}$ are a factor of $\sim$2 larger than our nominal values, as given in Extended Data Figure \ref{fig:growbumptocomparenominal}.}
 \end{subfigure}
\begin{subfigure}[t]{\textwidth}
    \centering
\includegraphics[scale=1.]{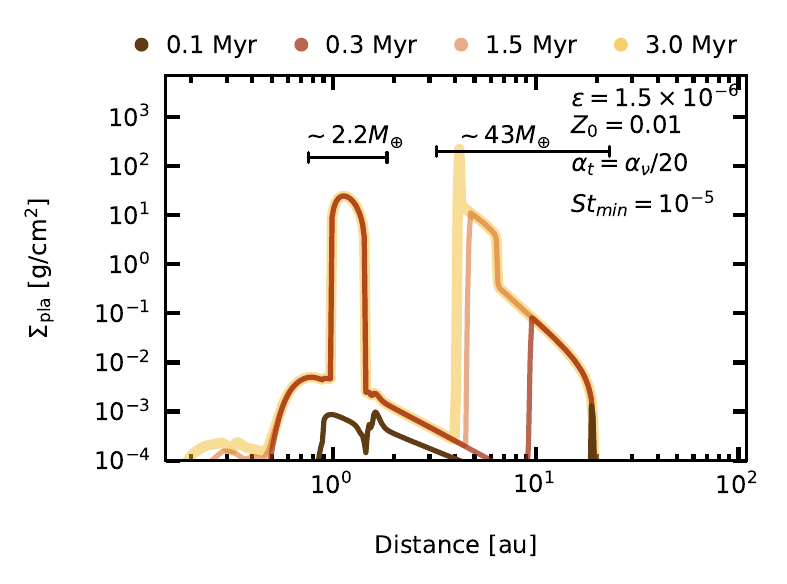}
\caption{Same as the panel above but relaxing the minimum stokes number for planetesimal formation to $St_{min}=10^{-5}$.}
 \end{subfigure}
\begin{subfigure}[t]{\textwidth}
    \centering
\includegraphics[scale=1.]{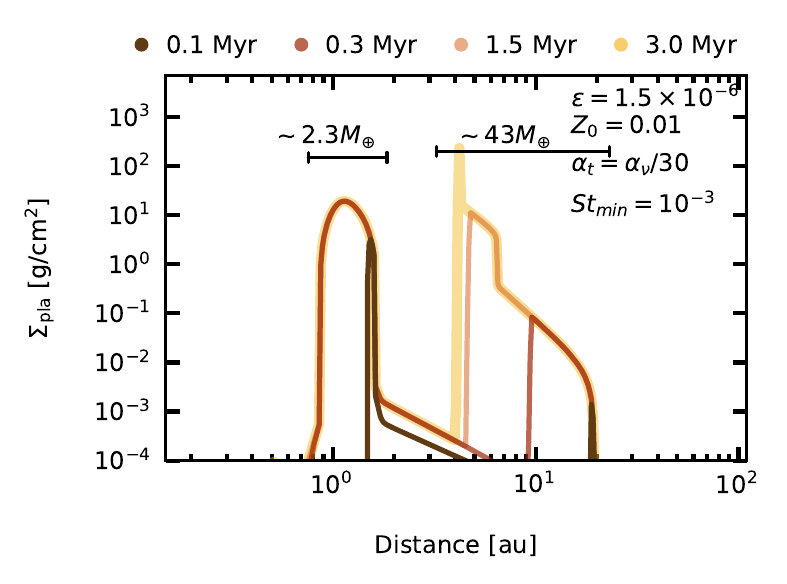}
\caption{Same as the top panel but reducing the turbulent level at the disk midplane  by a factor of 1.5.}
 \end{subfigure}
\caption{Additional simulations exploring the effects of different parameters, as in Supplementary Figure \ref{fig:starwithtbump} and Extended Data Figure \ref{fig:growbumptocomparenominal}. Panels show the final distributions of planetesimals. In all cases, the pressure bump forms at $\sim$0.1~Myr. Here we also show the effects of changing the minimtum Stokes number allowed for planetesimal formation~ ($St_{\rm min}$)}.
\label{fig:growbumpadditionalcases}
\end{figure*}

Supplementary Figure \ref{fig:growbumpadditionalcases} shows three simulations exploring the effects of different parameters in simulations with time dependent pressure bumps. In all cases the pressure bump at the snowline form at $\sim$0.1~Myr. Supplementary Figure \ref{fig:growbumpadditionalcases}a-b shows that  in simulations where $\alpha_t=\alpha_{\nu}/20$ (a factor 2 larger than nominal) do not form a ring of planetesimals in the inner disk if the minimum stokes number for planetesimal formation is set as $St_{min}=10^{-3}$. However, if one relax this restriction\cite{yangetal17} and assume that planetesimal formation is efficient when $St_{min}\geq10^{-5}$, a ring of planetesimals form in the inner ring with  $2.2M_{\oplus}$ in planetesimals. In Supplementary Figure \ref{fig:growbumpadditionalcases}c we recover the result of Supplementary Figure \ref{fig:growbumpadditionalcases}b by decreasing the level of turbulence in the disk and, again, by  increasing the minimum stokes number for planetesimal formation\cite{birnstieletal12}. These results shows that different combination of parameters may lead to planetesimal rings with similar structures (total mass and location). The minimum stokes number of pebbles required to streaming instability to operate is not strongly constrained, but recent simulations suggest that  it may require particles with $St_{min}\geq10^{-3}$ and local dust-to-gas ratio of at least a few percent\cite{yangetal17,liyoudin21}. 

The amount of mass in planetesimals in the inner ring correlates with the level of turbulence in the disk midplane controlling pebbles sizes.  If silicate pebbles are sufficiently large they pile-up efficiently at the inner ring. If they are sufficiently small, they are well coupled to gas and do not pile at the inner bump very efficiently. If pebbles do not pile up at the bump, planetesimal formation in the inner disk can only happen via assistance of zonal flows enhancing dust concentration in the disk.  This mechanism may require stokes number larger than $St_{min}\geq10^{-3}-10^{-2}$ to efficiently operate\cite{lenzetal19}. As shown in Supplementary Figure \ref{fig:starwithtbump}-e  The total planetesimal mass in the inner bump also depends on $\alpha_{MRI}$. If this parameters is assumed larger than in our simulations the inner bump may become ``stronger''  and pile-up small pebbles more efficiently.

Still, we have demonstrated in this section the sensibility of our results to poorly constrained parameters of our model. We showed that different combinations of parameters produce similar results. The main advantage of our comprehensive approach is that it is able to identity a part of the parameter space that can explain several solar system constraints in a fairly simplified model of planet formation.

\section{Additional Methods}

Our simulations of the late stage of accretion of terrestrial planets start from distributions of planetary embryos and planetesimals.  A sub-set of the simulations presented here is presented in previous studies~\cite{raymondizidoro17b}. Figure \ref{fig:terrestrial} (main text) is produced from:


\begin{itemize}
    \item 25 simulations starting with $\sim$2.5$M_{\oplus}$ in solids distributed between 0.7~au and 1.5~au following a disk with surface density profile proportional to $r^{-1}$. 2000 planetesimals receive about 30\% of the total available mass and 50 equal-mass planetary embryos receive the remaining mass. About 20\% (400) of the planetesimals and planetary embryos (10) were spread out between 1 and 1.5 AU.
    \item 15 simulations starting with 2.5$M_{\oplus}$ in solids distributed between 0.7~au and 1.5~au following a disk with surface density profile proportional to $r^{-5.5}$. 2000 planetesimals receive about 30\% of the total available mass and 50 equal-mass planetary embryos receive the remaining mass.   
    \item 20 simulations starting with 2.5$M_{\oplus}$ in solids distributed between 0.7~au and 1.5~au following a disk with surface density profile proportional to $r^{-1}$.  3000 planetesimals receive about 30\% of the total available mass and 20 planetary embryos receive the remaining mass. Planetary embryos' masses are randomly selected between 0.5$M_{\rm Mars}$ and 1.5$M_{\rm Mars}$.
    \item 20 simulations starting with  with 2.5$M_{\oplus}$ in solids distributed between 0.7~au and 1.5~au following a disk with surface density profile proportional to $r^{-5.5}$.  3000 planetesimals receive about 30\% of the total available mass and 20 planetary embryos receive the remaining mass. Planetary embryos' masses are randomly selected between 0.5$M_{\rm Mars}$ and 1.5$M_{\rm Mars}$.    
\end{itemize}

\subsection{Solar System analogues -} 
We define solar system analogues in Figure \ref{fig:feeding} as planetary systems forming at least three planets simultaneously satisfying the following conditions:
\begin{itemize}
\item \textbf{Venus-analogue}: $0.5~{\rm au}<a< 0.9~{\rm au}$ and $0.4M_{\oplus}<M<1.2M_{\oplus}$;
\item \textbf{Earth-analogue}: $0.7~{\rm au}<a< 1.25~{\rm au}$ and $0.7M_{\oplus}<M<1.4M_{\oplus}$;
\item \textbf{Mars-analogue}: $1.25~{\rm au}<a< 1.8~{\rm au}$ and $0.03M_{\oplus}<M<0.3M_{\oplus}$.
\end{itemize}

\begin{figure*}[h]
\centering
\includegraphics[scale=.6]{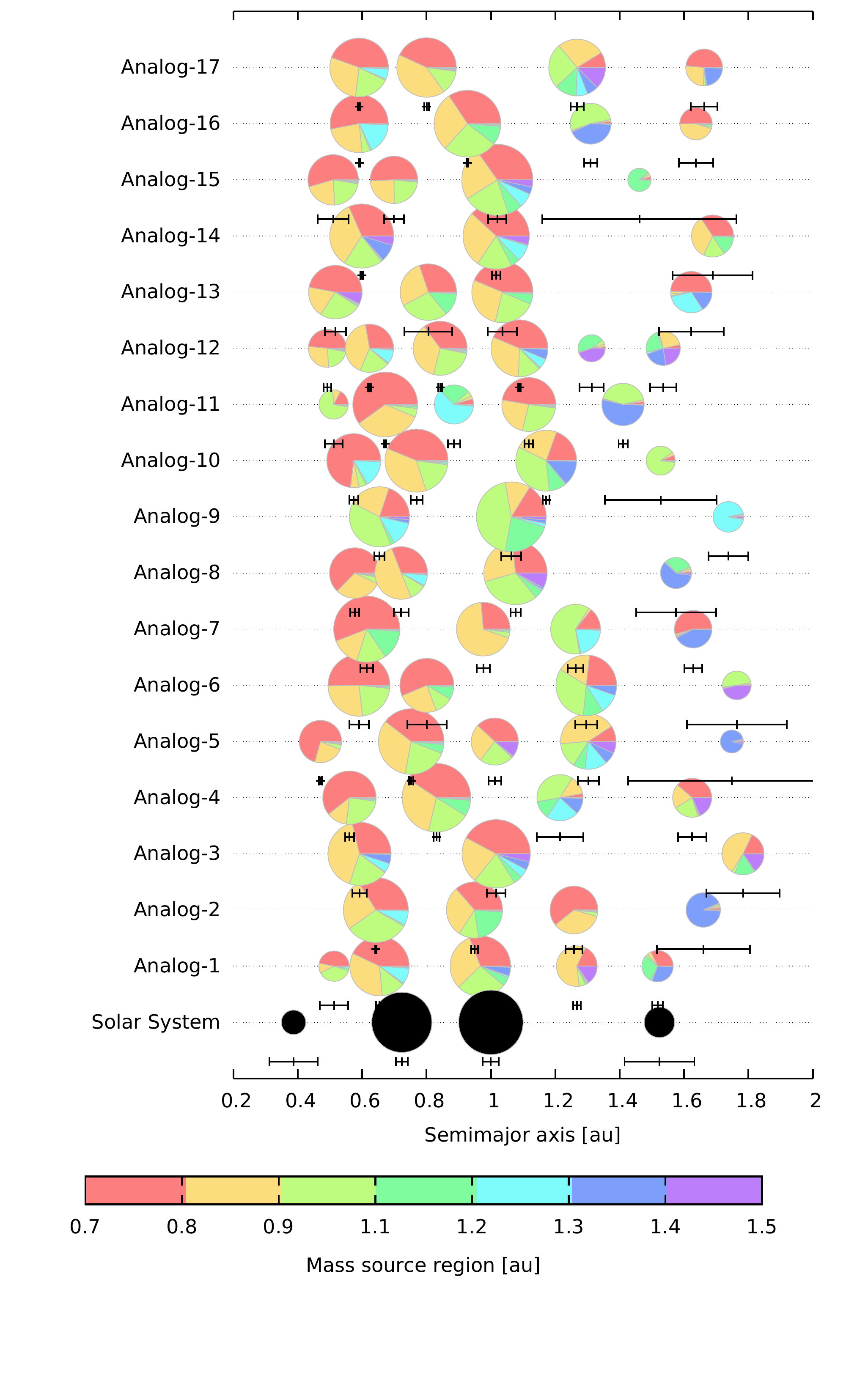}
\vspace{-1.5cm}
    \caption{Planetary systems produced in our simulations satisfying our definition of solar system analogue. Each row represents one system of terrestrial planets.  The solar system is shown at the bottom for reference. The size of the dot representing each planet scales with the planet mass as $m^{1/3}$, where $m$ is the mass. Pie charts show the relative feeding zones of each planet as indicated by the color-coding. The orbital eccentricity of each planet is represented by its horizontal bar showing the perihelion and aphelion of the orbit.}
    \label{fig:selectedsystems}
\end{figure*}

Supplementary Figure \ref{fig:selectedsystems} shows  seventeen selected solar system analogues extracted from  Figure  \ref{fig:terrestrial}. Supplementary Figure \ref{fig:selectedsystems}  shows that our best solar system analogues produce between 3 and 4 terrestrial planets. These individual systems also show that Earth and Mars-analogues can have very different feeding zones, which is in agreement with Figure \ref{fig:feeding} of the main paper. 

Our simulations produced a few Mercury-mass planets around 0.4-0.5~au (Figure \ref{fig:terrestrial}).  Mercury's small mass is most likely an effect of the inner edge of the terrestrial ring\cite{hansen09,walshetal11}.  None of our selected solar system analogues produced  good Mercury-analogues. This issue will be subject of a future work.

\subsection{Growth from planetesimal to planetary embryos -}

\begin{figure*}
\centering
\includegraphics[scale=0.7]{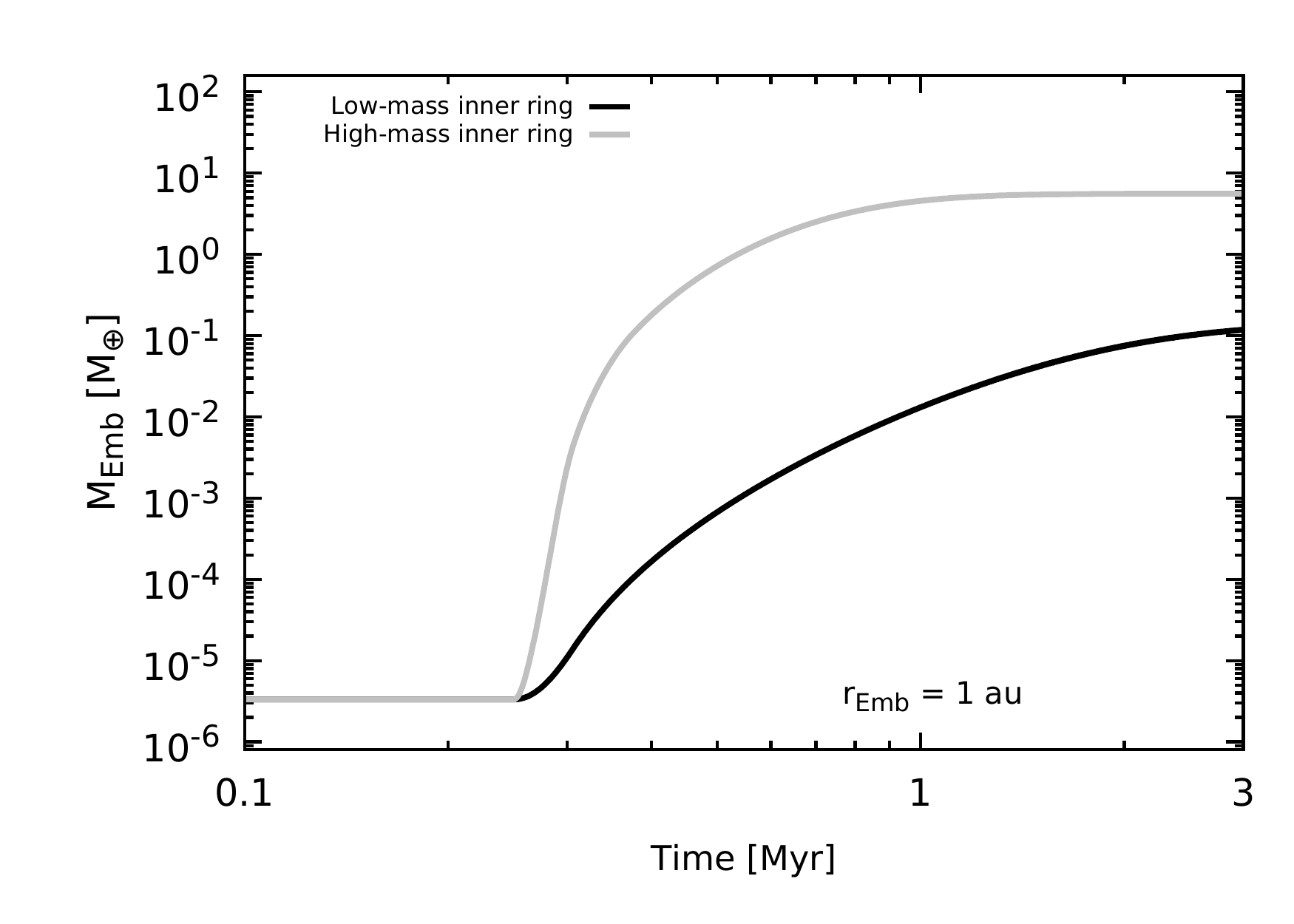}
\includegraphics[scale=0.7]{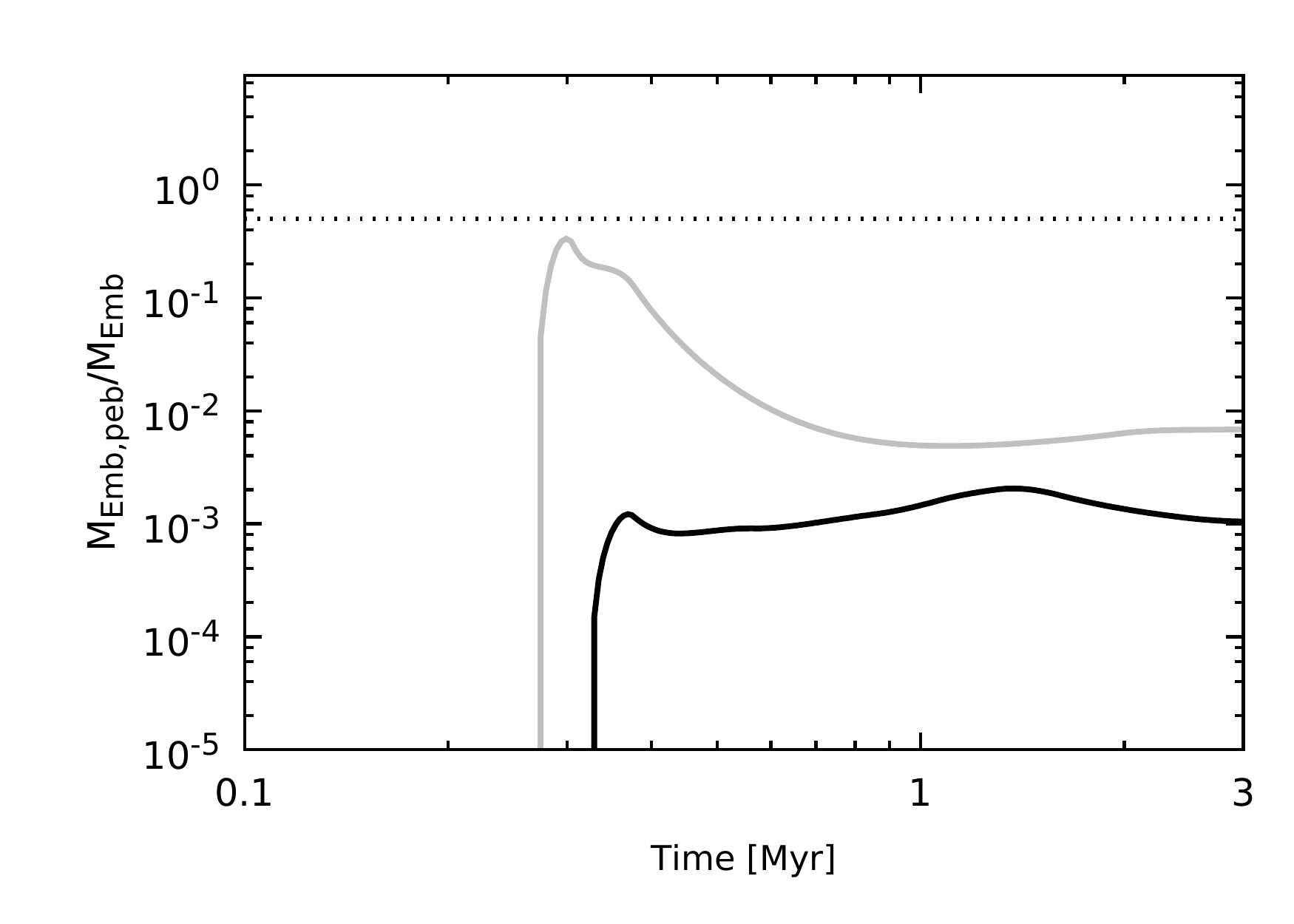}
    \caption{Growth of protoplanetary embryos at 1~au in inner rings of planetesimals with different masses. {\bf Top:} The black-curve shows a protoplanetary embryo growing in a disk with $\sim2.5M_{\oplus}$ (Extended Data Figure \ref{fig:twobumps_nominal}) and the grey-curve a planetary embryo in a ring with $\sim20M_{\oplus}$ (Supplementary Figure \ref{fig:starwithtbump}a). {\bf Bottom:} Fractional contribution of pebble accretion to the embryo’s mass. The horizontal dotted line marks a  50\% contribution of pebble accretion. We use analytical prescriptions to calculate growth via pebble and planetesimal accretion as described in previous studies\citep{izidoroetal21}. We self-consistently feed our calculations with the results from our dust coagulation simulations.}
    \label{fig:analytical}
\end{figure*}

We first estimate the growth mode of planetesimals in the inner ring using analytical prescription to model pebble and planetesimal accretion\citep{izidoroetal21}. Supplementary Figure \ref{fig:analytical} shows the growth of two protoplanetary embryos in inner rings with different total masses in planetesimals (top-panel), and the contribution of pebble accretion to their final masses (bottom-panel). The  black growth curve represents a 100-km sized planetesimal growing in the inner ring of Extended Data Figure \ref{fig:twobumps_nominal}, where the total mass in planetesimals in the inner ring is $2.5M_{\oplus}$. This planetesimal grow relatively slowly, reaching one Mars-mass in about 2-3~Myr. The grey growth curve  shows a  100-km sized planetesimal growing in a inner ring with $\sim20M_{\oplus}$ (Supplementary Figure \ref{fig:starwithtbump}a). The planetesimal in this  high-mass  inner ring grow relatively faster reaching mass of a few Earth-masses during the first 500~kyr of the simulation. We  do not model the migration of protoplanetary embryos in our simulations. Whereas a slowly growing Mars-mass planetary embryo would be unlikely to migrate substantially\cite{kleynelson12}, an Earth-mass (or more massive) protoplanetary embryo would strongly interact with the disk and potentially migrate inwards, reaching the inner edge of the disk\citep{lambrechtsetal19}. This evolution would make this system fully inconsistent with the current solar system. 

The bottom panel of Supplementary Figure \ref{fig:analytical} shows that both planetesimals grow to planetary embryos systematically by accreting other planetesimals rather than via pebble accretion (this is also the case for planetesimals in Figure \ref{fig:lowefficiency} of the main text) . This is consistent with the results of Izidoro et al\cite{izidoroetal21}, and reflects the interruption of the pebble supply to the inner disc caused by the pressure bump at the water snowline.

\begin{figure*}
\centering
\vspace{-1cm}
\includegraphics[scale=.31]{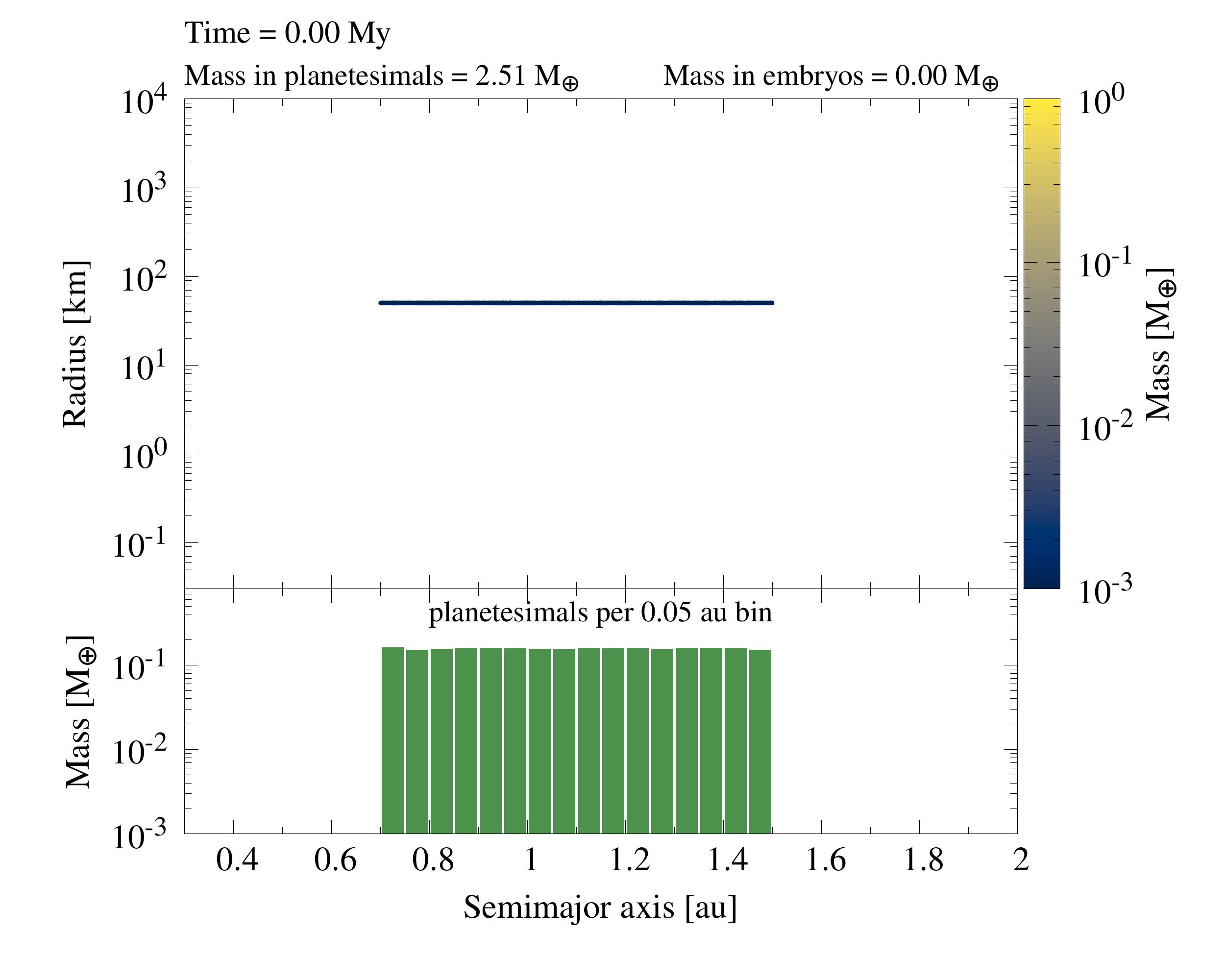}
\includegraphics[scale=.31]{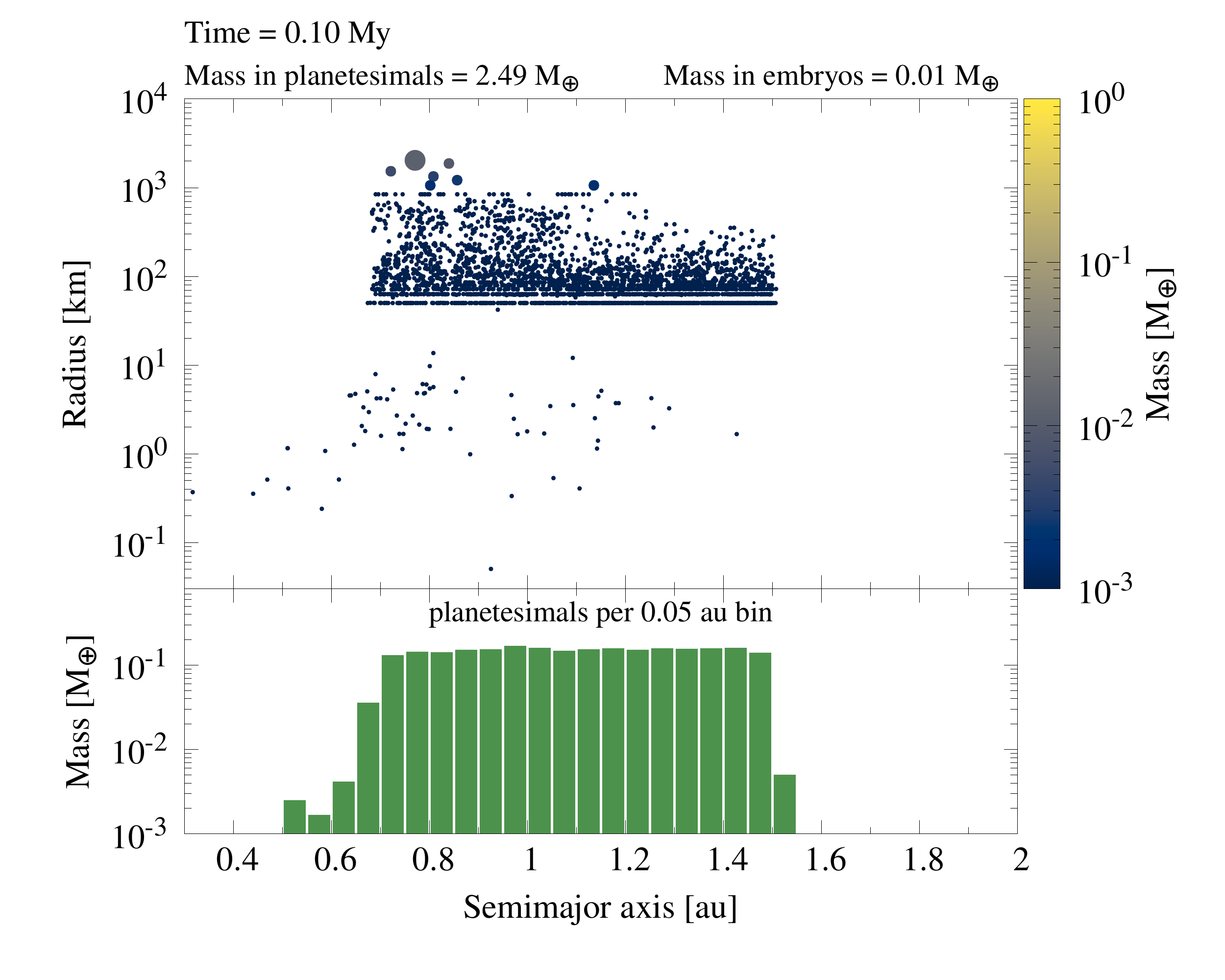}
\includegraphics[scale=.31]{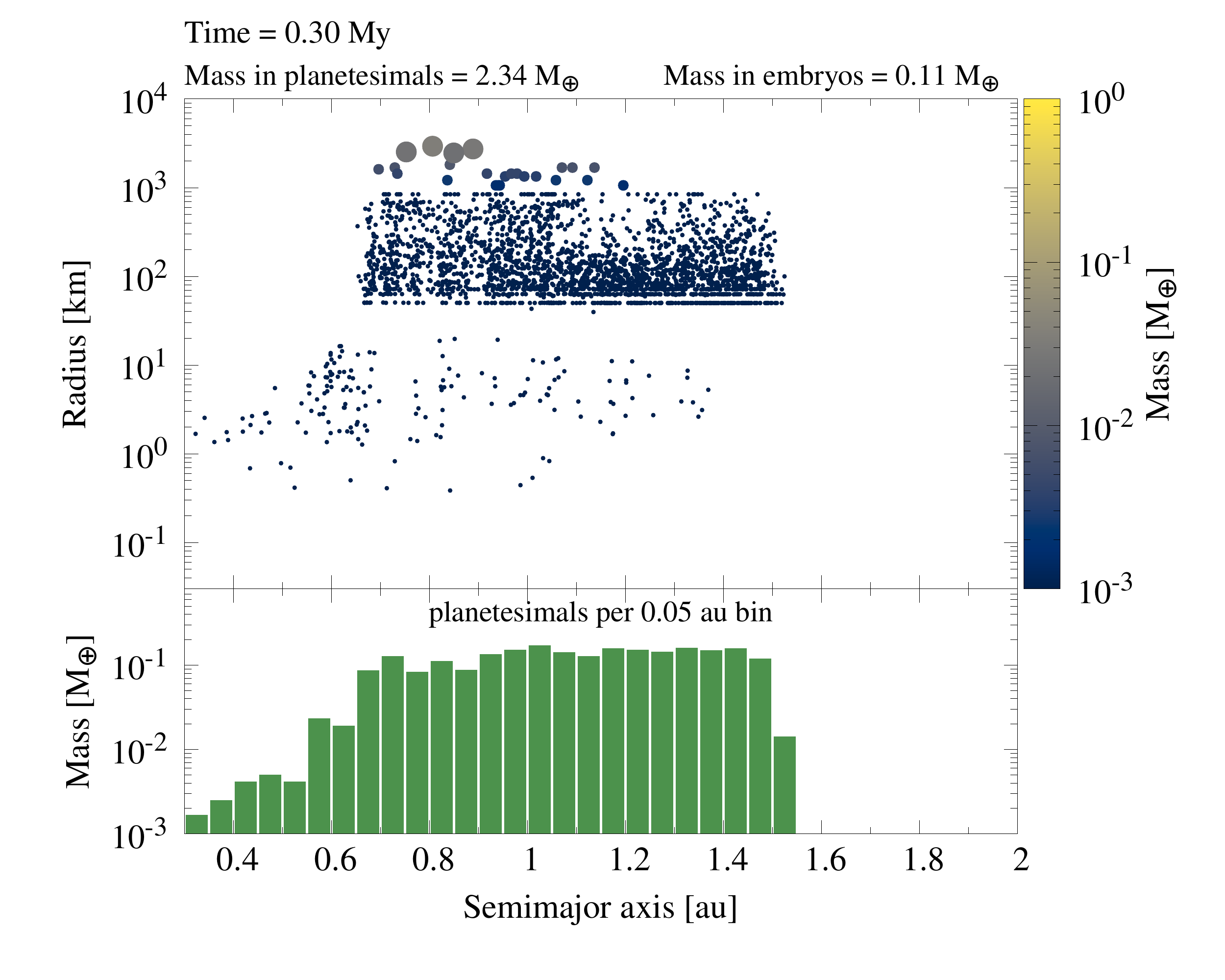}
\includegraphics[scale=.31]{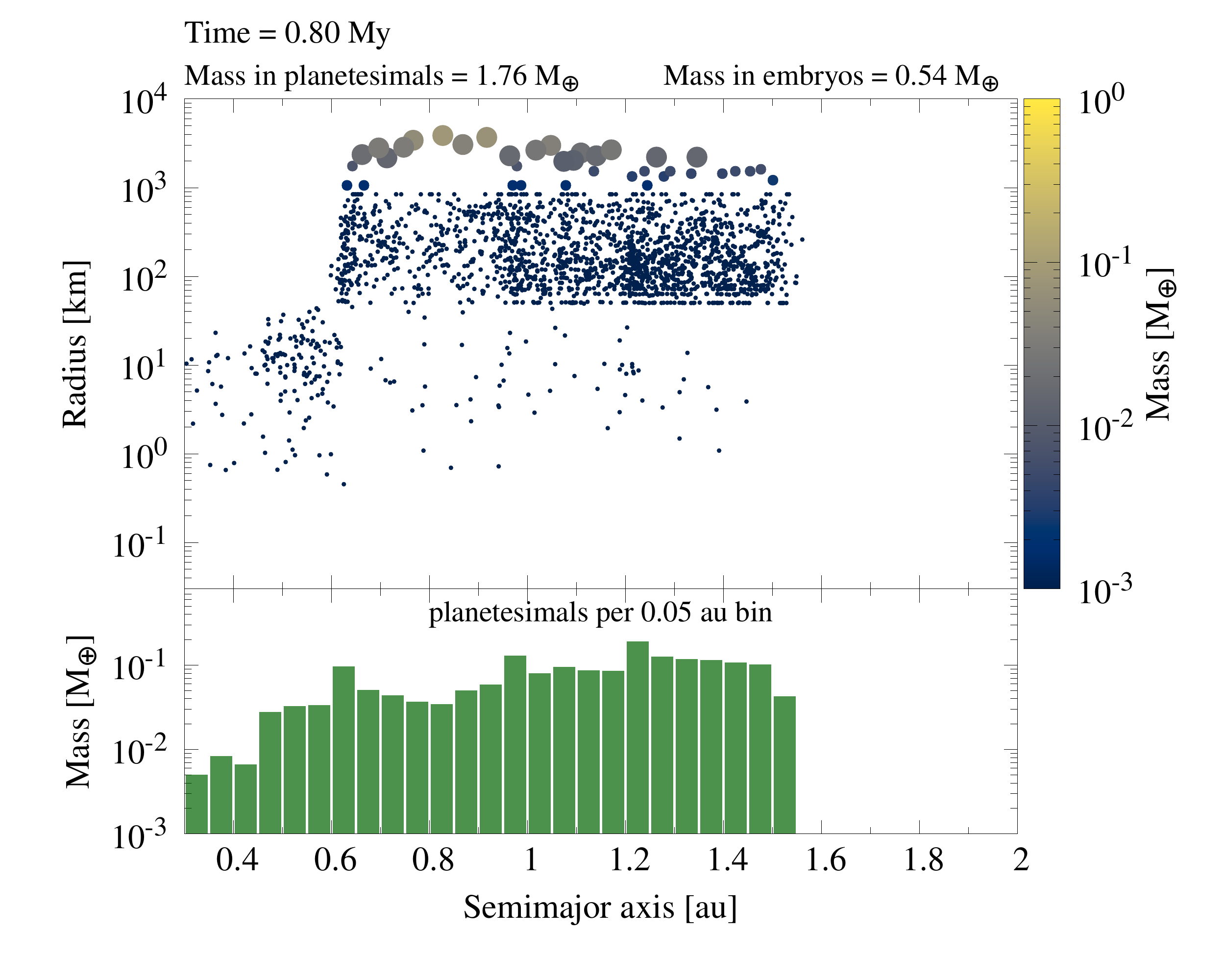}
\includegraphics[scale=.31]{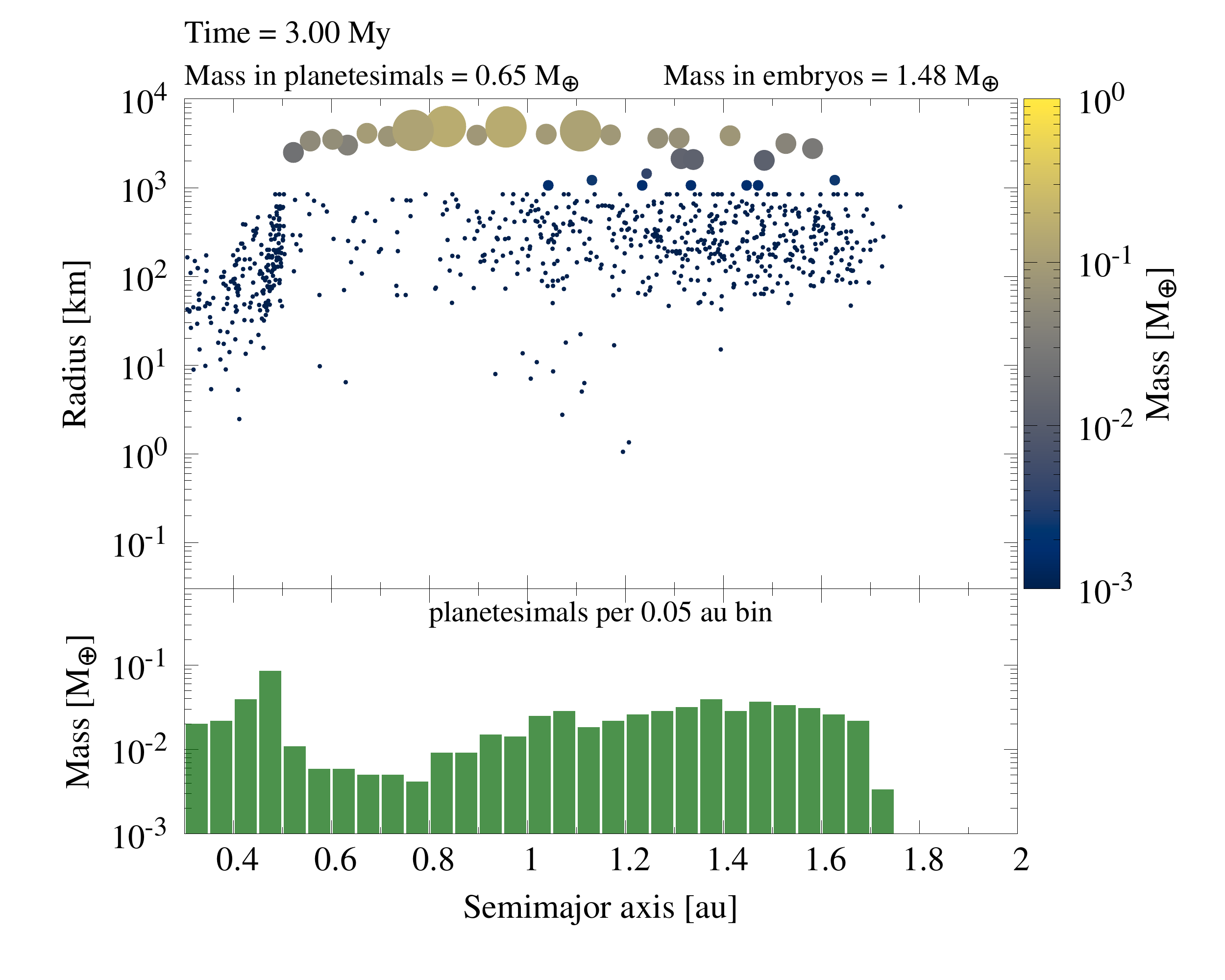}
\includegraphics[scale=.31]{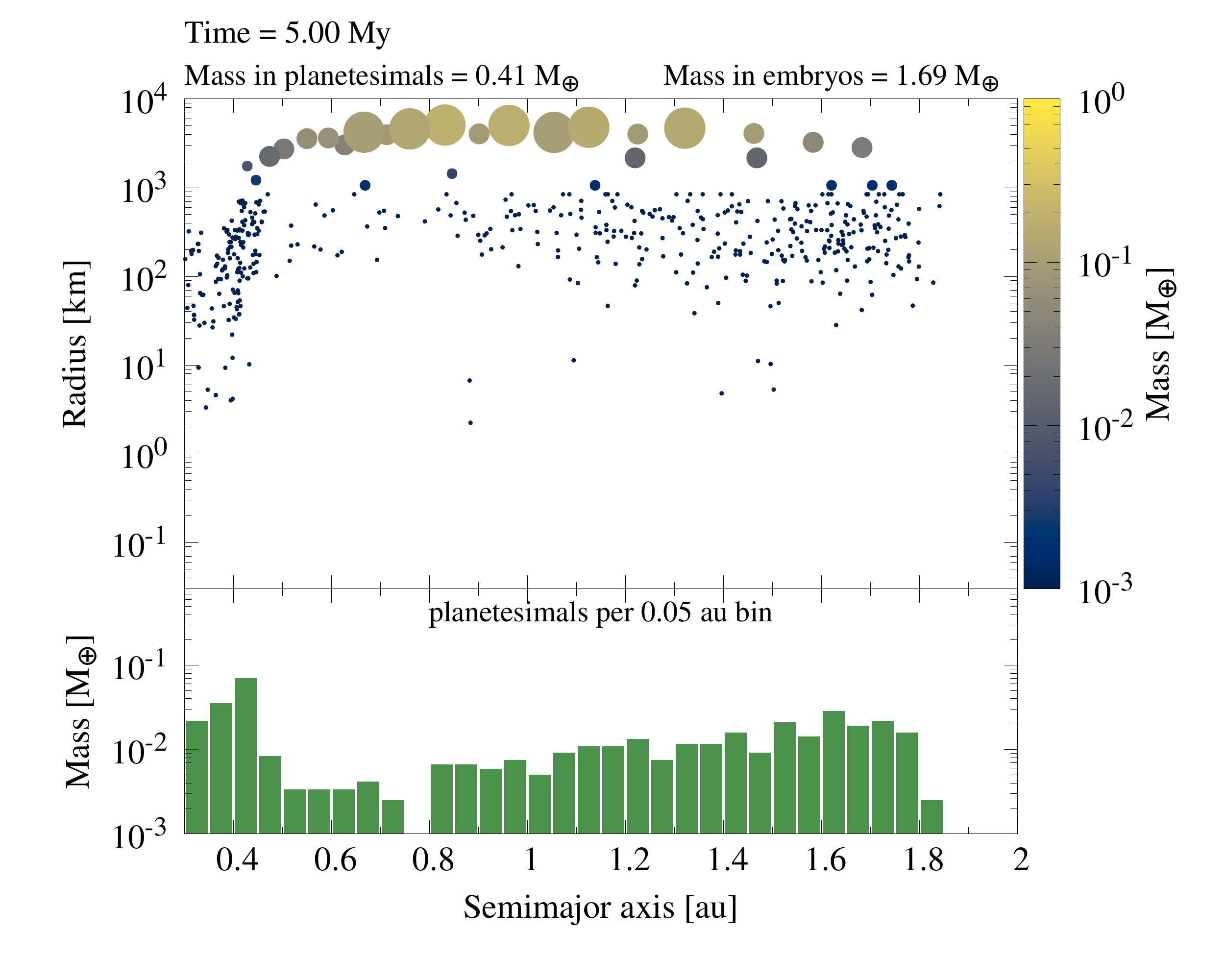}
\vspace{-0.5cm}
\caption{Snapshots of the evolution of growing planetesimals in a narrow ring around 1~au with surface density profile proportional to $r^{-1}$. Each snapshot shows the planetesimal sizes (top-panel) and  the planetesimal mass (bottom-panel), both as function of the semi-major axis.  The ring extends initially from $\sim$0.7 to $\sim$1.5~au. The initial total mass in planetesimals is $\sim$2.5$M_{\oplus}$. Planetesimals have initial diameters of 100~km. Planetesimals  gravitationally interact with each other\cite{levisonetal12} and also feel gas-drag effects\cite{brasseretal07}. Collisions may result in growth or fragmentation, as modeled in the LIPAD code\citep{benzasphaug99,levisonetal12}. The gas disk model used to conduct this simulation is described in the Methods section of the main paper. Color-coding show the mass of individual planetary objects.}
\label{fig:planetesimallipad}
\end{figure*}

\begin{figure*}
\centering
\includegraphics[scale=.31]{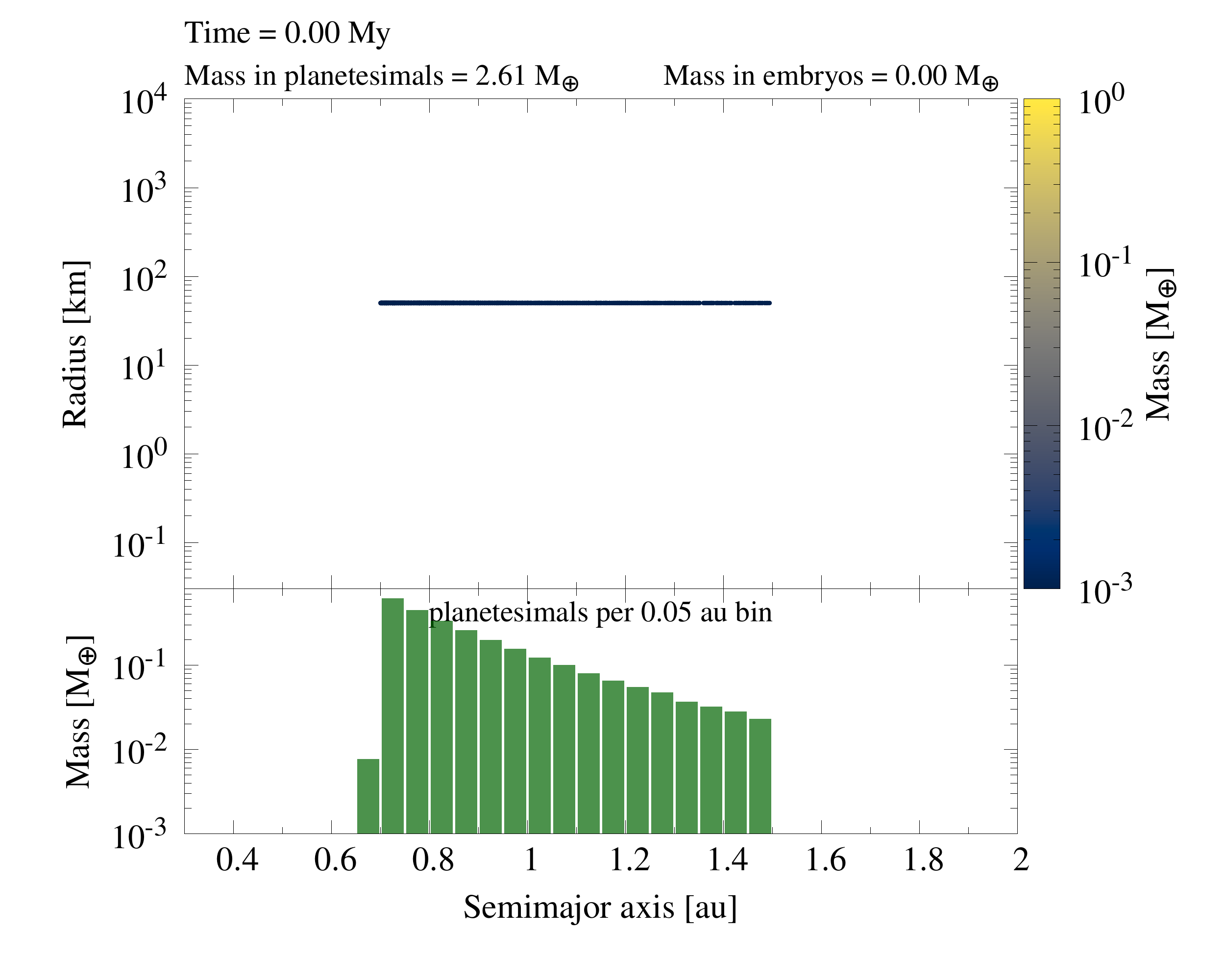}
\includegraphics[scale=.31]{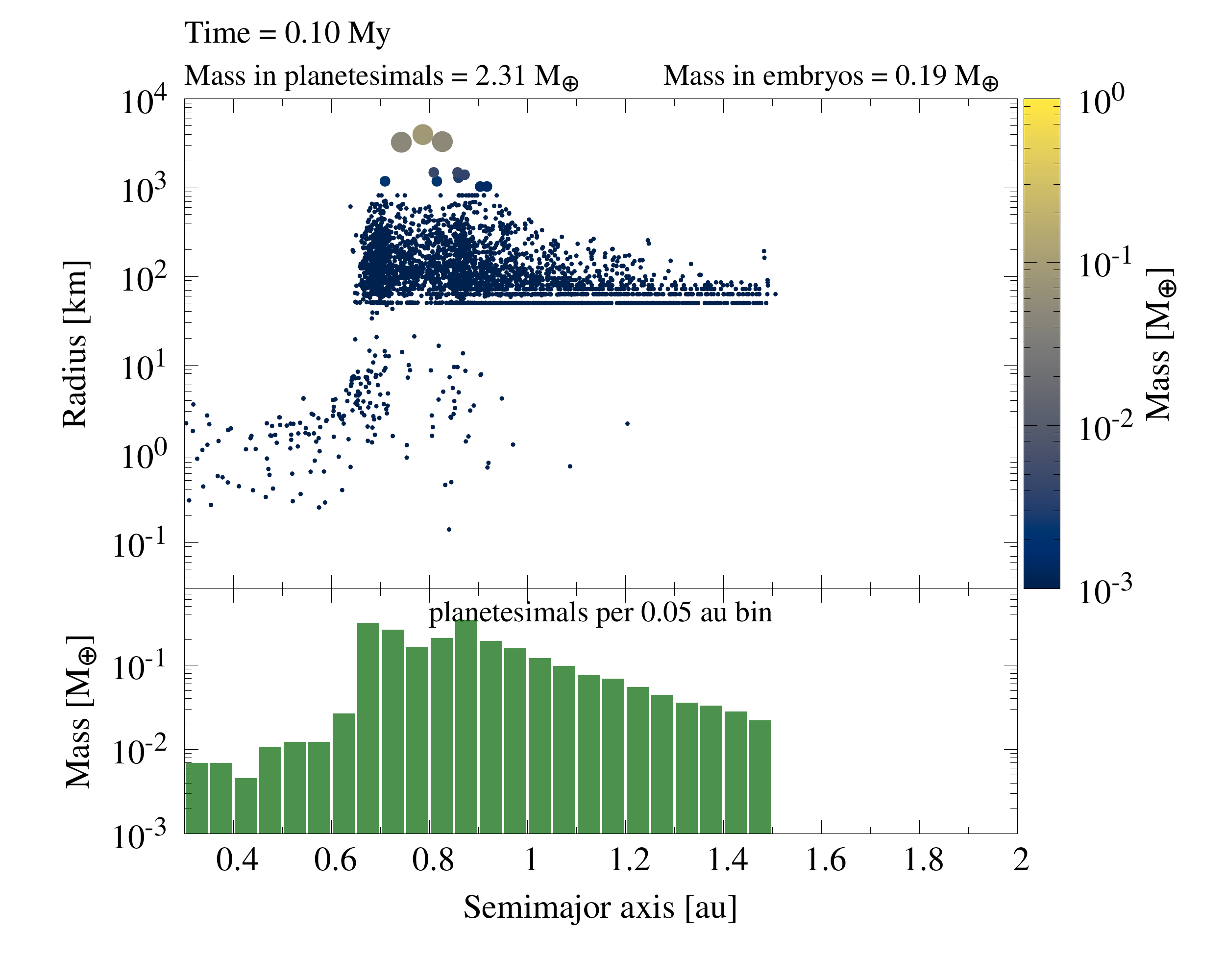}
\includegraphics[scale=.31]{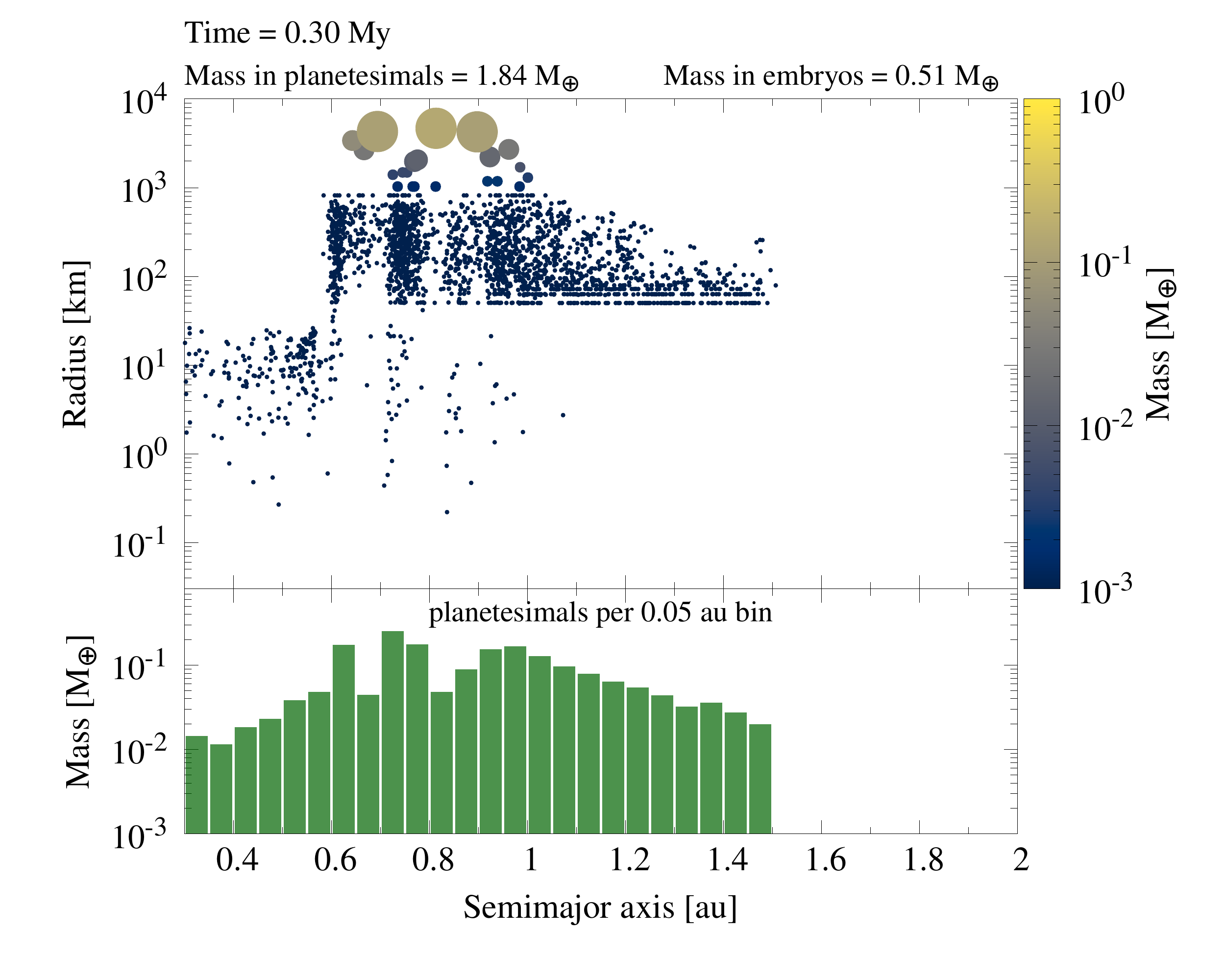}
\includegraphics[scale=.31]{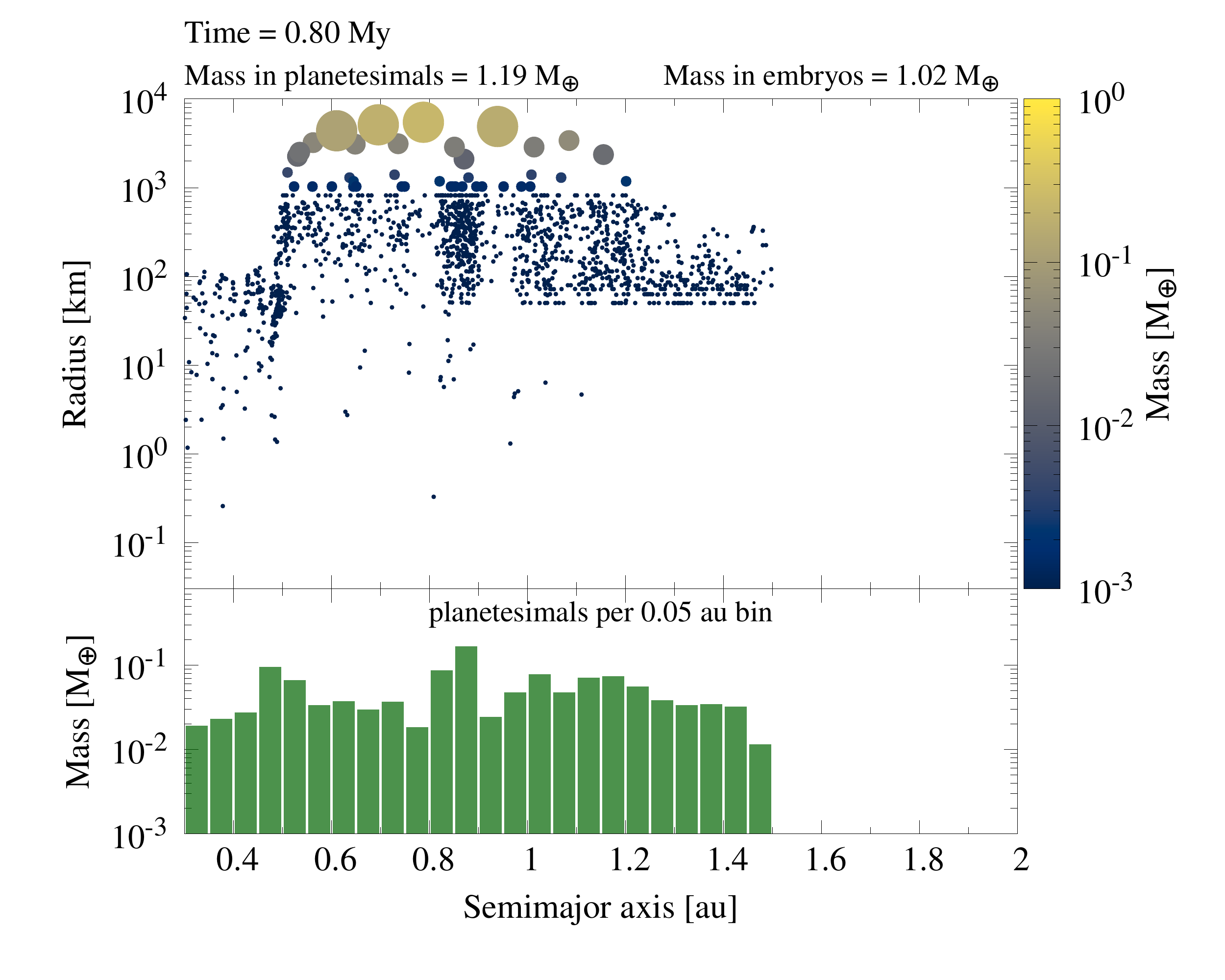}
\includegraphics[scale=.31]{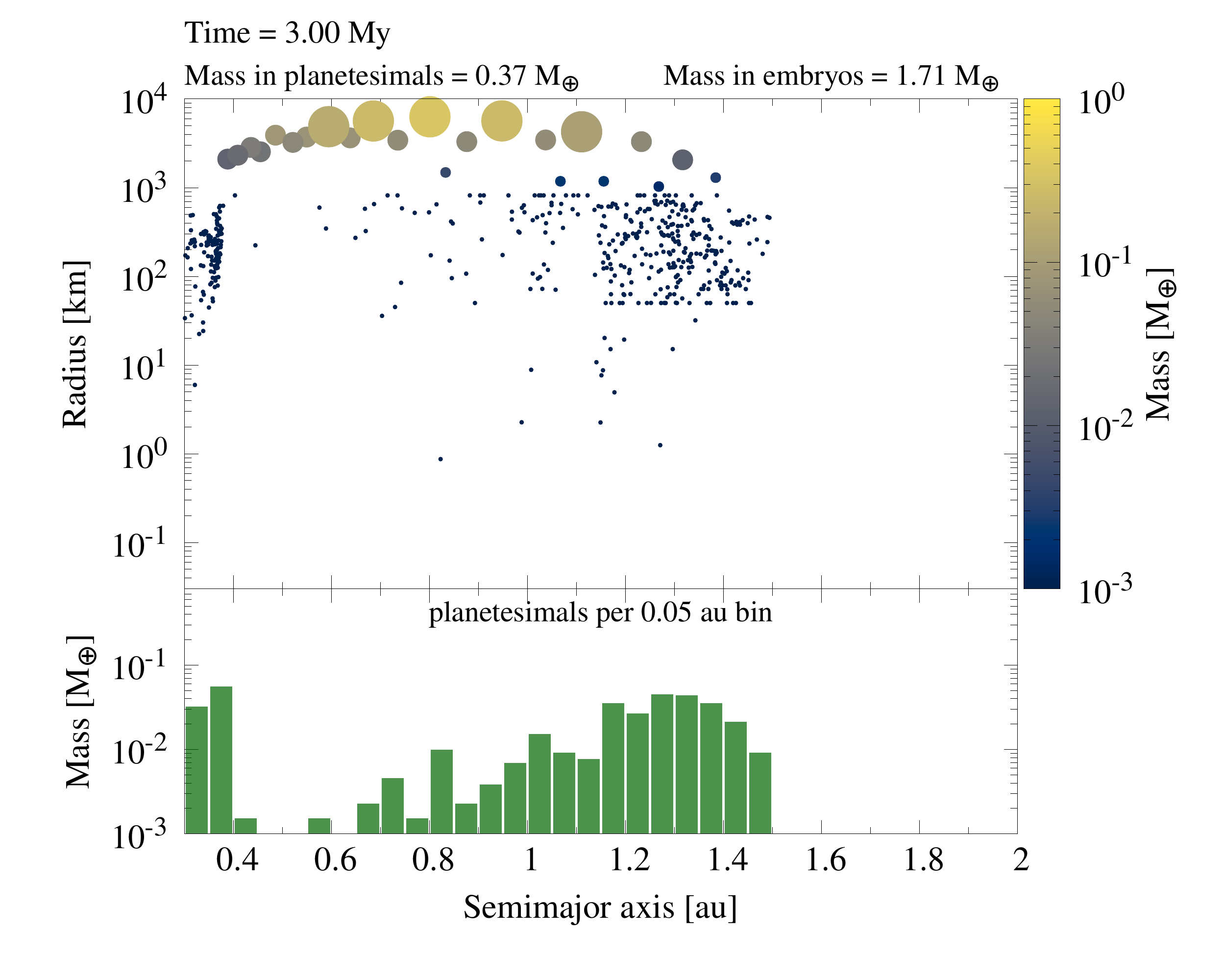}
\includegraphics[scale=.31]{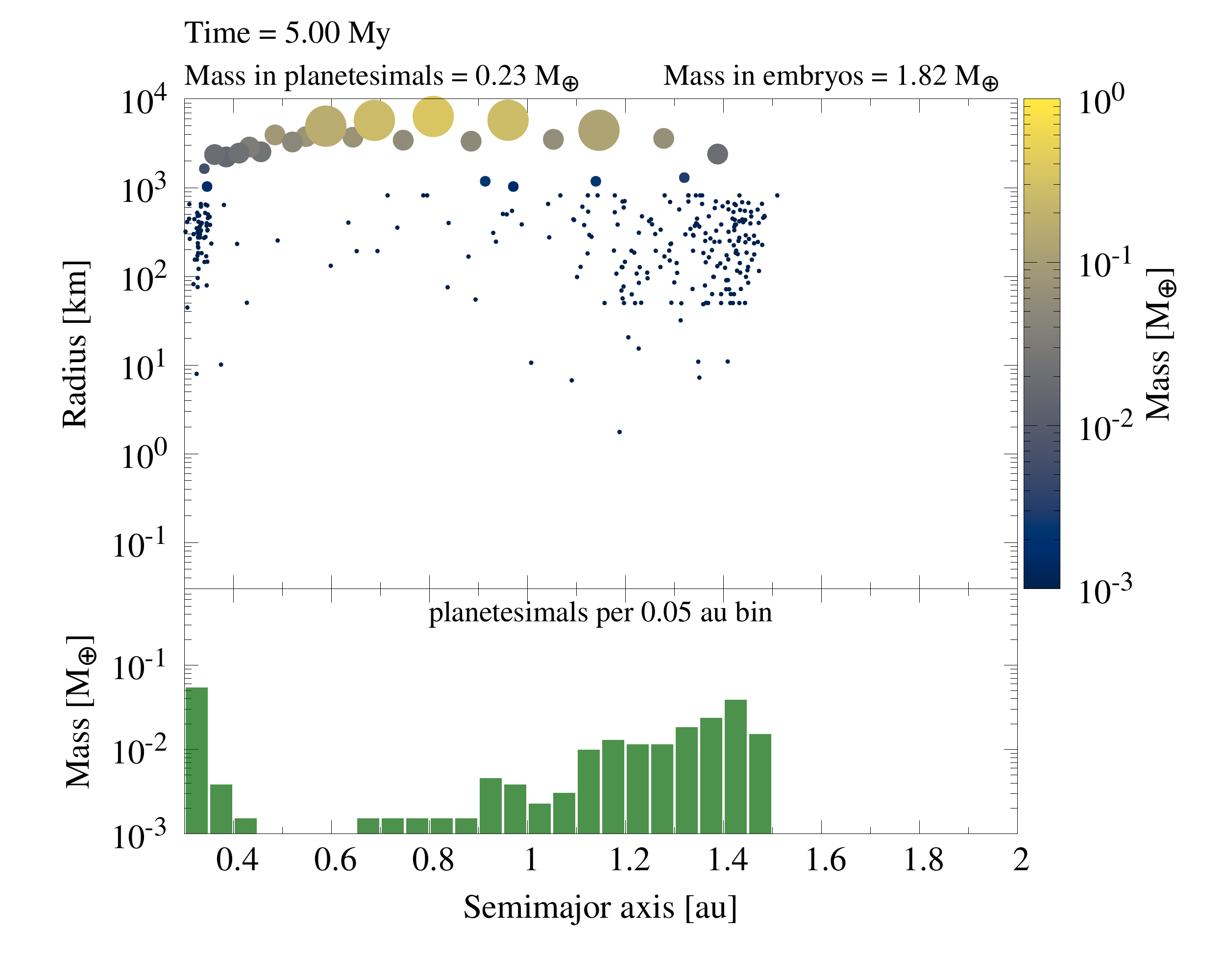}
\caption{Same as Figure \ref{fig:planetesimallipad} but for a ring with planetesimal surface density proportitional to $r^{-5.5}$.}
\label{fig:planetesimallipad2}
\end{figure*}

The results of Supplementary Figure \ref{fig:analytical} (bottom panel) allow us to  neglect the effects of pebble accretion in our N-body numerical simulations modeling planetary growth in the inner ring. Supplementary Figures \ref{fig:planetesimallipad} and \ref{fig:planetesimallipad2} show the growth of planetesimals in two N-body simulations using the LIPAD code\cite{levisonetal12}. These simulations start with 3000 objects which are treated as tracer particles\cite{walshlevison16,deiennoetal19}. Tracer particles are initially distributed following radial surface density profiles proportional to $r^{-1}$  (Supplementary Figure \ref{fig:planetesimallipad}) and $r^{-5.5}$  (Supplementary Figure \ref{fig:planetesimallipad2}).  Each tracer represents a collection of individual planetesimals of different sizes\citep{levisonetal12}. In the beginning of our simulations, 3000 tracers represent  $\sim$9,528,000 planetesimals with, diameters of 100~km. For simplicity we will refer to these particles as planetesimals as well. Their initial sizes are set D~$=$~100~km and they gravitationally interact with each other and evolve via collisions\citep{benzasphaug99}. Our model includes the effects  of gas drag on planetesimals, where the underlying gaseous protoplanetary disk is described in the Methods section. Once individual planetesimals inside tracer particles grow larger than the Moon, LIPAD intrinsic routines smoothly transition them to planetary embryos\cite{levisonetal12}.

Supplementary Figure \ref{fig:planetesimallipad} and \ref{fig:planetesimallipad2} show snapshots of the evolution of planetesimals in narrow rings around 1~au.  Planetesimals in the inner parts of the rings grow faster because of short local dynamical timescales. In Supplementary Figure \ref{fig:planetesimallipad}, at 0.8~Myr, Moon-mass planetary embryos have formed in the entire ring (note that this is consistent with the result of Supplementary Figure \ref{fig:analytical}). Planetary growth is slower in outer regions of the ring of Supplementary Figure \ref{fig:planetesimallipad2} because of the  steep initial distribution of mass in the ring. Collisions among objects induce fragmentation. This may be noticed by the existence of planetesimals with sizes smaller than the initial size. As planetary embryos grow, the inner ring spreads\cite{hansen09,walshetal11,raymondizidoro17b,deiennoetal19}. At 3~Myr and 5~Myr, the Earth-Mars region is heavily depleted of planetesimals.  In Supplementary Figure \ref{fig:planetesimallipad}, the total mass in planetesimals beyond 1~au at 5~Myr is about $0.23M_{\oplus}$ and the region inside 0.5~au contains about $0.14M_{\oplus}$ in planetesimals. Although the amount of mass in planetesimals in the innermost regions of the disk is comparable to the mass beyond 1~au, the probability of scattering and trapping into the belt planetesimals from the innermost most region of the disk (e.g. $<$0.5~au) is dramatically low\cite{bottkeetal06}. Planetesimals are mostly trapped into the belt after gas dispersal in a timescale of several tens of million years~\cite{raymondizidoro17b}. Once the gas disk dissipates, planetesimals inside 0.5-0.7~au are rapidly consumed or scattered by growing embryos, until they eventually collide with the Sun\cite{walshlevison19,deiennoetal19}. On the other hand, planetesimals originally near the outer edge of the ring are  pushed   outwards -- by the spreading of the planetesimals ring --  until they eventually reach the inner edge of the main asteroid belt, at $\sim$1.8~au (see Supplementary Figure \ref{fig:planetesimallipad}). These objects have the highest likelihood  to be subsequently scattered and trapped into the asteroid belt. In Supplementary Figure \ref{fig:planetesimallipad2}, most surviving planetesimals reside beyond 1.2~au.

In our simulations, we assume that the gas disk dissipates following a exponential decay timescale with an e-fold timescale of 2~Myr, and at 5~Myr the gas disk is assumed to fully dissipate. In our model, planetesimal formation in the inner ring typically ceases early, during the first 1 Myr of the solar system formation. If at 1 Myr, the terrestrial region is depleted of
gas due to efficient photo-evaporation, as may be the case for some transition disks~\cite[e.g.][]{vandermareletal16}, the subsequent growth of protoplanetary embryos would slow down because planetesimals would have higher random velocities due to the reduced effect of gas drag~\cite[e.g.][]{kokuboida96}.  By assuming that at 1Myr, the gas disk density drops instantaneously to 0.1\% of the nominal
density used in Supplementary Figure \ref{fig:analytical}, we found that the planetary embryo at 1 AU grows only to
about $\sim$3 Moon masses rather than $\sim$1 Mars-mass. This would set the onset of the late
stage of accretion of terrestrial from relatively lower-mass protoplanetary embryos. Our
simulations of the late stage of accretion of terrestrial planets start from different initial
distribution of masses for planetary embryos so we believe that these different scenarios
are  covered at some level by our current simulations.

\subsection{Dust production via impacts-} 
\begin{figure}
\centering
\includegraphics[scale=.7]{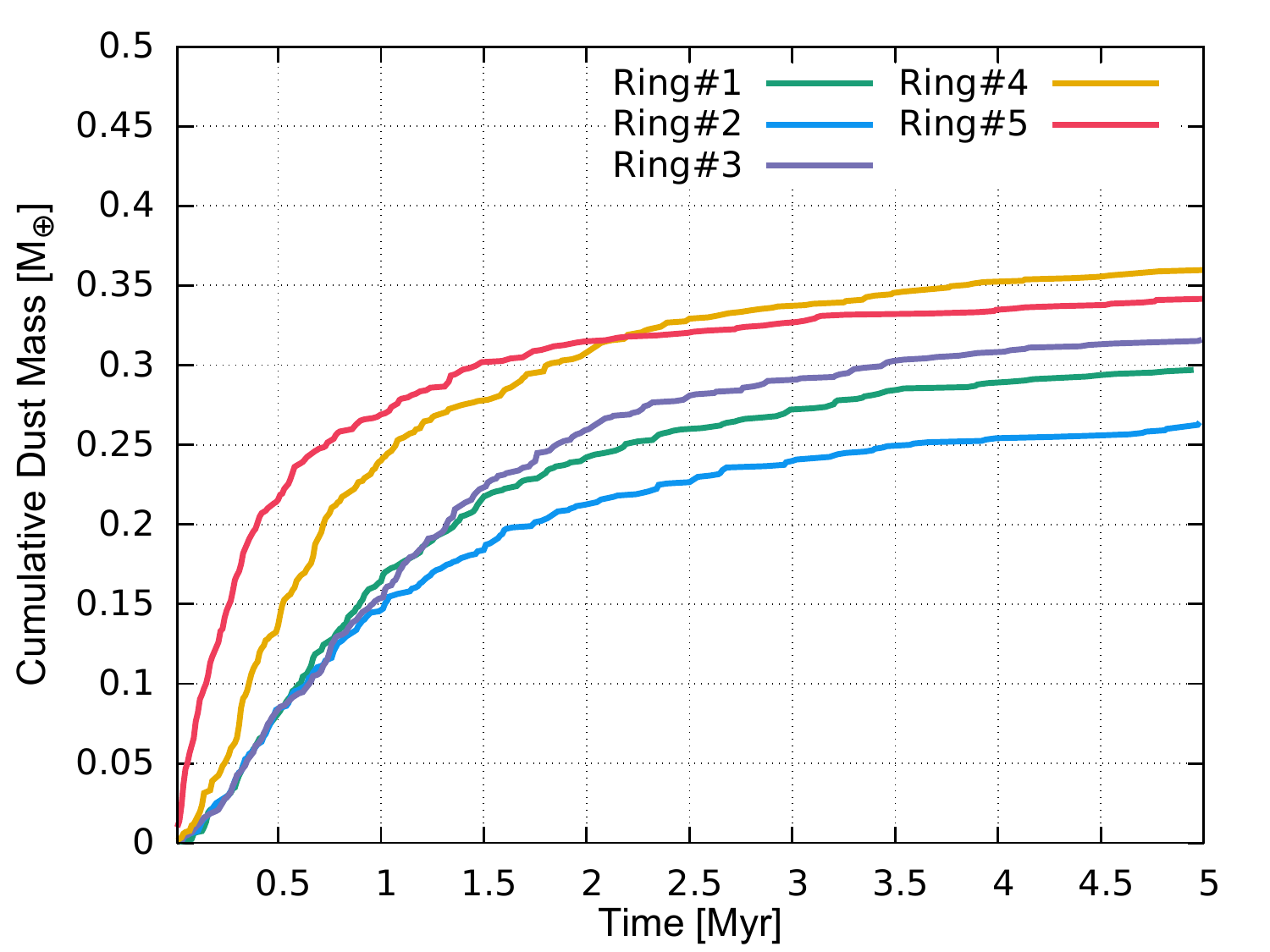}
\caption{Cumulative dust mass produced in N-body simulations modeling the growth of planetesimals in rings with different configurations (as in Supplementary Figures \ref{fig:planetesimallipad} and \ref{fig:planetesimallipad2}). Color-coded lines also show rings with different planetesimal surface density profiles, proportional to $r^0$ (green), $r^{-1}$ (blue), $r^{-2}$ (purple), $r^{-5.5}$ (pink), and $(-200(r/{\rm au}-1)^2+24)g/cm^2$ (yellow; downward quadratic function). }
\label{fig:dust}
\end{figure}

We have argued that a second generation of planetesimals could have formed from dust produced via impacts among planetary embryos during the gas disk phase.  In our simulations we do not model the formation of a second planetesimal population  from this dust reservoir. However, we have used our simulations to estimate the amount of dust produced via impacts during the gas disk phase. Supplementary Figure \ref{fig:dust} shows the cumulative dust mass produced in  simulations (similar to Supplementary Figure \ref{fig:planetesimallipad}) modeling different ring  configurations around 1~au (as in Figure \ref{fig:implantation} of the main paper). In this particular set of simulations, we flag fragments with radius smaller than 1~m\cite{walshlevison19,deiennoetal20} as dust. Particles that evolve down to $\sim$1~m rapidly evolve to dust sizes, due to collisional cascade\cite{nakagawaetal86,wallaceetal21}. Our simulations produce via impacts from $\sim$0.25 to $\sim$0.35$M_{\oplus}$ of dust in 5~Myr.


\subsection{Disk Temperature evolution}


At the beginning of our simulations, the water snowline is at $\sim$7~au (top panel of Supplementary Figure \ref{fig:temperature}; disk with $\beta=1$) and $\sim$17~au (bottom panel of of Supplementary Figure \ref{fig:temperature}; disk with $\beta=0.7$). In each of these disks, the silicate sublimation line is initially at about $\sim$0.8-0.9~au, which is consistent with previous disk models\cite{zhangjin15,drazkowskadullemond18,baillieetal19}. The transition in the disk viscosity due to the thermal ionization of the gas is initially at about $\sim$1.3-1.4au. We mimic the disk temperature evolution by using a simple exponential decay (see Methods). After 3Myr, the water snowline is at $\sim$3-4~au, and the disk silicate sublimation line is at $\sim$0.15-0.35~au. A disk model invoking much colder disks with pressure bumps would also produce two or three rings of planetesimals as in our nominal simulations. However, the position of these rings would be shifted. If the pressure bump near the silicate sublimation line (at the interface between the strongly and weakly ionized regions of the disk)  is initially too close to the star (e.g. $\ll0.5$~au), planetesimals would probably form in the innermost parts of the disk and it would be difficult to explain the lack of planets inside Mercury's orbit in the solar system\citep{izidoroetal21} (see discussion in the main paper).

\begin{figure}
\centering
\includegraphics[scale=1.5]{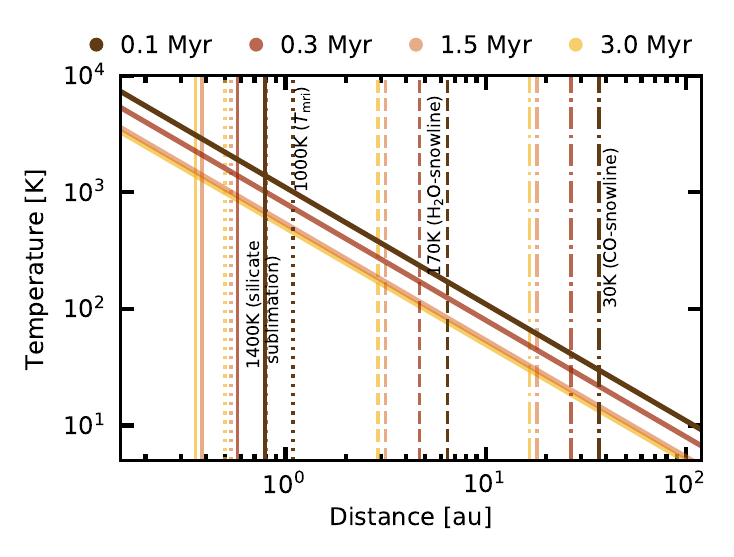}
\includegraphics[scale=1.5]{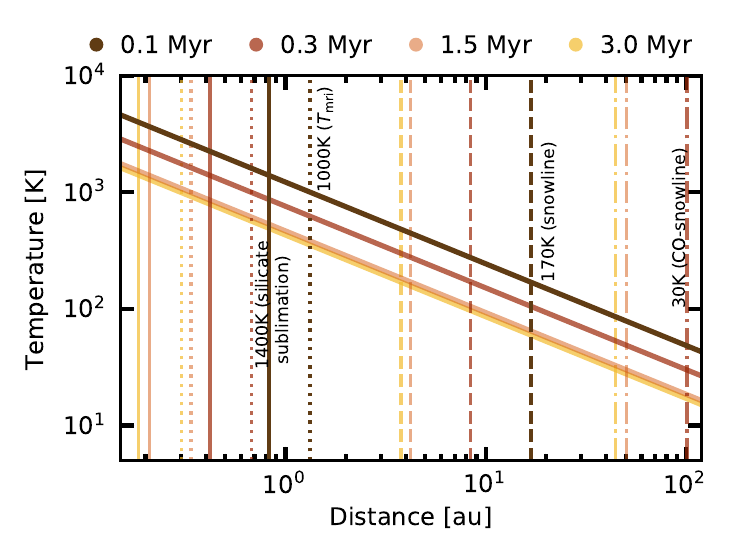}
    \caption{{\bf Top:} Evolution of the disk temperature in the simulation of Figure \ref{fig:lowefficiency}. {\bf Bottom:} Evolution of the disk temperature in all simulations presented in the Supplementary Information that produced two rings of planetesimals. In both panels, the dotted vertical lines shows the location of the transition in the disk viscosity caused by the thermal ionization of the gas. The dashed vertical lines shows the location of  the water-snowline. The dot-dashed vertical lines shows the location of the CO-snowline. The solid vertical lines shows the location of the disk silicate sublimation line. The color-coding represent different ages of the disk.}
    \label{fig:temperature}
\end{figure}

\subsection{Stokes Number -} 

\begin{figure*}
\centering
\includegraphics[scale=1.05]{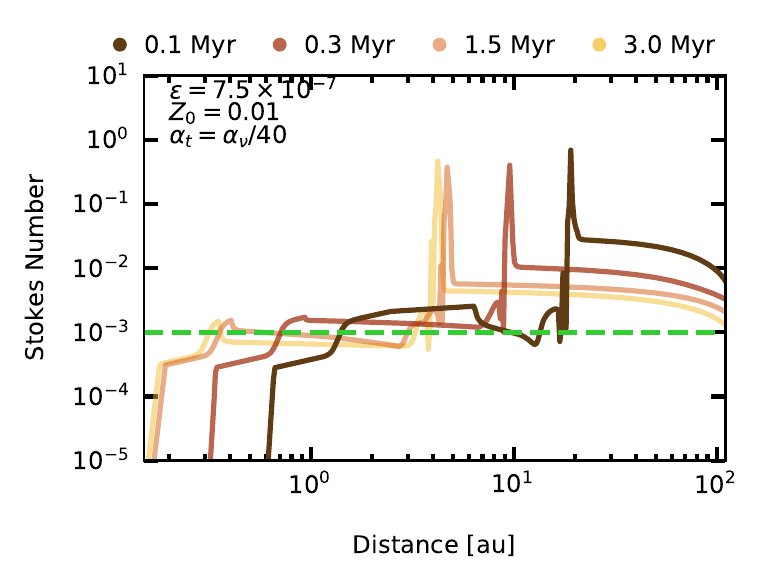}
\includegraphics[scale=1.05]{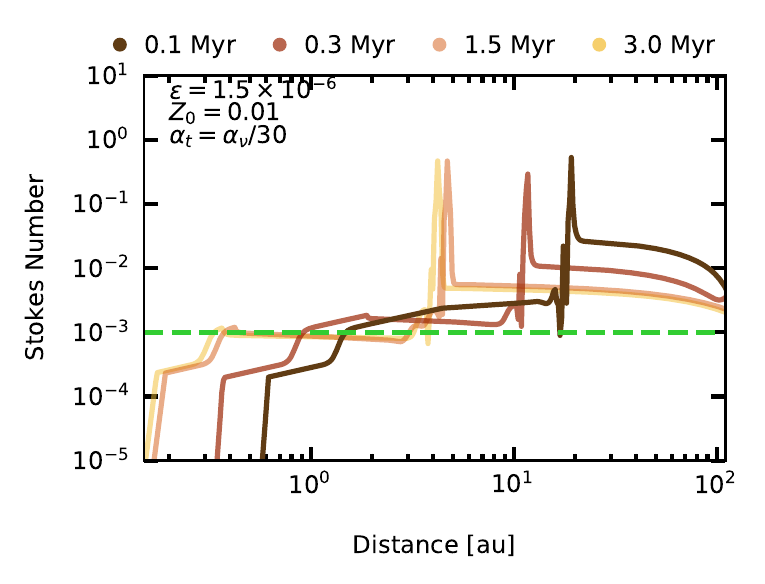}

\includegraphics[scale=1.05]{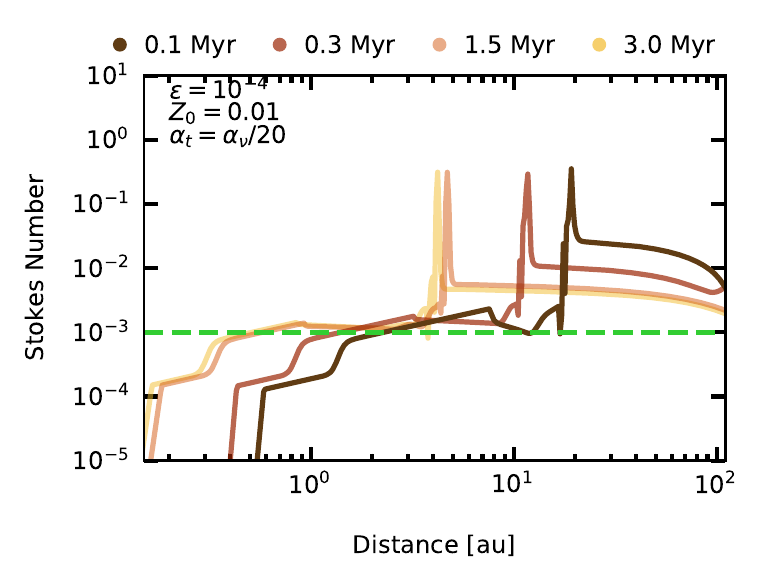}
\includegraphics[scale=1.05]{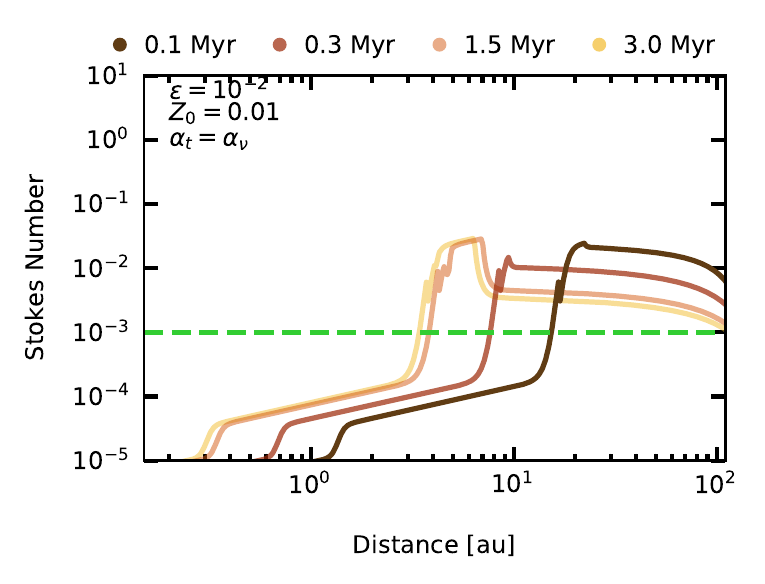}
    \caption{Temporal evolution of Stokes number in selected simulations with two pressure bumps. Each panel represents one different simulation. The different parameters that each simulation was performed is indicated at the top-left corner of the respective panel. Colored lines represent different times as indicated a the top of each panel. The green-dashed horizontal line shows the nominal minimum Stokes number ($St_{min}$) for planetesimal formation.}
    \label{fig:stokes}
\end{figure*}

Supplementary Figure \ref{fig:stokes} shows the maximum Stokes number of particles in  simulations considering different parameters. The dashed horizontal line shows our nominal minimum stokes number for planetesimal formation ($St_{\rm min}$). The stokes number of pebbles beyond the water snowline are typically larger than those inside. This reflects the transition in pebbles sizes at the snowline~\cite{blum18}. The sharp-drop in the stokes number at the innermost parts (e.g. $<1$~au) reflects the sublimation of silicate grains at regions of the disk where the temperature is higher than 1400~K. One can also see that  disks with higher levels of turbulence at the disk midplane ($\alpha_{\rm t}$) have pebbles with relatively smaller stokes number.

\section{Comparison of our model with existing Solar System formation  models}

Models of solar system formation from continuous disks of planetesimals have invoked different mechanisms to deplete the region beyound 1~au and account for the low-mass of Mars and dynamical structure of the asteroid belt. The so-called Grand-Tack model invokes  large-scale inward-then-outward migration  of Jupiter and Saturn in a gas disk to explain the inner solar system~\citep{walshetal11}. While this evolution represents one specific migration history of Jupiter and Saturn seen in hydrodynamical simulations, Jupiter and Saturn may have simply slowly migrated always-inwards or may have just avoided migration due to their specific mass ratios once they open a common gap in the disk~\cite{massetsnellgrove01,pierensetal14,zhangzhou10}, which may cancel out torques from the gas disk. Our model does  not require a Grand-Tack like evolution of the giant planets to explain the inner solar system, but it is broadly consistent with our current understanding of gas-driven giant planet migration~\citep{schneideretal21,nduguetal21}. As an alternative model for the formation of the terrestrial planets, the early instability scenario~\cite{clementetal18} proposes that the low-mass of Mars is the outcome of an early dynamical instability among the giant planets that scattered and depleted -- via gravitational interactions -- the Mars and the asteroid belt regions. This model requires that a  dynamical instability~\cite{morbidellinesvorny12} capable of depleting the primordial asteroid belt region at a level of 99.9\%~\cite{clementetal18} takes place no later than a few million of years after the formation of the solar system. The timing of the solar system giant planet  instability is not strongly constrained, but dynamical and geochemical analysis suggest that it must have happened during the first $\sim$100~Myr of the solar system history~\cite{morbidellietal18c,nesvornyetal18b}.  Our model is consistent with an early (e.g. $\sim$5~Myr after solar system formation) or relatively late dynamical instability (e.g. $\sim$100~Myr after solar system formation). In our model there is no need to deplete the belt very early as in the early instability scenario, because the belt is initially low mass or even empty.

In our simulations in the main paper (e.g. Figure \ref{fig:feeding}), Jupiter and Saturn are assumed to have a pre-instability dynamical configuration - namely almost coplanar and circular orbits~\cite{morbidellietal07}. The giant planet dynamical instability may excite the orbits of planetesimals in the belt~\cite{izidoroetal16,deiennoetal18} and induce small eccentricities and inclinations in the orbits of planetary objects in the terrestrial region due to secular and resonant interactions~\cite{brasseretal13b,nesvornyetal21}. Planetary embryos with larger orbital eccentricities may have wider feeding zones than those with relatively lower eccentricity orbits. We have tested the effects of the orbits of the giant planets on the feeding zones of Venus, Earth and Mars analogues in our model.

 We have performed 2 sets of 50 simulations each from planetesimal rings extending from 0.7 to 1.5~au and containing 2.5$M_{\oplus}$ in planetesimals. These sets come in two flavors. In the first one, we invoke a planetesimal ring with initial surface density profile proportional to $r^{-1}$. In the second one, we invoke a planetesimal ring with initial surface density profile proportional to $r^{-5.5}$. To roughly mimic the evolution of the giant planets in the dynamical instability phase, we assume that the giant planets instantaneously evolve from pre-instability orbits to their current ones at the end of the gas disk phase (5 Myr). This approach is reasonable and has been validated in a previous study as an adequate proxy for the effect of the instability on the terrestrial region\citep{nesvornyetal21},  although the very effect of the instability on the asteroid belt region may be far more dramatic~\citep{izidoroetal16,deiennoetal16,clementetal18}.

We compute the feeding zones of Venus, Earth and Mars analogues by tracking all collisions of planetesimals and protoplanetary embryos during the gas disk phase up to $\sim$200 Myr. Extended Data Figure \ref{fig:feeding_2rings} shows our results. The planetesimal ring represented by  $\Sigma_{pla}\propto r^{-1}$ (Extended Data Figure \ref{fig:feeding_2rings}a) produced 6 solar system analogues, and the ring represented by  $\Sigma_{pla}\propto r^{-5.5}$ produced 12 solar system analogues (Extended Data Figure \ref{fig:feeding_2rings}b). We have used Kolmogorov-Smirnov tests to compare the feeding zones of Earth and Mars analogues in both set of simulations. Our tests returned p-values much smaller than 0.001 for both sets of simulations which allow us to reject the null hypothesis that Earth and Mars analogues have statistically similar feeding zones. Extended Data Figure \ref{fig:feeding_2rings} also shows that a planetesimal ring with a steep surface density profile tend to produce Earth and Mars analogues with even more distinct feeding zones than a relatively shallow inner ring. This is caused by the fact that in the steep ring case, most of mass in planetesimals is initially located in the innermost parts of the ring (Supplementary Figure \ref{fig:planetesimallipad2}) which force Earth/Venus analogues to grow mostly from material inside 1~au~\citep{mahbrasseretal21}. Mars-analogues, on the other hand, have to grow from the outermost parts of the disk to avoid becoming too massive. As shown also in Supplementary Figure \ref{fig:selectedsystems} and in Extended Data Figure \ref{fig:feeding_2rings}, Mars-analogues  do not always grow exclusively from planetesimals from the outermost parts of the disk, and some Mars analogues may have more similar feeding zones to Earth-analogues than others~\cite{hansen09}. Results of Extended Data Figure \ref{fig:feeding_2rings} (and Figure \ref{fig:feeding} in the main paper) suggest, however, that the different feeding zones of Mars and Earth analogues in our simulations is generally robust and statistically significant regardless of the invoked giant planet configuration and initial distribution of planetesimals in the inner ring.

\bibliography{mybib.bib}

\begin{thebibliography}{100}
\expandafter\ifx\csname url\endcsname\relax
  \def\url#1{\texttt{#1}}\fi
\expandafter\ifx\csname urlprefix\endcsname\relax\def\urlprefix{URL }\fi
\providecommand{\bibinfo}[2]{#2}
\providecommand{\eprint}[2][]{\url{#2}}

\bibitem{demeocarry14}
\bibinfo{author}{DeMeo, F.~E.} \& \bibinfo{author}{Carry, B.}
\newblock \bibinfo{title}{{Solar System evolution from compositional mapping of
  the asteroid belt.}}
\newblock \emph{\bibinfo{journal}{Nature}} \textbf{\bibinfo{volume}{505}},
  \bibinfo{pages}{629--34} (\bibinfo{year}{2014}).

\bibitem{kruijeretal20}
\bibinfo{author}{{Kruijer}, T.~S.}, \bibinfo{author}{{Kleine}, T.} \&
  \bibinfo{author}{{Borg}, L.~E.}
\newblock \bibinfo{title}{{The great isotopic dichotomy of the early Solar
  System}}.
\newblock \emph{\bibinfo{journal}{Nat. Astron.}} \textbf{\bibinfo{volume}{4}},
  \bibinfo{pages}{32--40} (\bibinfo{year}{2020}).

\bibitem{grewaletal21}
\bibinfo{author}{{Grewal}, D.~S.}, \bibinfo{author}{{Dasgupta}, R.} \&
  \bibinfo{author}{{Marty}, B.}
\newblock \bibinfo{title}{{A very early origin of isotopically distinct
  nitrogen in inner Solar System protoplanets}}.
\newblock \emph{\bibinfo{journal}{Nat. Astron.}}  (\bibinfo{year}{2021}).

\bibitem{brassermojzsis20}
\bibinfo{author}{{Brasser}, R.} \& \bibinfo{author}{{Mojzsis}, S.~J.}
\newblock \bibinfo{title}{{The partitioning of the inner and outer Solar System
  by a structured protoplanetary disk}}.
\newblock \emph{\bibinfo{journal}{Nat. Astron.}} \textbf{\bibinfo{volume}{4}},
  \bibinfo{pages}{492--499} (\bibinfo{year}{2020}).

\bibitem{birnstieletal12}
\bibinfo{author}{{Birnstiel}, T.}, \bibinfo{author}{{Klahr}, H.} \&
  \bibinfo{author}{{Ercolano}, B.}
\newblock \bibinfo{title}{{A simple model for the evolution of the dust
  population in protoplanetary disks}}.
\newblock \emph{\bibinfo{journal}{\aap}} \textbf{\bibinfo{volume}{539}},
  \bibinfo{pages}{A148} (\bibinfo{year}{2012}).

\bibitem{walshetal11}
\bibinfo{author}{Walsh, K.~J.}, \bibinfo{author}{Morbidelli, A.},
  \bibinfo{author}{Raymond, S.~N.}, \bibinfo{author}{O'Brien, D.~P.} \&
  \bibinfo{author}{Mandell, A.~M.}
\newblock \bibinfo{title}{{A low mass for Mars from Jupiter's early gas-driven
  migration.}}
\newblock \emph{\bibinfo{journal}{Nature}} \textbf{\bibinfo{volume}{475}},
  \bibinfo{pages}{206--209} (\bibinfo{year}{2011}).

\bibitem{raymondizidoro17a}
\bibinfo{author}{{Raymond}, S.~N.} \& \bibinfo{author}{{Izidoro}, A.}
\newblock \bibinfo{title}{{Origin of water in the inner Solar System:
  Planetesimals scattered inward during Jupiter and Saturn's rapid gas
  accretion}}.
\newblock \emph{\bibinfo{journal}{\icarus}} \textbf{\bibinfo{volume}{297}},
  \bibinfo{pages}{134--148} (\bibinfo{year}{2017}).

\bibitem{huangetal18}
\bibinfo{author}{{Huang}, J.} \emph{et~al.}
\newblock \bibinfo{title}{{The Disk Substructures at High Angular Resolution
  Project (DSHARP). II. Characteristics of Annular Substructures}}.
\newblock \emph{\bibinfo{journal}{\apjl}} \textbf{\bibinfo{volume}{869}},
  \bibinfo{pages}{L42} (\bibinfo{year}{2018}).

\bibitem{dullemondetal18}
\bibinfo{author}{{Dullemond}, C.~P.} \emph{et~al.}
\newblock \bibinfo{title}{{The Disk Substructures at High Angular Resolution
  Project (DSHARP). VI. Dust Trapping in Thin-ringed Protoplanetary Disks}}.
\newblock \emph{\bibinfo{journal}{\apjl}} \textbf{\bibinfo{volume}{869}},
  \bibinfo{pages}{L46} (\bibinfo{year}{2018}).

\bibitem{johansenetal07}
\bibinfo{author}{{Johansen}, A.} \emph{et~al.}
\newblock \bibinfo{title}{{Rapid planetesimal formation in turbulent
  circumstellar disks}}.
\newblock \emph{\bibinfo{journal}{\nat}} \textbf{\bibinfo{volume}{448}},
  \bibinfo{pages}{1022--1025} (\bibinfo{year}{2007}).

\bibitem{mulleretal21}
\bibinfo{author}{{M{\"u}ller}, J.}, \bibinfo{author}{{Savvidou}, S.} \&
  \bibinfo{author}{{Bitsch}, B.}
\newblock \bibinfo{title}{{The water-ice line as a birthplace of planets:
  implications of a species-dependent dust fragmentation threshold}}.
\newblock \emph{\bibinfo{journal}{\aap}} \textbf{\bibinfo{volume}{650}},
  \bibinfo{pages}{A185} (\bibinfo{year}{2021}).

\bibitem{charnozetal21}
\bibinfo{author}{{Charnoz}, S.}, \bibinfo{author}{{Avice}, G.},
  \bibinfo{author}{{Hyodo}, R.}, \bibinfo{author}{{Pignatale}, F.~C.} \&
  \bibinfo{author}{{Chaussidon}, M.}
\newblock \bibinfo{title}{{Forming pressure traps at the snow line to isolate
  isotopic reservoirs in the absence of a planet}}.
\newblock \emph{\bibinfo{journal}{\aap}} \textbf{\bibinfo{volume}{652}},
  \bibinfo{pages}{A35} (\bibinfo{year}{2021}).

\bibitem{gundlachblum}
\bibinfo{author}{{Gundlach}, B.} \& \bibinfo{author}{{Blum}, J.}
\newblock \bibinfo{title}{{The Stickiness of Micrometer-sized Water-ice
  Particles}}.
\newblock \emph{\bibinfo{journal}{\apj}} \textbf{\bibinfo{volume}{798}},
  \bibinfo{pages}{34} (\bibinfo{year}{2015}).

\bibitem{deschturneretal15}
\bibinfo{author}{{Desch}, S.~J.} \& \bibinfo{author}{{Turner}, N.~J.}
\newblock \bibinfo{title}{{High-temperature Ionization in Protoplanetary
  Disks}}.
\newblock \emph{\bibinfo{journal}{\apj}} \textbf{\bibinfo{volume}{811}},
  \bibinfo{pages}{156} (\bibinfo{year}{2015}).

\bibitem{flocketal17}
\bibinfo{author}{{Flock}, M.}, \bibinfo{author}{{Fromang}, S.},
  \bibinfo{author}{{Turner}, N.~J.} \& \bibinfo{author}{{Benisty}, M.}
\newblock \bibinfo{title}{{3D Radiation Nonideal Magnetohydrodynamical
  Simulations of the Inner Rim in Protoplanetary Disks}}.
\newblock \emph{\bibinfo{journal}{\apj}} \textbf{\bibinfo{volume}{835}},
  \bibinfo{pages}{230} (\bibinfo{year}{2017}).

\bibitem{pinillaetal12}
\bibinfo{author}{{Pinilla}, P.} \emph{et~al.}
\newblock \bibinfo{title}{{Trapping dust particles in the outer regions of
  protoplanetary disks}}.
\newblock \emph{\bibinfo{journal}{\aap}} \textbf{\bibinfo{volume}{538}},
  \bibinfo{pages}{A114} (\bibinfo{year}{2012}).

\bibitem{dittrichetal13}
\bibinfo{author}{{Dittrich}, K.}, \bibinfo{author}{{Klahr}, H.} \&
  \bibinfo{author}{{Johansen}, A.}
\newblock \bibinfo{title}{{Gravoturbulent Planetesimal Formation: The Positive
  Effect of Long-lived Zonal Flows}}.
\newblock \emph{\bibinfo{journal}{\apj}} \textbf{\bibinfo{volume}{763}},
  \bibinfo{pages}{117} (\bibinfo{year}{2013}).

\bibitem{izidoroetal21}
\bibinfo{author}{{Izidoro}, A.}, \bibinfo{author}{{Bitsch}, B.} \&
  \bibinfo{author}{{Dasgupta}, R.}
\newblock \bibinfo{title}{{The Effect of a Strong Pressure Bump in the Sun's
  Natal Disk: Terrestrial Planet Formation via Planetesimal Accretion Rather
  than Pebble Accretion}}.
\newblock \emph{\bibinfo{journal}{\apj}} \textbf{\bibinfo{volume}{915}},
  \bibinfo{pages}{62} (\bibinfo{year}{2021}).

\bibitem{drazkowskaalibert17}
\bibinfo{author}{{Dr{\c a}{\.z}kowska}, J.} \& \bibinfo{author}{{Alibert}, Y.}
\newblock \bibinfo{title}{{Planetesimal formation starts at the snow line}}.
\newblock \emph{\bibinfo{journal}{\aap}} \textbf{\bibinfo{volume}{608}},
  \bibinfo{pages}{A92} (\bibinfo{year}{2017}).

\bibitem{simonetal16}
\bibinfo{author}{{Simon}, J.~B.}, \bibinfo{author}{{Armitage}, P.~J.},
  \bibinfo{author}{{Li}, R.} \& \bibinfo{author}{{Youdin}, A.~N.}
\newblock \bibinfo{title}{{The Mass and Size Distribution of Planetesimals
  Formed by the Streaming Instability. I. The Role of Self-gravity}}.
\newblock \emph{\bibinfo{journal}{\apj}} \textbf{\bibinfo{volume}{822}},
  \bibinfo{pages}{55} (\bibinfo{year}{2016}).

\bibitem{chambers06}
\bibinfo{author}{Chambers, J.}
\newblock \bibinfo{title}{{A semi-analytic model for oligarchic growth}}.
\newblock \emph{\bibinfo{journal}{Icarus}} \textbf{\bibinfo{volume}{180}},
  \bibinfo{pages}{496--513} (\bibinfo{year}{2006}).

\bibitem{tanakaetal02}
\bibinfo{author}{Tanaka, H.}, \bibinfo{author}{Takeuchi, T.} \&
  \bibinfo{author}{Ward, W.~R.}
\newblock \bibinfo{title}{{Three-dimensional Interaction between a Planet and
  an Isothermal Gaseous Disk. I. Corotation and Lindblad Torques and Planet
  Migration}}.
\newblock \emph{\bibinfo{journal}{\apj}} \textbf{\bibinfo{volume}{565}},
  \bibinfo{pages}{1257--1274} (\bibinfo{year}{2002}).

\bibitem{lambrechtsetal19}
\bibinfo{author}{{Lambrechts}, M.} \emph{et~al.}
\newblock \bibinfo{title}{{Formation of planetary systems by pebble accretion
  and migration. How the radial pebble flux determines a terrestrial-planet or
  super-Earth growth mode}}.
\newblock \emph{\bibinfo{journal}{\aap}} \textbf{\bibinfo{volume}{627}},
  \bibinfo{pages}{A83} (\bibinfo{year}{2019}).

\bibitem{morbidellicrida07}
\bibinfo{author}{Morbidelli, A.} \& \bibinfo{author}{Crida, A.}
\newblock \bibinfo{title}{{The dynamics of Jupiter and Saturn in the gaseous
  protoplanetary disk}}.
\newblock \emph{\bibinfo{journal}{Icarus}} \textbf{\bibinfo{volume}{191}},
  \bibinfo{pages}{158--171} (\bibinfo{year}{2007}).

\bibitem{hansen09}
\bibinfo{author}{Hansen, B. M.~S.}
\newblock \bibinfo{title}{{Formation of the Terrestrial Planets From a Narrow
  Annulus}}.
\newblock \emph{\bibinfo{journal}{\apj}} \textbf{\bibinfo{volume}{703}},
  \bibinfo{pages}{1131--1140} (\bibinfo{year}{2009}).

\bibitem{izidoroetal15b}
\bibinfo{author}{Izidoro, A.}, \bibinfo{author}{Raymond, S.~N.},
  \bibinfo{author}{Morbidelli, A.} \& \bibinfo{author}{Winter, O.~C.}
\newblock \bibinfo{title}{{Terrestrial planet formation constrained by Mars and
  the structure of the asteroid belt}}.
\newblock \emph{\bibinfo{journal}{\mnras}} \textbf{\bibinfo{volume}{453}},
  \bibinfo{pages}{3619--3634} (\bibinfo{year}{2015}).

\bibitem{levisonetal15b}
\bibinfo{author}{{Levison}, H.~F.}, \bibinfo{author}{{Kretke}, K.~A.} \&
  \bibinfo{author}{{Duncan}, M.~J.}
\newblock \bibinfo{title}{{Growing the gas-giant planets by the gradual
  accumulation of pebbles}}.
\newblock \emph{\bibinfo{journal}{\nat}} \textbf{\bibinfo{volume}{524}},
  \bibinfo{pages}{322--324} (\bibinfo{year}{2015}).

\bibitem{raymondizidoro17b}
\bibinfo{author}{{Raymond}, S.~N.} \& \bibinfo{author}{{Izidoro}, A.}
\newblock \bibinfo{title}{{The empty primordial asteroid belt}}.
\newblock \emph{\bibinfo{journal}{Science Advances}}
  \textbf{\bibinfo{volume}{3}}, \bibinfo{pages}{e1701138}
  (\bibinfo{year}{2017}).

\bibitem{morbidellietal16}
\bibinfo{author}{{Morbidelli}, A.} \emph{et~al.}
\newblock \bibinfo{title}{{Fossilized condensation lines in the Solar System
  protoplanetary disk}}.
\newblock \emph{\bibinfo{journal}{\icarus}} \textbf{\bibinfo{volume}{267}},
  \bibinfo{pages}{368--376} (\bibinfo{year}{2016}).

\bibitem{warrenetal11}
\bibinfo{author}{{Warren}, P.~H.}
\newblock \bibinfo{title}{{Stable-isotopic anomalies and the accretionary
  assemblage of the Earth and Mars: A subordinate role for carbonaceous
  chondrites}}.
\newblock \emph{\bibinfo{journal}{Earth and Planetary Science Letters}}
  \textbf{\bibinfo{volume}{311}}, \bibinfo{pages}{93--100}
  (\bibinfo{year}{2011}).

\bibitem{dauphasetal14}
\bibinfo{author}{{Dauphas}, N.} \emph{et~al.}
\newblock \bibinfo{title}{{Calcium-48 isotopic anomalies in bulk chondrites and
  achondrites: Evidence for a uniform isotopic reservoir in the inner
  protoplanetary disk}}.
\newblock \emph{\bibinfo{journal}{Earth and Planetary Science Letters}}
  \textbf{\bibinfo{volume}{407}}, \bibinfo{pages}{96--108}
  (\bibinfo{year}{2014}).

\bibitem{wittmannetal15}
\bibinfo{author}{{Wittmann}, A.} \emph{et~al.}
\newblock \bibinfo{title}{{Petrography and composition of Martian regolith
  breccia meteorite Northwest Africa 7475}}.
\newblock \emph{\bibinfo{journal}{Meteoritics and Planetary Science}}
  \textbf{\bibinfo{volume}{50}}, \bibinfo{pages}{326--352}
  (\bibinfo{year}{2015}).

\bibitem{lodders00}
\bibinfo{author}{{Lodders}, K.}
\newblock \bibinfo{title}{{An Oxygen Isotope Mixing Model for the Accretion and
  Composition of Rocky Planets}}.
\newblock \emph{\bibinfo{journal}{\ssr}} \textbf{\bibinfo{volume}{92}},
  \bibinfo{pages}{341--354} (\bibinfo{year}{2000}).

\bibitem{dauphas17}
\bibinfo{author}{{Dauphas}, N.}
\newblock \bibinfo{title}{{The isotopic nature of the Earth{\textquoteright}s
  accreting material through time}}.
\newblock \emph{\bibinfo{journal}{\nat}} \textbf{\bibinfo{volume}{541}},
  \bibinfo{pages}{521--524} (\bibinfo{year}{2017}).

\bibitem{brasseretal17}
\bibinfo{author}{{Brasser}, R.}, \bibinfo{author}{{Mojzsis}, S.~J.},
  \bibinfo{author}{{Matsumura}, S.} \& \bibinfo{author}{{Ida}, S.}
\newblock \bibinfo{title}{{The cool and distant formation of Mars}}.
\newblock \emph{\bibinfo{journal}{Earth and Planetary Science Letters}}
  \textbf{\bibinfo{volume}{468}}, \bibinfo{pages}{85--93}
  (\bibinfo{year}{2017}).

\bibitem{bottkeetal06}
\bibinfo{author}{{Bottke}, W.~F.}, \bibinfo{author}{{Nesvorn{\'y}}, D.},
  \bibinfo{author}{{Grimm}, R.~E.}, \bibinfo{author}{{Morbidelli}, A.} \&
  \bibinfo{author}{{O'Brien}, D.~P.}
\newblock \bibinfo{title}{{Iron meteorites as remnants of planetesimals formed
  in the terrestrial planet region}}.
\newblock \emph{\bibinfo{journal}{\nat}} \textbf{\bibinfo{volume}{439}},
  \bibinfo{pages}{821--824} (\bibinfo{year}{2006}).

\bibitem{chambers01}
\bibinfo{author}{Chambers, J.~E.}
\newblock \bibinfo{title}{{Making More Terrestrial Planets}}.
\newblock \emph{\bibinfo{journal}{Icarus}} \textbf{\bibinfo{volume}{152}},
  \bibinfo{pages}{205} (\bibinfo{year}{2001}).

\bibitem{kokuboida02}
\bibinfo{author}{{Kokubo}, E.} \& \bibinfo{author}{{Ida}, S.}
\newblock \bibinfo{title}{{Formation of Protoplanet Systems and Diversity of
  Planetary Systems}}.
\newblock \emph{\bibinfo{journal}{\apj}} \textbf{\bibinfo{volume}{581}},
  \bibinfo{pages}{666--680} (\bibinfo{year}{2002}).

\bibitem{busbinzel02}
\bibinfo{author}{{Bus}, S.~J.} \& \bibinfo{author}{{Binzel}, R.~P.}
\newblock \bibinfo{title}{{Phase II of the Small Main-Belt Asteroid
  Spectroscopic Survey. A Feature-Based Taxonomy}}.
\newblock \emph{\bibinfo{journal}{\icarus}} \textbf{\bibinfo{volume}{158}},
  \bibinfo{pages}{146--177} (\bibinfo{year}{2002}).

\bibitem{urey55}
\bibinfo{author}{{Urey}, H.~C.}
\newblock \bibinfo{title}{{The Cosmic Abundances of Potassium, Uranium, and
  Thorium and the Heat Balances of the Earth, the Moon, and Mars}}.
\newblock \emph{\bibinfo{journal}{Proceedings of the National Academy of
  Science}} \textbf{\bibinfo{volume}{41}}, \bibinfo{pages}{127--144}
  (\bibinfo{year}{1955}).

\bibitem{vernazzaetal15}
\bibinfo{author}{{Vernazza}, P.}, \bibinfo{author}{{Zanda}, B.},
  \bibinfo{author}{{Nakamura}, T.}, \bibinfo{author}{{Scott}, E.~R.~D.} \&
  \bibinfo{author}{{Russell}, S.}
\newblock \emph{\bibinfo{title}{{The Formation and Evolution of Ordinary
  Chondrite Parent Bodies}}}, \bibinfo{pages}{617--634} (\bibinfo{year}{2015}).

\bibitem{weisselkins-tanton13}
\bibinfo{author}{{Weiss}, B.~P.} \& \bibinfo{author}{{Elkins-Tanton}, L.~T.}
\newblock \bibinfo{title}{{Differentiated Planetesimals and the Parent Bodies
  of Chondrites}}.
\newblock \emph{\bibinfo{journal}{Annu. Rev. Earth Planet. Sci.}}
  \textbf{\bibinfo{volume}{41}}, \bibinfo{pages}{529--560}
  (\bibinfo{year}{2013}).

\bibitem{neumannetal18}
\bibinfo{author}{{Neumann}, W.}, \bibinfo{author}{{Kruijer}, T.~S.},
  \bibinfo{author}{{Breuer}, D.} \& \bibinfo{author}{{Kleine}, T.}
\newblock \bibinfo{title}{{Multistage Core Formation in Planetesimals Revealed
  by Numerical Modeling and Hf-W Chronometry of Iron Meteorites}}.
\newblock \emph{\bibinfo{journal}{Journal of Geophysical Research (Planets)}}
  \textbf{\bibinfo{volume}{123}}, \bibinfo{pages}{421--444}
  (\bibinfo{year}{2018}).

\bibitem{sandersscott12}
\bibinfo{author}{{Sanders}, I.~S.} \& \bibinfo{author}{{Scott}, E. R.~D.}
\newblock \bibinfo{title}{{The origin of chondrules and chondrites: Debris from
  low-velocity impacts between molten planetesimals?}}
\newblock \emph{\bibinfo{journal}{Meteoritics and Planetary Science}}
  \textbf{\bibinfo{volume}{47}}, \bibinfo{pages}{2170--2192}
  (\bibinfo{year}{2012}).

\bibitem{moskovitzgaidos11}
\bibinfo{author}{{Moskovitz}, N.} \& \bibinfo{author}{{Gaidos}, E.}
\newblock \bibinfo{title}{{Differentiation of planetesimals and the thermal
  consequences of melt migration}}.
\newblock \emph{\bibinfo{journal}{Meteoritics and Planetary Science}}
  \textbf{\bibinfo{volume}{46}}, \bibinfo{pages}{903--918}
  (\bibinfo{year}{2011}).

\bibitem{asphaugmovshovitz11}
\bibinfo{author}{{Asphaug}, E.}, \bibinfo{author}{{Jutzi}, M.} \&
  \bibinfo{author}{{Movshovitz}, N.}
\newblock \bibinfo{title}{{Chondrule formation during planetesimal accretion}}.
\newblock \emph{\bibinfo{journal}{Earth and Planetary Science Letters}}
  \textbf{\bibinfo{volume}{308}}, \bibinfo{pages}{369--379}
  (\bibinfo{year}{2011}).

\bibitem{deschconnoly02}
\bibinfo{author}{{Desch}, S.~J.} \& \bibinfo{author}{{Connolly}, J., H.~C.}
\newblock \bibinfo{title}{{A model of the thermal processing of particles in
  solar nebula shocks: Application to the cooling rates of chondrules}}.
\newblock \emph{\bibinfo{journal}{Meteoritics and Planetary Science}}
  \textbf{\bibinfo{volume}{37}}, \bibinfo{pages}{183--207}
  (\bibinfo{year}{2002}).

\bibitem{yangetal17}
\bibinfo{author}{{Yang}, C.~C.}, \bibinfo{author}{{Johansen}, A.} \&
  \bibinfo{author}{{Carrera}, D.}
\newblock \bibinfo{title}{{Concentrating small particles in protoplanetary
  disks through the streaming instability}}.
\newblock \emph{\bibinfo{journal}{\aap}} \textbf{\bibinfo{volume}{606}},
  \bibinfo{pages}{A80} (\bibinfo{year}{2017}).

\bibitem{kunitomoetal18}
\bibinfo{author}{{Kunitomo}, M.}, \bibinfo{author}{{Guillot}, T.},
  \bibinfo{author}{{Ida}, S.} \& \bibinfo{author}{{Takeuchi}, T.}
\newblock \bibinfo{title}{{Revisiting the pre-main-sequence evolution of stars.
  II. Consequences of planet formation on stellar surface composition}}.
\newblock \emph{\bibinfo{journal}{\aap}} \textbf{\bibinfo{volume}{618}},
  \bibinfo{pages}{A132} (\bibinfo{year}{2018}).

\bibitem{izidoroetal15c}
\bibinfo{author}{{Izidoro}, A.}, \bibinfo{author}{{Morbidelli}, A.},
  \bibinfo{author}{{Raymond}, S.~N.}, \bibinfo{author}{{Hersant}, F.} \&
  \bibinfo{author}{{Pierens}, A.}
\newblock \bibinfo{title}{{Accretion of Uranus and Neptune from
  inward-migrating planetary embryos blocked by Jupiter and Saturn}}.
\newblock \emph{\bibinfo{journal}{Astron. Astrophysics.}}
  \textbf{\bibinfo{volume}{582}}, \bibinfo{pages}{A99} (\bibinfo{year}{2015}).

\bibitem{tsiganisetal05}
\bibinfo{author}{Tsiganis, K.}, \bibinfo{author}{Gomes, R.},
  \bibinfo{author}{Morbidelli, A.} \& \bibinfo{author}{Levison, H.~F.}
\newblock \bibinfo{title}{{Origin of the orbital architecture of the giant
  planets of the Solar System.}}
\newblock \emph{\bibinfo{journal}{Nature}} \textbf{\bibinfo{volume}{435}},
  \bibinfo{pages}{459--61} (\bibinfo{year}{2005}).

\bibitem{nesvorny18}
\bibinfo{author}{{Nesvorn{\'y}}, D.}
\newblock \bibinfo{title}{{Dynamical Evolution of the Early Solar System}}.
\newblock \emph{\bibinfo{journal}{\araa}} \textbf{\bibinfo{volume}{56}},
  \bibinfo{pages}{137--174} (\bibinfo{year}{2018}).

\bibitem{deiennoetal17}
\bibinfo{author}{{Deienno}, R.}, \bibinfo{author}{{Morbidelli}, A.},
  \bibinfo{author}{{Gomes}, R.~S.} \& \bibinfo{author}{{Nesvorn{\'y}}, D.}
\newblock \bibinfo{title}{{Constraining the Giant Planets{\textquoteright}
  Initial Configuration from Their Evolution: Implications for the Timing of
  the Planetary Instability}}.
\newblock \emph{\bibinfo{journal}{\aj}} \textbf{\bibinfo{volume}{153}},
  \bibinfo{pages}{153} (\bibinfo{year}{2017}).

\bibitem{nesvornyetal20}
\bibinfo{author}{{Nesvorn{\'y}}, D.} \emph{et~al.}
\newblock \bibinfo{title}{{OSSOS XX: The Meaning of Kuiper Belt Colors}}.
\newblock \emph{\bibinfo{journal}{\aj}} \textbf{\bibinfo{volume}{160}},
  \bibinfo{pages}{46} (\bibinfo{year}{2020}).

\bibitem{gladmanetal08}
\bibinfo{author}{{Gladman}, B.}, \bibinfo{author}{{Marsden}, B.~G.} \&
  \bibinfo{author}{{Vanlaerhoven}, C.}
\newblock \emph{\bibinfo{title}{{Nomenclature in the Outer Solar System}}},
  \bibinfo{pages}{43} (\bibinfo{year}{2008}).

\bibitem{fressin2013}
\bibinfo{author}{{Fressin}, F.} \emph{et~al.}
\newblock \bibinfo{title}{{The False Positive Rate of Kepler and the Occurrence
  of Planets}}.
\newblock \emph{\bibinfo{journal}{\apj}} \textbf{\bibinfo{volume}{766}},
  \bibinfo{pages}{81} (\bibinfo{year}{2013}).

\bibitem{mayoretal11}
\bibinfo{author}{{Mayor}, M.} \emph{et~al.}
\newblock \bibinfo{title}{{The HARPS search for southern extra-solar planets
  XXXIV. Occurrence, mass distribution and orbital properties of super-Earths
  and Neptune-mass planets}}.
\newblock \emph{\bibinfo{journal}{ArXiv e-prints}}  (\bibinfo{year}{2011}).

\bibitem{izidoroetal19}
\bibinfo{author}{{Izidoro}, A.} \emph{et~al.}
\newblock \bibinfo{title}{{Formation of planetary systems by pebble accretion
  and migration. Hot super-Earth systems from breaking compact resonant
  chains}}.
\newblock \emph{\bibinfo{journal}{\aap}} \textbf{\bibinfo{volume}{650}},
  \bibinfo{pages}{A152} (\bibinfo{year}{2021}).

\bibitem{morbidelli20}
\bibinfo{author}{{Morbidelli}, A.}
\newblock \bibinfo{title}{{Planet formation by pebble accretion in ringed
  disks}}.
\newblock \emph{\bibinfo{journal}{\aap}} \textbf{\bibinfo{volume}{638}},
  \bibinfo{pages}{A1} (\bibinfo{year}{2020}).

\bibitem{lambrechtsetal14}
\bibinfo{author}{Lambrechts, M.}, \bibinfo{author}{Johansen, A.} \&
  \bibinfo{author}{Morbidelli, A.}
\newblock \bibinfo{title}{{Separating gas-giant and ice-giant planets by
  halting pebble accretion}}.
\newblock \emph{\bibinfo{journal}{Astron. Astrophysics.}}
  \textbf{\bibinfo{volume}{572}}, \bibinfo{pages}{A35} (\bibinfo{year}{2014}).

\bibitem{raymondetal18c}
\bibinfo{author}{{Raymond}, S.~N.}, \bibinfo{author}{{Izidoro}, A.} \&
  \bibinfo{author}{{Morbidelli}, A.}
\newblock \bibinfo{title}{{Solar System Formation in the Context of Extra-Solar
  Planets}}.
\newblock \emph{\bibinfo{journal}{arXiv e-prints}}
  \bibinfo{pages}{arXiv:1812.01033} (\bibinfo{year}{2018}).

\bibitem{pinillaetal17}
\bibinfo{author}{{Pinilla}, P.}, \bibinfo{author}{{Pohl}, A.},
  \bibinfo{author}{{Stammler}, S.~M.} \& \bibinfo{author}{{Birnstiel}, T.}
\newblock \bibinfo{title}{{Dust Density Distribution and Imaging Analysis of
  Different Ice Lines in Protoplanetary Disks}}.
\newblock \emph{\bibinfo{journal}{\apj}} \textbf{\bibinfo{volume}{845}},
  \bibinfo{pages}{68} (\bibinfo{year}{2017}).

\bibitem{drazkowskadullemond18}
\bibinfo{author}{{Dr{\c a}{\.z}kowska}, J.} \& \bibinfo{author}{{Dullemond},
  C.~P.}
\newblock \bibinfo{title}{{Planetesimal formation during protoplanetary disk
  buildup}}.
\newblock \emph{\bibinfo{journal}{\aap}} \textbf{\bibinfo{volume}{614}},
  \bibinfo{pages}{A62} (\bibinfo{year}{2018}).

\bibitem{uedaetal19}
\bibinfo{author}{{Ueda}, T.}, \bibinfo{author}{{Flock}, M.} \&
  \bibinfo{author}{{Okuzumi}, S.}
\newblock \bibinfo{title}{{Dust Pileup at the Dead-zone Inner Edge and
  Implications for the Disk Shadow}}.
\newblock \emph{\bibinfo{journal}{\apj}} \textbf{\bibinfo{volume}{871}},
  \bibinfo{pages}{10} (\bibinfo{year}{2019}).

\bibitem{idaetal16b}
\bibinfo{author}{{Ida}, S.}, \bibinfo{author}{{Guillot}, T.} \&
  \bibinfo{author}{{Morbidelli}, A.}
\newblock \bibinfo{title}{{The radial dependence of pebble accretion rates: A
  source of diversity in planetary systems. I. Analytical formulation}}.
\newblock \emph{\bibinfo{journal}{\aap}} \textbf{\bibinfo{volume}{591}},
  \bibinfo{pages}{A72} (\bibinfo{year}{2016}).

\bibitem{zhangjin15}
\bibinfo{author}{{Zhang}, Y.} \& \bibinfo{author}{{Jin}, L.}
\newblock \bibinfo{title}{{The Evolution of the Snow Line in a Protoplanetary
  Disk}}.
\newblock \emph{\bibinfo{journal}{\apj}} \textbf{\bibinfo{volume}{802}},
  \bibinfo{pages}{58} (\bibinfo{year}{2015}).

\bibitem{zhangetal15}
\bibinfo{author}{{Zhang}, K.}, \bibinfo{author}{{Blake}, G.~A.} \&
  \bibinfo{author}{{Bergin}, E.~A.}
\newblock \bibinfo{title}{{Evidence of Fast Pebble Growth Near Condensation
  Fronts in the HL Tau Protoplanetary Disk}}.
\newblock \emph{\bibinfo{journal}{\apjl}} \textbf{\bibinfo{volume}{806}},
  \bibinfo{pages}{L7} (\bibinfo{year}{2015}).

\bibitem{baillieetal19}
\bibinfo{author}{{Bailli{\'e}}, K.}, \bibinfo{author}{{Marques}, J.} \&
  \bibinfo{author}{{Piau}, L.}
\newblock \bibinfo{title}{{Building protoplanetary disks from the molecular
  cloud: redefining the disk timeline}}.
\newblock \emph{\bibinfo{journal}{\aap}} \textbf{\bibinfo{volume}{624}},
  \bibinfo{pages}{A93} (\bibinfo{year}{2019}).

\bibitem{bitschetal14}
\bibinfo{author}{{Bitsch}, B.}, \bibinfo{author}{{Morbidelli}, A.},
  \bibinfo{author}{{Lega}, E.} \& \bibinfo{author}{{Crida}, A.}
\newblock \bibinfo{title}{{Stellar irradiated discs and implications on
  migration of embedded planets. II. Accreting-discs}}.
\newblock \emph{\bibinfo{journal}{\aap}} \textbf{\bibinfo{volume}{564}},
  \bibinfo{pages}{A135} (\bibinfo{year}{2014}).

\bibitem{ziamprasetal20}
\bibinfo{author}{{Ziampras}, A.}, \bibinfo{author}{{Ataiee}, S.},
  \bibinfo{author}{{Kley}, W.}, \bibinfo{author}{{Dullemond}, C.~P.} \&
  \bibinfo{author}{{Baruteau}, C.}
\newblock \bibinfo{title}{{The impact of planet wakes on the location and shape
  of the water ice line in a protoplanetary disk}}.
\newblock \emph{\bibinfo{journal}{\aap}} \textbf{\bibinfo{volume}{633}},
  \bibinfo{pages}{A29} (\bibinfo{year}{2020}).

\bibitem{birnstieletal15}
\bibinfo{author}{{Birnstiel}, T.}, \bibinfo{author}{{Andrews}, S.~M.},
  \bibinfo{author}{{Pinilla}, P.} \& \bibinfo{author}{{Kama}, M.}
\newblock \bibinfo{title}{{Dust Evolution Can Produce Scattered Light Gaps in
  Protoplanetary Disks}}.
\newblock \emph{\bibinfo{journal}{\apjl}} \textbf{\bibinfo{volume}{813}},
  \bibinfo{pages}{L14} (\bibinfo{year}{2015}).

\bibitem{drazkowskaetal16}
\bibinfo{author}{{Dr{\k{a}}{\.z}kowska}, J.}, \bibinfo{author}{{Alibert}, Y.}
  \& \bibinfo{author}{{Moore}, B.}
\newblock \bibinfo{title}{{Close-in planetesimal formation by pile-up of
  drifting pebbles}}.
\newblock \emph{\bibinfo{journal}{\aap}} \textbf{\bibinfo{volume}{594}},
  \bibinfo{pages}{A105} (\bibinfo{year}{2016}).

\bibitem{asplundetal09}
\bibinfo{author}{{Asplund}, M.}, \bibinfo{author}{{Grevesse}, N.},
  \bibinfo{author}{{Sauval}, A.~J.} \& \bibinfo{author}{{Scott}, P.}
\newblock \bibinfo{title}{{The Chemical Composition of the Sun}}.
\newblock \emph{\bibinfo{journal}{\araa}} \textbf{\bibinfo{volume}{47}},
  \bibinfo{pages}{481--522} (\bibinfo{year}{2009}).

\bibitem{deschetal18}
\bibinfo{author}{{Desch}, S.~J.}, \bibinfo{author}{{Kalyaan}, A.} \&
  \bibinfo{author}{{O'D. Alexander}, C.~M.}
\newblock \bibinfo{title}{{The Effect of Jupiter's Formation on the
  Distribution of Refractory Elements and Inclusions in Meteorites}}.
\newblock \emph{\bibinfo{journal}{\apjs}} \textbf{\bibinfo{volume}{238}},
  \bibinfo{pages}{11} (\bibinfo{year}{2018}).

\bibitem{pinillaetal21}
\bibinfo{author}{{Pinilla}, P.}, \bibinfo{author}{{Lenz}, C.~T.} \&
  \bibinfo{author}{{Stammler}, S.~M.}
\newblock \bibinfo{title}{{Growing and trapping pebbles with fragile collisions
  of particles in protoplanetary disks}}.
\newblock \emph{\bibinfo{journal}{\aap}} \textbf{\bibinfo{volume}{645}},
  \bibinfo{pages}{A70} (\bibinfo{year}{2021}).

\bibitem{schneideretal21}
\bibinfo{author}{{Schneider}, A.~D.} \& \bibinfo{author}{{Bitsch}, B.}
\newblock \bibinfo{title}{{How drifting and evaporating pebbles shape giant
  planets. I. Heavy element content and atmospheric C/O}}.
\newblock \emph{\bibinfo{journal}{\aap}} \textbf{\bibinfo{volume}{654}},
  \bibinfo{pages}{A71} (\bibinfo{year}{2021}).

\bibitem{lenzetal20}
\bibinfo{author}{{Lenz}, C.~T.}, \bibinfo{author}{{Klahr}, H.},
  \bibinfo{author}{{Birnstiel}, T.}, \bibinfo{author}{{Kretke}, K.} \&
  \bibinfo{author}{{Stammler}, S.}
\newblock \bibinfo{title}{{Constraining the parameter space for the solar
  nebula. The effect of disk properties on planetesimal formation}}.
\newblock \emph{\bibinfo{journal}{\aap}} \textbf{\bibinfo{volume}{640}},
  \bibinfo{pages}{A61} (\bibinfo{year}{2020}).

\bibitem{lenzetal19}
\bibinfo{author}{{Lenz}, C.~T.}, \bibinfo{author}{{Klahr}, H.} \&
  \bibinfo{author}{{Birnstiel}, T.}
\newblock \bibinfo{title}{{Planetesimal Population Synthesis: Pebble
  Flux-regulated Planetesimal Formation}}.
\newblock \emph{\bibinfo{journal}{\apj}} \textbf{\bibinfo{volume}{874}},
  \bibinfo{pages}{36} (\bibinfo{year}{2019}).

\bibitem{okuzumietal12}
\bibinfo{author}{{Okuzumi}, S.} \& \bibinfo{author}{{Hirose}, S.}
\newblock \bibinfo{title}{{Planetesimal Formation in Magnetorotationally Dead
  Zones: Critical Dependence on the Net Vertical Magnetic Flux}}.
\newblock \emph{\bibinfo{journal}{\apjl}} \textbf{\bibinfo{volume}{753}},
  \bibinfo{pages}{L8} (\bibinfo{year}{2012}).

\bibitem{lyndenbellpringle74}
\bibinfo{author}{{Lynden-Bell}, D.} \& \bibinfo{author}{{Pringle}, J.~E.}
\newblock \bibinfo{title}{{The evolution of viscous discs and the origin of the
  nebular variables.}}
\newblock \emph{\bibinfo{journal}{\mnras}} \textbf{\bibinfo{volume}{168}},
  \bibinfo{pages}{603--637} (\bibinfo{year}{1974}).

\bibitem{shakurasunyaev73}
\bibinfo{author}{Shakura, N.~I.} \& \bibinfo{author}{Sunyaev, R.~A.}
\newblock \bibinfo{title}{{Black holes in binary systems. Observational
  appearance.}}
\newblock \emph{\bibinfo{journal}{Astron. Astrophysics.}}
  \textbf{\bibinfo{volume}{24}} (\bibinfo{year}{1973}).

\bibitem{baistone14}
\bibinfo{author}{{Bai}, X.-N.} \& \bibinfo{author}{{Stone}, J.~M.}
\newblock \bibinfo{title}{{Magnetic Flux Concentration and Zonal Flows in
  Magnetorotational Instability Turbulence}}.
\newblock \emph{\bibinfo{journal}{\apj}} \textbf{\bibinfo{volume}{796}},
  \bibinfo{pages}{31} (\bibinfo{year}{2014}).

\bibitem{gerbigetal19}
\bibinfo{author}{{Gerbig}, K.}, \bibinfo{author}{{Lenz}, C.~T.} \&
  \bibinfo{author}{{Klahr}, H.}
\newblock \bibinfo{title}{{Linking planetesimal and dust content in
  protoplanetary disks via a local toy model}}.
\newblock \emph{\bibinfo{journal}{\aap}} \textbf{\bibinfo{volume}{629}},
  \bibinfo{pages}{A116} (\bibinfo{year}{2019}).

\bibitem{ormelklahr10}
\bibinfo{author}{{Ormel}, C.~W.} \& \bibinfo{author}{{Klahr}, H.~H.}
\newblock \bibinfo{title}{{The effect of gas drag on the growth of
  protoplanets. Analytical expressions for the accretion of small bodies in
  laminar disks}}.
\newblock \emph{\bibinfo{journal}{\aap}} \textbf{\bibinfo{volume}{520}},
  \bibinfo{pages}{A43} (\bibinfo{year}{2010}).

\bibitem{johansenlambrechts17}
\bibinfo{author}{{Johansen}, A.} \& \bibinfo{author}{{Lambrechts}, M.}
\newblock \bibinfo{title}{{Forming Planets via Pebble Accretion}}.
\newblock \emph{\bibinfo{journal}{Annual Review of Earth and Planetary
  Sciences}} \textbf{\bibinfo{volume}{45}}, \bibinfo{pages}{359--387}
  (\bibinfo{year}{2017}).

\bibitem{walshlevison19}
\bibinfo{author}{{Walsh}, K.~J.} \& \bibinfo{author}{{Levison}, H.~F.}
\newblock \bibinfo{title}{{Planetesimals to terrestrial planets: Collisional
  evolution amidst a dissipating gas disk}}.
\newblock \emph{\bibinfo{journal}{\icarus}} \textbf{\bibinfo{volume}{329}},
  \bibinfo{pages}{88--100} (\bibinfo{year}{2019}).

\bibitem{deiennoetal19}
\bibinfo{author}{{Deienno}, R.}, \bibinfo{author}{{Walsh}, K.~J.},
  \bibinfo{author}{{Kretke}, K.~A.} \& \bibinfo{author}{{Levison}, H.~F.}
\newblock \bibinfo{title}{{Energy Dissipation in Large
  Collisions{\textemdash}No Change in Planet Formation Outcomes}}.
\newblock \emph{\bibinfo{journal}{\apj}} \textbf{\bibinfo{volume}{876}},
  \bibinfo{pages}{103} (\bibinfo{year}{2019}).

\bibitem{raymondetal04}
\bibinfo{author}{Raymond, S.~N.}, \bibinfo{author}{Quinn, T.} \&
  \bibinfo{author}{Lunine, J.~I.}
\newblock \bibinfo{title}{{Making other earths: Dynamical simulations of
  terrestrial planet formation and water delivery}}.
\newblock \emph{\bibinfo{journal}{Icarus}} \textbf{\bibinfo{volume}{168}},
  \bibinfo{pages}{1--17} (\bibinfo{year}{2004}).

\bibitem{obrienetal06}
\bibinfo{author}{O'Brien, D.~P.}, \bibinfo{author}{Morbidelli, A.} \&
  \bibinfo{author}{Levison, H.~F.}
\newblock \bibinfo{title}{{Terrestrial planet formation with strong dynamical
  friction}}.
\newblock \emph{\bibinfo{journal}{Icarus}} \textbf{\bibinfo{volume}{184}},
  \bibinfo{pages}{39--58} (\bibinfo{year}{2006}).

\bibitem{levisonetal12}
\bibinfo{author}{{Levison}, H.~F.}, \bibinfo{author}{{Duncan}, M.~J.} \&
  \bibinfo{author}{{Thommes}, E.}
\newblock \bibinfo{title}{{A Lagrangian Integrator for Planetary Accretion and
  Dynamics (LIPAD)}}.
\newblock \emph{\bibinfo{journal}{\aj}} \textbf{\bibinfo{volume}{144}},
  \bibinfo{pages}{119} (\bibinfo{year}{2012}).

\bibitem{chambers99}
\bibinfo{author}{Chambers, J.~E.}
\newblock \bibinfo{title}{{A hybrid symplectic integrator that permits close
  encounters between massive bodies}}.
\newblock \emph{\bibinfo{journal}{\mnras}} \textbf{\bibinfo{volume}{304}},
  \bibinfo{pages}{793--799} (\bibinfo{year}{1999}).

\bibitem{Seguracoxetal20}
\bibinfo{author}{{Segura-Cox}, D.~M.} \emph{et~al.}
\newblock \bibinfo{title}{{Four annular structures in a protostellar disk less
  than 500,000 years old}}.
\newblock \emph{\bibinfo{journal}{\nat}} \textbf{\bibinfo{volume}{586}},
  \bibinfo{pages}{228--231} (\bibinfo{year}{2020}).

\bibitem{youdinshu02}
\bibinfo{author}{{Youdin}, A.~N.} \& \bibinfo{author}{{Shu}, F.~H.}
\newblock \bibinfo{title}{{Planetesimal Formation by Gravitational
  Instability}}.
\newblock \emph{\bibinfo{journal}{\apj}} \textbf{\bibinfo{volume}{580}},
  \bibinfo{pages}{494--505} (\bibinfo{year}{2002}).

\bibitem{Kruijeretal17}
\bibinfo{author}{Kruijer, T.~S.}, \bibinfo{author}{Burkhardt, C.},
  \bibinfo{author}{Budde, G.} \& \bibinfo{author}{Kleine, T.}
\newblock \bibinfo{title}{Age of jupiter inferred from the distinct genetics
  and formation times of meteorites}.
\newblock \emph{\bibinfo{journal}{Proceedings of the National Academy of
  Sciences}} \textbf{\bibinfo{volume}{114}}, \bibinfo{pages}{6712--6716}
  (\bibinfo{year}{2017}).

\bibitem{lambrechtsjohansen12}
\bibinfo{author}{{Lambrechts}, M.} \& \bibinfo{author}{{Johansen}, A.}
\newblock \bibinfo{title}{{Rapid growth of gas-giant cores by pebble
  accretion}}.
\newblock \emph{\bibinfo{journal}{\aap}} \textbf{\bibinfo{volume}{544}},
  \bibinfo{pages}{A32} (\bibinfo{year}{2012}).

\bibitem{birnstieletal10}
\bibinfo{author}{{Birnstiel}, T.}, \bibinfo{author}{{Dullemond}, C.~P.} \&
  \bibinfo{author}{{Brauer}, F.}
\newblock \bibinfo{title}{{Gas- and dust evolution in protoplanetary disks}}.
\newblock \emph{\bibinfo{journal}{\aap}} \textbf{\bibinfo{volume}{513}},
  \bibinfo{pages}{A79} (\bibinfo{year}{2010}).

\bibitem{liyoudin21}
\bibinfo{author}{{Li}, R.} \& \bibinfo{author}{{Youdin}, A.~N.}
\newblock \bibinfo{title}{{Thresholds for Particle Clumping by the Streaming
  Instability}}.
\newblock \emph{\bibinfo{journal}{\apj}} \textbf{\bibinfo{volume}{919}},
  \bibinfo{pages}{107} (\bibinfo{year}{2021}).

\bibitem{kleynelson12}
\bibinfo{author}{{Kley}, W.} \& \bibinfo{author}{{Nelson}, R.~P.}
\newblock \bibinfo{title}{{Planet-Disk Interaction and Orbital Evolution}}.
\newblock \emph{\bibinfo{journal}{\araa}} \textbf{\bibinfo{volume}{50}},
  \bibinfo{pages}{211--249} (\bibinfo{year}{2012}).

\bibitem{walshlevison16}
\bibinfo{author}{{Walsh}, K.~J.} \& \bibinfo{author}{{Levison}, H.~F.}
\newblock \bibinfo{title}{{Terrestrial Planet Formation from an Annulus}}.
\newblock \emph{\bibinfo{journal}{\aj}} \textbf{\bibinfo{volume}{152}},
  \bibinfo{pages}{68} (\bibinfo{year}{2016}).

\bibitem{benzasphaug99}
\bibinfo{author}{{Benz}, W.} \& \bibinfo{author}{{Asphaug}, E.}
\newblock \bibinfo{title}{{Catastrophic Disruptions Revisited}}.
\newblock \emph{\bibinfo{journal}{\icarus}} \textbf{\bibinfo{volume}{142}},
  \bibinfo{pages}{5--20} (\bibinfo{year}{1999}).

\bibitem{vandermareletal16}
\bibinfo{author}{{van der Marel}, N.} \emph{et~al.}
\newblock \bibinfo{title}{{The (w)hole survey: An unbiased sample study of
  transition disk candidates based on Spitzer catalogs}}.
\newblock \emph{\bibinfo{journal}{\aap}} \textbf{\bibinfo{volume}{592}},
  \bibinfo{pages}{A126} (\bibinfo{year}{2016}).

\bibitem{kokuboida96}
\bibinfo{author}{{Kokubo}, E.} \& \bibinfo{author}{{Ida}, S.}
\newblock \bibinfo{title}{{On Runaway Growth of Planetesimals}}.
\newblock \emph{\bibinfo{journal}{\icarus}} \textbf{\bibinfo{volume}{123}},
  \bibinfo{pages}{180--191} (\bibinfo{year}{1996}).

\bibitem{brasseretal07}
\bibinfo{author}{Brasser, R.}, \bibinfo{author}{Duncan, M.} \&
  \bibinfo{author}{Levison, H.}
\newblock \bibinfo{title}{{Embedded star clusters and the formation of the Oort
  cloud}}.
\newblock \emph{\bibinfo{journal}{Icarus}} \textbf{\bibinfo{volume}{191}},
  \bibinfo{pages}{413--433} (\bibinfo{year}{2007}).

\bibitem{deiennoetal20}
\bibinfo{author}{{Deienno}, R.}, \bibinfo{author}{{Walsh}, K.~J.},
  \bibinfo{author}{{Levison}, H.~F.} \& \bibinfo{author}{{Kretke}, K.~A.}
\newblock \bibinfo{title}{{Collisional Evolution of Meter- to Kilometer-sized
  Planetesimals in Mean Motion Resonances: Implications for Inward Planet
  Shepherding}}.
\newblock \emph{\bibinfo{journal}{\apj}} \textbf{\bibinfo{volume}{890}},
  \bibinfo{pages}{170} (\bibinfo{year}{2020}).

\bibitem{nakagawaetal86}
\bibinfo{author}{{Nakagawa}, Y.}, \bibinfo{author}{{Sekiya}, M.} \&
  \bibinfo{author}{{Hayashi}, C.}
\newblock \bibinfo{title}{{Settling and growth of dust particles in a laminar
  phase of a low-mass solar nebula}}.
\newblock \emph{\bibinfo{journal}{\icarus}} \textbf{\bibinfo{volume}{67}},
  \bibinfo{pages}{375--390} (\bibinfo{year}{1986}).

\bibitem{wallaceetal21}
\bibinfo{author}{{Wallace}, S.~C.}, \bibinfo{author}{{Quinn}, T.~R.} \&
  \bibinfo{author}{{Boley}, A.~C.}
\newblock \bibinfo{title}{{Collision rates of planetesimals near mean-motion
  resonances}}.
\newblock \emph{\bibinfo{journal}{\mnras}} \textbf{\bibinfo{volume}{503}},
  \bibinfo{pages}{5409--5424} (\bibinfo{year}{2021}).

\bibitem{blum18}
\bibinfo{author}{{Blum}, J.}
\newblock \bibinfo{title}{{Dust Evolution in Protoplanetary Discs and the
  Formation of Planetesimals. What Have We Learned from Laboratory
  Experiments?}}
\newblock \emph{\bibinfo{journal}{\ssr}} \textbf{\bibinfo{volume}{214}},
  \bibinfo{pages}{52} (\bibinfo{year}{2018}).

\bibitem{massetsnellgrove01}
\bibinfo{author}{Masset, F.} \& \bibinfo{author}{Snellgrove, M.}
\newblock \bibinfo{title}{{Reversing type II migration: Resonance trapping of a
  lighter giant protoplanet}}.
\newblock \emph{\bibinfo{journal}{\mnras}} \textbf{\bibinfo{volume}{320}},
  \bibinfo{pages}{L55--L59} (\bibinfo{year}{2001}).

\bibitem{pierensetal14}
\bibinfo{author}{Pierens, A.}, \bibinfo{author}{Raymond, S.~N.},
  \bibinfo{author}{Nesvorny, D.} \& \bibinfo{author}{Morbidelli, A.}
\newblock \bibinfo{title}{{Outward Migration of Jupiter and Saturn in 3:2 or
  2:1 Resonance in Radiative Disks: Implications for the Grand Tack and Nice
  Models}}.
\newblock \emph{\bibinfo{journal}{\apj}} \textbf{\bibinfo{volume}{795}},
  \bibinfo{pages}{L11} (\bibinfo{year}{2014}).

\bibitem{zhangzhou10}
\bibinfo{author}{Zhang, H.} \& \bibinfo{author}{Zhou, J.-L.}
\newblock \bibinfo{title}{{on the Orbital Evolution of a Giant Planet Pair
  Embedded in a Gaseous Disk. I. Jupiter-Saturn Configuration}}.
\newblock \emph{\bibinfo{journal}{\apj}} \textbf{\bibinfo{volume}{714}},
  \bibinfo{pages}{532--548} (\bibinfo{year}{2010}).

\bibitem{nduguetal21}
\bibinfo{author}{{Ndugu}, N.}, \bibinfo{author}{{Bitsch}, B.},
  \bibinfo{author}{{Morbidelli}, A.}, \bibinfo{author}{{Crida}, A.} \&
  \bibinfo{author}{{Jurua}, E.}
\newblock \bibinfo{title}{{Probing the impact of varied migration and gas
  accretion rates for the formation of giant planets in the pebble accretion
  scenario}}.
\newblock \emph{\bibinfo{journal}{\mnras}} \textbf{\bibinfo{volume}{501}},
  \bibinfo{pages}{2017--2028} (\bibinfo{year}{2021}).

\bibitem{clementetal18}
\bibinfo{author}{{Clement}, M.~S.}, \bibinfo{author}{{Kaib}, N.~A.},
  \bibinfo{author}{{Raymond}, S.~N.} \& \bibinfo{author}{{Walsh}, K.~J.}
\newblock \bibinfo{title}{{Mars' growth stunted by an early giant planet
  instability}}.
\newblock \emph{\bibinfo{journal}{\icarus}} \textbf{\bibinfo{volume}{311}},
  \bibinfo{pages}{340--356} (\bibinfo{year}{2018}).

\bibitem{morbidellinesvorny12}
\bibinfo{author}{{Morbidelli}, A.} \& \bibinfo{author}{{Nesvorny}, D.}
\newblock \bibinfo{title}{{Dynamics of pebbles in the vicinity of a growing
  planetary embryo: hydro-dynamical simulations}}.
\newblock \emph{\bibinfo{journal}{\aap}} \textbf{\bibinfo{volume}{546}},
  \bibinfo{pages}{A18} (\bibinfo{year}{2012}).

\bibitem{morbidellietal18c}
\bibinfo{author}{{Morbidelli}, A.} \emph{et~al.}
\newblock \bibinfo{title}{{The timeline of the lunar bombardment: Revisited}}.
\newblock \emph{\bibinfo{journal}{\icarus}} \textbf{\bibinfo{volume}{305}},
  \bibinfo{pages}{262--276} (\bibinfo{year}{2018}).

\bibitem{nesvornyetal18b}
\bibinfo{author}{{Nesvorn{\'y}}, D.}, \bibinfo{author}{{Vokrouhlick{\'y}}, D.},
  \bibinfo{author}{{Bottke}, W.~F.} \& \bibinfo{author}{{Levison}, H.~F.}
\newblock \bibinfo{title}{{Evidence for very early migration of the Solar
  System planets from the Patroclus-Menoetius binary Jupiter Trojan}}.
\newblock \emph{\bibinfo{journal}{Nat. Astron.}} \textbf{\bibinfo{volume}{2}},
  \bibinfo{pages}{878--882} (\bibinfo{year}{2018}).

\bibitem{morbidellietal07}
\bibinfo{author}{Morbidelli, A.}, \bibinfo{author}{Tsiganis, K.},
  \bibinfo{author}{Crida, A.}, \bibinfo{author}{Levison, H.~F.} \&
  \bibinfo{author}{Gomes, R.}
\newblock \bibinfo{title}{{Dynamics of the Giant Planets of the Solar System in
  the Gaseous Protoplanetary Disk and Their Relationship to the Current Orbital
  Architecture}}.
\newblock \emph{\bibinfo{journal}{Astron. J.}} \textbf{\bibinfo{volume}{134}},
  \bibinfo{pages}{1790--1798} (\bibinfo{year}{2007}).

\bibitem{izidoroetal16}
\bibinfo{author}{{Izidoro}, A.} \emph{et~al.}
\newblock \bibinfo{title}{{The Asteroid Belt as a Relic from a Chaotic Early
  Solar System}}.
\newblock \emph{\bibinfo{journal}{\apj}} \textbf{\bibinfo{volume}{833}},
  \bibinfo{pages}{40} (\bibinfo{year}{2016}).

\bibitem{deiennoetal18}
\bibinfo{author}{{Deienno}, R.} \emph{et~al.}
\newblock \bibinfo{title}{{Excitation of a Primordial Cold Asteroid Belt as an
  Outcome of Planetary Instability}}.
\newblock \emph{\bibinfo{journal}{\apj}} \textbf{\bibinfo{volume}{864}},
  \bibinfo{pages}{50} (\bibinfo{year}{2018}).

\bibitem{brasseretal13b}
\bibinfo{author}{{Brasser}, R.}, \bibinfo{author}{{Walsh}, K.~J.} \&
  \bibinfo{author}{{Nesvorn{\'y}}, D.}
\newblock \bibinfo{title}{{Constraining the primordial orbits of the
  terrestrial planets}}.
\newblock \emph{\bibinfo{journal}{\mnras}} \textbf{\bibinfo{volume}{433}},
  \bibinfo{pages}{3417--3427} (\bibinfo{year}{2013}).

\bibitem{nesvornyetal21}
\bibinfo{author}{{Nesvorn{\'y}}, D.}, \bibinfo{author}{{Roig}, F.~V.} \&
  \bibinfo{author}{{Deienno}, R.}
\newblock \bibinfo{title}{{The Role of Early Giant-planet Instability in
  Terrestrial Planet Formation}}.
\newblock \emph{\bibinfo{journal}{\aj}} \textbf{\bibinfo{volume}{161}},
  \bibinfo{pages}{50} (\bibinfo{year}{2021}).

\bibitem{deiennoetal16}
\bibinfo{author}{Deienno, R.}, \bibinfo{author}{Gomes, R.~S.},
  \bibinfo{author}{Walsh, K.~J.}, \bibinfo{author}{Morbidelli, A.} \&
  \bibinfo{author}{Nesvorn{\'{y}}, D.}
\newblock \bibinfo{title}{{Is the Grand Tack model compatible with the orbital
  distribution of main belt asteroids?}}
\newblock \emph{\bibinfo{journal}{Icarus}} \textbf{\bibinfo{volume}{272}},
  \bibinfo{pages}{114--124} (\bibinfo{year}{2016}).

\bibitem{mahbrasseretal21}
\bibinfo{author}{{Mah}, J.} \& \bibinfo{author}{{Brasser}, R.}
\newblock \bibinfo{title}{{Isotopically distinct terrestrial planets via local
  accretion}}.
\newblock \emph{\bibinfo{journal}{\icarus}} \textbf{\bibinfo{volume}{354}},
  \bibinfo{pages}{114052} (\bibinfo{year}{2021}).

\end{thebibliography}


\begin{thebibliography}{10}
\expandafter\ifx\csname url\endcsname\relax
  \def\url#1{\texttt{#1}}\fi
\expandafter\ifx\csname urlprefix\endcsname\relax\def\urlprefix{URL }\fi
\providecommand{\bibinfo}[2]{#2}
\providecommand{\eprint}[2][]{\url{#2}}

\bibitem{Seguracoxetal20}
\bibinfo{author}{{Segura-Cox}, D.~M.} \emph{et~al.}
\newblock \bibinfo{title}{{Four annular structures in a protostellar disk less
  than 500,000 years old}}.
\newblock \emph{\bibinfo{journal}{\nat}} \textbf{\bibinfo{volume}{586}},
  \bibinfo{pages}{228--231} (\bibinfo{year}{2020}).

\bibitem{zhangetal15}
\bibinfo{author}{{Zhang}, K.}, \bibinfo{author}{{Blake}, G.~A.} \&
  \bibinfo{author}{{Bergin}, E.~A.}
\newblock \bibinfo{title}{{Evidence of Fast Pebble Growth Near Condensation
  Fronts in the HL Tau Protoplanetary Disk}}.
\newblock \emph{\bibinfo{journal}{\apjl}} \textbf{\bibinfo{volume}{806}},
  \bibinfo{pages}{L7} (\bibinfo{year}{2015}).

\bibitem{baillieetal19}
\bibinfo{author}{{Bailli{\'e}}, K.}, \bibinfo{author}{{Marques}, J.} \&
  \bibinfo{author}{{Piau}, L.}
\newblock \bibinfo{title}{{Building protoplanetary disks from the molecular
  cloud: redefining the disk timeline}}.
\newblock \emph{\bibinfo{journal}{\aap}} \textbf{\bibinfo{volume}{624}},
  \bibinfo{pages}{A93} (\bibinfo{year}{2019}).

\bibitem{asplundetal09}
\bibinfo{author}{{Asplund}, M.}, \bibinfo{author}{{Grevesse}, N.},
  \bibinfo{author}{{Sauval}, A.~J.} \& \bibinfo{author}{{Scott}, P.}
\newblock \bibinfo{title}{{The Chemical Composition of the Sun}}.
\newblock \emph{\bibinfo{journal}{\araa}} \textbf{\bibinfo{volume}{47}},
  \bibinfo{pages}{481--522} (\bibinfo{year}{2009}).

\bibitem{youdinshu02}
\bibinfo{author}{{Youdin}, A.~N.} \& \bibinfo{author}{{Shu}, F.~H.}
\newblock \bibinfo{title}{{Planetesimal Formation by Gravitational
  Instability}}.
\newblock \emph{\bibinfo{journal}{\apj}} \textbf{\bibinfo{volume}{580}},
  \bibinfo{pages}{494--505} (\bibinfo{year}{2002}).

\bibitem{drazkowskaalibert17}
\bibinfo{author}{{Dr{\c a}{\.z}kowska}, J.} \& \bibinfo{author}{{Alibert}, Y.}
\newblock \bibinfo{title}{{Planetesimal formation starts at the snow line}}.
\newblock \emph{\bibinfo{journal}{\aap}} \textbf{\bibinfo{volume}{608}},
  \bibinfo{pages}{A92} (\bibinfo{year}{2017}).

\bibitem{drazkowskadullemond18}
\bibinfo{author}{{Dr{\c a}{\.z}kowska}, J.} \& \bibinfo{author}{{Dullemond},
  C.~P.}
\newblock \bibinfo{title}{{Planetesimal formation during protoplanetary disk
  buildup}}.
\newblock \emph{\bibinfo{journal}{\aap}} \textbf{\bibinfo{volume}{614}},
  \bibinfo{pages}{A62} (\bibinfo{year}{2018}).

\bibitem{gerbigetal19}
\bibinfo{author}{{Gerbig}, K.}, \bibinfo{author}{{Lenz}, C.~T.} \&
  \bibinfo{author}{{Klahr}, H.}
\newblock \bibinfo{title}{{Linking planetesimal and dust content in
  protoplanetary disks via a local toy model}}.
\newblock \emph{\bibinfo{journal}{\aap}} \textbf{\bibinfo{volume}{629}},
  \bibinfo{pages}{A116} (\bibinfo{year}{2019}).

\bibitem{lenzetal19}
\bibinfo{author}{{Lenz}, C.~T.}, \bibinfo{author}{{Klahr}, H.} \&
  \bibinfo{author}{{Birnstiel}, T.}
\newblock \bibinfo{title}{{Planetesimal Population Synthesis: Pebble
  Flux-regulated Planetesimal Formation}}.
\newblock \emph{\bibinfo{journal}{\apj}} \textbf{\bibinfo{volume}{874}},
  \bibinfo{pages}{36} (\bibinfo{year}{2019}).

\bibitem{Kruijeretal17}
\bibinfo{author}{Kruijer, T.~S.}, \bibinfo{author}{Burkhardt, C.},
  \bibinfo{author}{Budde, G.} \& \bibinfo{author}{Kleine, T.}
\newblock \bibinfo{title}{Age of jupiter inferred from the distinct genetics
  and formation times of meteorites}.
\newblock \emph{\bibinfo{journal}{Proceedings of the National Academy of
  Sciences}} \textbf{\bibinfo{volume}{114}}, \bibinfo{pages}{6712--6716}
  (\bibinfo{year}{2017}).

\bibitem{kruijeretal20}
\bibinfo{author}{{Kruijer}, T.~S.}, \bibinfo{author}{{Kleine}, T.} \&
  \bibinfo{author}{{Borg}, L.~E.}
\newblock \bibinfo{title}{{The great isotopic dichotomy of the early Solar
  System}}.
\newblock \emph{\bibinfo{journal}{Nat. Astron.}} \textbf{\bibinfo{volume}{4}},
  \bibinfo{pages}{32--40} (\bibinfo{year}{2020}).

\bibitem{lambrechtsjohansen12}
\bibinfo{author}{{Lambrechts}, M.} \& \bibinfo{author}{{Johansen}, A.}
\newblock \bibinfo{title}{{Rapid growth of gas-giant cores by pebble
  accretion}}.
\newblock \emph{\bibinfo{journal}{\aap}} \textbf{\bibinfo{volume}{544}},
  \bibinfo{pages}{A32} (\bibinfo{year}{2012}).

\bibitem{uedaetal19}
\bibinfo{author}{{Ueda}, T.}, \bibinfo{author}{{Flock}, M.} \&
  \bibinfo{author}{{Okuzumi}, S.}
\newblock \bibinfo{title}{{Dust Pileup at the Dead-zone Inner Edge and
  Implications for the Disk Shadow}}.
\newblock \emph{\bibinfo{journal}{\apj}} \textbf{\bibinfo{volume}{871}},
  \bibinfo{pages}{10} (\bibinfo{year}{2019}).

\bibitem{birnstieletal10}
\bibinfo{author}{{Birnstiel}, T.}, \bibinfo{author}{{Dullemond}, C.~P.} \&
  \bibinfo{author}{{Brauer}, F.}
\newblock \bibinfo{title}{{Gas- and dust evolution in protoplanetary disks}}.
\newblock \emph{\bibinfo{journal}{\aap}} \textbf{\bibinfo{volume}{513}},
  \bibinfo{pages}{A79} (\bibinfo{year}{2010}).

\bibitem{yangetal17}
\bibinfo{author}{{Yang}, C.~C.}, \bibinfo{author}{{Johansen}, A.} \&
  \bibinfo{author}{{Carrera}, D.}
\newblock \bibinfo{title}{{Concentrating small particles in protoplanetary
  disks through the streaming instability}}.
\newblock \emph{\bibinfo{journal}{\aap}} \textbf{\bibinfo{volume}{606}},
  \bibinfo{pages}{A80} (\bibinfo{year}{2017}).

\bibitem{birnstieletal12}
\bibinfo{author}{{Birnstiel}, T.}, \bibinfo{author}{{Klahr}, H.} \&
  \bibinfo{author}{{Ercolano}, B.}
\newblock \bibinfo{title}{{A simple model for the evolution of the dust
  population in protoplanetary disks}}.
\newblock \emph{\bibinfo{journal}{\aap}} \textbf{\bibinfo{volume}{539}},
  \bibinfo{pages}{A148} (\bibinfo{year}{2012}).

\bibitem{liyoudin21}
\bibinfo{author}{{Li}, R.} \& \bibinfo{author}{{Youdin}, A.~N.}
\newblock \bibinfo{title}{{Thresholds for Particle Clumping by the Streaming
  Instability}}.
\newblock \emph{\bibinfo{journal}{\apj}} \textbf{\bibinfo{volume}{919}},
  \bibinfo{pages}{107} (\bibinfo{year}{2021}).

\bibitem{raymondizidoro17b}
\bibinfo{author}{{Raymond}, S.~N.} \& \bibinfo{author}{{Izidoro}, A.}
\newblock \bibinfo{title}{{The empty primordial asteroid belt}}.
\newblock \emph{\bibinfo{journal}{Science Advances}}
  \textbf{\bibinfo{volume}{3}}, \bibinfo{pages}{e1701138}
  (\bibinfo{year}{2017}).

\bibitem{hansen09}
\bibinfo{author}{Hansen, B. M.~S.}
\newblock \bibinfo{title}{{Formation of the Terrestrial Planets From a Narrow
  Annulus}}.
\newblock \emph{\bibinfo{journal}{\apj}} \textbf{\bibinfo{volume}{703}},
  \bibinfo{pages}{1131--1140} (\bibinfo{year}{2009}).

\bibitem{walshetal11}
\bibinfo{author}{Walsh, K.~J.}, \bibinfo{author}{Morbidelli, A.},
  \bibinfo{author}{Raymond, S.~N.}, \bibinfo{author}{O'Brien, D.~P.} \&
  \bibinfo{author}{Mandell, A.~M.}
\newblock \bibinfo{title}{{A low mass for Mars from Jupiter's early gas-driven
  migration.}}
\newblock \emph{\bibinfo{journal}{Nature}} \textbf{\bibinfo{volume}{475}},
  \bibinfo{pages}{206--209} (\bibinfo{year}{2011}).

\bibitem{izidoroetal21}
\bibinfo{author}{{Izidoro}, A.}, \bibinfo{author}{{Bitsch}, B.} \&
  \bibinfo{author}{{Dasgupta}, R.}
\newblock \bibinfo{title}{{The Effect of a Strong Pressure Bump in the Sun's
  Natal Disk: Terrestrial Planet Formation via Planetesimal Accretion Rather
  than Pebble Accretion}}.
\newblock \emph{\bibinfo{journal}{\apj}} \textbf{\bibinfo{volume}{915}},
  \bibinfo{pages}{62} (\bibinfo{year}{2021}).

\bibitem{kleynelson12}
\bibinfo{author}{{Kley}, W.} \& \bibinfo{author}{{Nelson}, R.~P.}
\newblock \bibinfo{title}{{Planet-Disk Interaction and Orbital Evolution}}.
\newblock \emph{\bibinfo{journal}{\araa}} \textbf{\bibinfo{volume}{50}},
  \bibinfo{pages}{211--249} (\bibinfo{year}{2012}).

\bibitem{lambrechtsetal19}
\bibinfo{author}{{Lambrechts}, M.} \emph{et~al.}
\newblock \bibinfo{title}{{Formation of planetary systems by pebble accretion
  and migration. How the radial pebble flux determines a terrestrial-planet or
  super-Earth growth mode}}.
\newblock \emph{\bibinfo{journal}{\aap}} \textbf{\bibinfo{volume}{627}},
  \bibinfo{pages}{A83} (\bibinfo{year}{2019}).

\bibitem{levisonetal12}
\bibinfo{author}{{Levison}, H.~F.}, \bibinfo{author}{{Duncan}, M.~J.} \&
  \bibinfo{author}{{Thommes}, E.}
\newblock \bibinfo{title}{{A Lagrangian Integrator for Planetary Accretion and
  Dynamics (LIPAD)}}.
\newblock \emph{\bibinfo{journal}{\aj}} \textbf{\bibinfo{volume}{144}},
  \bibinfo{pages}{119} (\bibinfo{year}{2012}).

\bibitem{walshlevison16}
\bibinfo{author}{{Walsh}, K.~J.} \& \bibinfo{author}{{Levison}, H.~F.}
\newblock \bibinfo{title}{{Terrestrial Planet Formation from an Annulus}}.
\newblock \emph{\bibinfo{journal}{\aj}} \textbf{\bibinfo{volume}{152}},
  \bibinfo{pages}{68} (\bibinfo{year}{2016}).

\bibitem{deiennoetal19}
\bibinfo{author}{{Deienno}, R.}, \bibinfo{author}{{Walsh}, K.~J.},
  \bibinfo{author}{{Kretke}, K.~A.} \& \bibinfo{author}{{Levison}, H.~F.}
\newblock \bibinfo{title}{{Energy Dissipation in Large
  Collisions{\textemdash}No Change in Planet Formation Outcomes}}.
\newblock \emph{\bibinfo{journal}{\apj}} \textbf{\bibinfo{volume}{876}},
  \bibinfo{pages}{103} (\bibinfo{year}{2019}).

\bibitem{benzasphaug99}
\bibinfo{author}{{Benz}, W.} \& \bibinfo{author}{{Asphaug}, E.}
\newblock \bibinfo{title}{{Catastrophic Disruptions Revisited}}.
\newblock \emph{\bibinfo{journal}{\icarus}} \textbf{\bibinfo{volume}{142}},
  \bibinfo{pages}{5--20} (\bibinfo{year}{1999}).

\bibitem{bottkeetal06}
\bibinfo{author}{{Bottke}, W.~F.}, \bibinfo{author}{{Nesvorn{\'y}}, D.},
  \bibinfo{author}{{Grimm}, R.~E.}, \bibinfo{author}{{Morbidelli}, A.} \&
  \bibinfo{author}{{O'Brien}, D.~P.}
\newblock \bibinfo{title}{{Iron meteorites as remnants of planetesimals formed
  in the terrestrial planet region}}.
\newblock \emph{\bibinfo{journal}{\nat}} \textbf{\bibinfo{volume}{439}},
  \bibinfo{pages}{821--824} (\bibinfo{year}{2006}).

\bibitem{walshlevison19}
\bibinfo{author}{{Walsh}, K.~J.} \& \bibinfo{author}{{Levison}, H.~F.}
\newblock \bibinfo{title}{{Planetesimals to terrestrial planets: Collisional
  evolution amidst a dissipating gas disk}}.
\newblock \emph{\bibinfo{journal}{\icarus}} \textbf{\bibinfo{volume}{329}},
  \bibinfo{pages}{88--100} (\bibinfo{year}{2019}).

\bibitem{vandermareletal16}
\bibinfo{author}{{van der Marel}, N.} \emph{et~al.}
\newblock \bibinfo{title}{{The (w)hole survey: An unbiased sample study of
  transition disk candidates based on Spitzer catalogs}}.
\newblock \emph{\bibinfo{journal}{\aap}} \textbf{\bibinfo{volume}{592}},
  \bibinfo{pages}{A126} (\bibinfo{year}{2016}).

\bibitem{kokuboida96}
\bibinfo{author}{{Kokubo}, E.} \& \bibinfo{author}{{Ida}, S.}
\newblock \bibinfo{title}{{On Runaway Growth of Planetesimals}}.
\newblock \emph{\bibinfo{journal}{\icarus}} \textbf{\bibinfo{volume}{123}},
  \bibinfo{pages}{180--191} (\bibinfo{year}{1996}).

\bibitem{brasseretal07}
\bibinfo{author}{Brasser, R.}, \bibinfo{author}{Duncan, M.} \&
  \bibinfo{author}{Levison, H.}
\newblock \bibinfo{title}{{Embedded star clusters and the formation of the Oort
  cloud}}.
\newblock \emph{\bibinfo{journal}{Icarus}} \textbf{\bibinfo{volume}{191}},
  \bibinfo{pages}{413--433} (\bibinfo{year}{2007}).

\bibitem{deiennoetal20}
\bibinfo{author}{{Deienno}, R.}, \bibinfo{author}{{Walsh}, K.~J.},
  \bibinfo{author}{{Levison}, H.~F.} \& \bibinfo{author}{{Kretke}, K.~A.}
\newblock \bibinfo{title}{{Collisional Evolution of Meter- to Kilometer-sized
  Planetesimals in Mean Motion Resonances: Implications for Inward Planet
  Shepherding}}.
\newblock \emph{\bibinfo{journal}{\apj}} \textbf{\bibinfo{volume}{890}},
  \bibinfo{pages}{170} (\bibinfo{year}{2020}).

\bibitem{nakagawaetal86}
\bibinfo{author}{{Nakagawa}, Y.}, \bibinfo{author}{{Sekiya}, M.} \&
  \bibinfo{author}{{Hayashi}, C.}
\newblock \bibinfo{title}{{Settling and growth of dust particles in a laminar
  phase of a low-mass solar nebula}}.
\newblock \emph{\bibinfo{journal}{\icarus}} \textbf{\bibinfo{volume}{67}},
  \bibinfo{pages}{375--390} (\bibinfo{year}{1986}).

\bibitem{wallaceetal21}
\bibinfo{author}{{Wallace}, S.~C.}, \bibinfo{author}{{Quinn}, T.~R.} \&
  \bibinfo{author}{{Boley}, A.~C.}
\newblock \bibinfo{title}{{Collision rates of planetesimals near mean-motion
  resonances}}.
\newblock \emph{\bibinfo{journal}{\mnras}} \textbf{\bibinfo{volume}{503}},
  \bibinfo{pages}{5409--5424} (\bibinfo{year}{2021}).

\bibitem{zhangjin15}
\bibinfo{author}{{Zhang}, Y.} \& \bibinfo{author}{{Jin}, L.}
\newblock \bibinfo{title}{{The Evolution of the Snow Line in a Protoplanetary
  Disk}}.
\newblock \emph{\bibinfo{journal}{\apj}} \textbf{\bibinfo{volume}{802}},
  \bibinfo{pages}{58} (\bibinfo{year}{2015}).

\bibitem{blum18}
\bibinfo{author}{{Blum}, J.}
\newblock \bibinfo{title}{{Dust Evolution in Protoplanetary Discs and the
  Formation of Planetesimals. What Have We Learned from Laboratory
  Experiments?}}
\newblock \emph{\bibinfo{journal}{\ssr}} \textbf{\bibinfo{volume}{214}},
  \bibinfo{pages}{52} (\bibinfo{year}{2018}).

\bibitem{massetsnellgrove01}
\bibinfo{author}{Masset, F.} \& \bibinfo{author}{Snellgrove, M.}
\newblock \bibinfo{title}{{Reversing type II migration: Resonance trapping of a
  lighter giant protoplanet}}.
\newblock \emph{\bibinfo{journal}{\mnras}} \textbf{\bibinfo{volume}{320}},
  \bibinfo{pages}{L55--L59} (\bibinfo{year}{2001}).

\bibitem{pierensetal14}
\bibinfo{author}{Pierens, A.}, \bibinfo{author}{Raymond, S.~N.},
  \bibinfo{author}{Nesvorny, D.} \& \bibinfo{author}{Morbidelli, A.}
\newblock \bibinfo{title}{{Outward Migration of Jupiter and Saturn in 3:2 or
  2:1 Resonance in Radiative Disks: Implications for the Grand Tack and Nice
  Models}}.
\newblock \emph{\bibinfo{journal}{\apj}} \textbf{\bibinfo{volume}{795}},
  \bibinfo{pages}{L11} (\bibinfo{year}{2014}).

\bibitem{zhangzhou10}
\bibinfo{author}{Zhang, H.} \& \bibinfo{author}{Zhou, J.-L.}
\newblock \bibinfo{title}{{on the Orbital Evolution of a Giant Planet Pair
  Embedded in a Gaseous Disk. I. Jupiter-Saturn Configuration}}.
\newblock \emph{\bibinfo{journal}{\apj}} \textbf{\bibinfo{volume}{714}},
  \bibinfo{pages}{532--548} (\bibinfo{year}{2010}).

\bibitem{schneideretal21}
\bibinfo{author}{{Schneider}, A.~D.} \& \bibinfo{author}{{Bitsch}, B.}
\newblock \bibinfo{title}{{How drifting and evaporating pebbles shape giant
  planets. I. Heavy element content and atmospheric C/O}}.
\newblock \emph{\bibinfo{journal}{\aap}} \textbf{\bibinfo{volume}{654}},
  \bibinfo{pages}{A71} (\bibinfo{year}{2021}).

\bibitem{nduguetal21}
\bibinfo{author}{{Ndugu}, N.}, \bibinfo{author}{{Bitsch}, B.},
  \bibinfo{author}{{Morbidelli}, A.}, \bibinfo{author}{{Crida}, A.} \&
  \bibinfo{author}{{Jurua}, E.}
\newblock \bibinfo{title}{{Probing the impact of varied migration and gas
  accretion rates for the formation of giant planets in the pebble accretion
  scenario}}.
\newblock \emph{\bibinfo{journal}{\mnras}} \textbf{\bibinfo{volume}{501}},
  \bibinfo{pages}{2017--2028} (\bibinfo{year}{2021}).

\bibitem{clementetal18}
\bibinfo{author}{{Clement}, M.~S.}, \bibinfo{author}{{Kaib}, N.~A.},
  \bibinfo{author}{{Raymond}, S.~N.} \& \bibinfo{author}{{Walsh}, K.~J.}
\newblock \bibinfo{title}{{Mars' growth stunted by an early giant planet
  instability}}.
\newblock \emph{\bibinfo{journal}{\icarus}} \textbf{\bibinfo{volume}{311}},
  \bibinfo{pages}{340--356} (\bibinfo{year}{2018}).

\bibitem{morbidellinesvorny12}
\bibinfo{author}{{Morbidelli}, A.} \& \bibinfo{author}{{Nesvorny}, D.}
\newblock \bibinfo{title}{{Dynamics of pebbles in the vicinity of a growing
  planetary embryo: hydro-dynamical simulations}}.
\newblock \emph{\bibinfo{journal}{\aap}} \textbf{\bibinfo{volume}{546}},
  \bibinfo{pages}{A18} (\bibinfo{year}{2012}).

\bibitem{morbidellietal18c}
\bibinfo{author}{{Morbidelli}, A.} \emph{et~al.}
\newblock \bibinfo{title}{{The timeline of the lunar bombardment: Revisited}}.
\newblock \emph{\bibinfo{journal}{\icarus}} \textbf{\bibinfo{volume}{305}},
  \bibinfo{pages}{262--276} (\bibinfo{year}{2018}).

\bibitem{nesvornyetal18b}
\bibinfo{author}{{Nesvorn{\'y}}, D.}, \bibinfo{author}{{Vokrouhlick{\'y}}, D.},
  \bibinfo{author}{{Bottke}, W.~F.} \& \bibinfo{author}{{Levison}, H.~F.}
\newblock \bibinfo{title}{{Evidence for very early migration of the Solar
  System planets from the Patroclus-Menoetius binary Jupiter Trojan}}.
\newblock \emph{\bibinfo{journal}{Nat. Astron.}} \textbf{\bibinfo{volume}{2}},
  \bibinfo{pages}{878--882} (\bibinfo{year}{2018}).

\bibitem{morbidellietal07}
\bibinfo{author}{Morbidelli, A.}, \bibinfo{author}{Tsiganis, K.},
  \bibinfo{author}{Crida, A.}, \bibinfo{author}{Levison, H.~F.} \&
  \bibinfo{author}{Gomes, R.}
\newblock \bibinfo{title}{{Dynamics of the Giant Planets of the Solar System in
  the Gaseous Protoplanetary Disk and Their Relationship to the Current Orbital
  Architecture}}.
\newblock \emph{\bibinfo{journal}{Astron. J.}} \textbf{\bibinfo{volume}{134}},
  \bibinfo{pages}{1790--1798} (\bibinfo{year}{2007}).

\bibitem{izidoroetal16}
\bibinfo{author}{{Izidoro}, A.} \emph{et~al.}
\newblock \bibinfo{title}{{The Asteroid Belt as a Relic from a Chaotic Early
  Solar System}}.
\newblock \emph{\bibinfo{journal}{\apj}} \textbf{\bibinfo{volume}{833}},
  \bibinfo{pages}{40} (\bibinfo{year}{2016}).

\bibitem{deiennoetal18}
\bibinfo{author}{{Deienno}, R.} \emph{et~al.}
\newblock \bibinfo{title}{{Excitation of a Primordial Cold Asteroid Belt as an
  Outcome of Planetary Instability}}.
\newblock \emph{\bibinfo{journal}{\apj}} \textbf{\bibinfo{volume}{864}},
  \bibinfo{pages}{50} (\bibinfo{year}{2018}).

\bibitem{brasseretal13b}
\bibinfo{author}{{Brasser}, R.}, \bibinfo{author}{{Walsh}, K.~J.} \&
  \bibinfo{author}{{Nesvorn{\'y}}, D.}
\newblock \bibinfo{title}{{Constraining the primordial orbits of the
  terrestrial planets}}.
\newblock \emph{\bibinfo{journal}{\mnras}} \textbf{\bibinfo{volume}{433}},
  \bibinfo{pages}{3417--3427} (\bibinfo{year}{2013}).

\bibitem{nesvornyetal21}
\bibinfo{author}{{Nesvorn{\'y}}, D.}, \bibinfo{author}{{Roig}, F.~V.} \&
  \bibinfo{author}{{Deienno}, R.}
\newblock \bibinfo{title}{{The Role of Early Giant-planet Instability in
  Terrestrial Planet Formation}}.
\newblock \emph{\bibinfo{journal}{\aj}} \textbf{\bibinfo{volume}{161}},
  \bibinfo{pages}{50} (\bibinfo{year}{2021}).

\bibitem{deiennoetal16}
\bibinfo{author}{Deienno, R.}, \bibinfo{author}{Gomes, R.~S.},
  \bibinfo{author}{Walsh, K.~J.}, \bibinfo{author}{Morbidelli, A.} \&
  \bibinfo{author}{Nesvorn{\'{y}}, D.}
\newblock \bibinfo{title}{{Is the Grand Tack model compatible with the orbital
  distribution of main belt asteroids?}}
\newblock \emph{\bibinfo{journal}{Icarus}} \textbf{\bibinfo{volume}{272}},
  \bibinfo{pages}{114--124} (\bibinfo{year}{2016}).

\bibitem{mahbrasseretal21}
\bibinfo{author}{{Mah}, J.} \& \bibinfo{author}{{Brasser}, R.}
\newblock \bibinfo{title}{{Isotopically distinct terrestrial planets via local
  accretion}}.
\newblock \emph{\bibinfo{journal}{\icarus}} \textbf{\bibinfo{volume}{354}},
  \bibinfo{pages}{114052} (\bibinfo{year}{2021}).

\end{thebibliography}

\end{document}